\documentclass[12pt,epsf,amssymb]{article}

\usepackage{graphicx}
\usepackage{amsfonts}
\usepackage{amsmath}
\usepackage[usenames]{color}
\usepackage{amssymb}
\usepackage[mathscr]{eucal} 
\usepackage{url}

\def\N{\mathbb{N}}
\def\Z{\mathbb{Z}}
\def\Q{\mathbb{Q}}
\def\R{\mathbb{R}}
\def\C{\mathbb{C}}
\def\P{\mathbb{P}}

\def\Im{\mathrm{Im}}

\usepackage{here}

\newcommand{\Ltil}{\widetilde{\Lambda}_S}

\renewcommand{\Im}{\operatorname{Im}} 

\usepackage{amsthm}
\theoremstyle{definition}

\begin{document}

\baselineskip 0.6cm
\newcommand{\vev}[1]{ \left\langle {#1} \right\rangle }
\newcommand{\bra}[1]{ \langle {#1} | }
\newcommand{\ket}[1]{ | {#1} \rangle }
\newcommand{\Dsl}{\mbox{\ooalign{\hfil/\hfil\crcr$D$}}}
\newcommand{\nequiv}{\mbox{\ooalign{\hfil/\hfil\crcr$\equiv$}}}
\newcommand{\nsupset}{\mbox{\ooalign{\hfil/\hfil\crcr$\supset$}}}
\newcommand{\nni}{\mbox{\ooalign{\hfil/\hfil\crcr$\ni$}}}
\newcommand{\nin}{\mbox{\ooalign{\hfil/\hfil\crcr$\in$}}}
\newcommand{\Slash}[1]{{\ooalign{\hfil/\hfil\crcr$#1$}}}
\newcommand{\EV}{ {\rm eV} }
\newcommand{\KEV}{ {\rm keV} }
\newcommand{\MEV}{ {\rm MeV} }
\newcommand{\GEV}{ {\rm GeV} }
\newcommand{\TEV}{ {\rm TeV} }

\def\diag{\mathop{\rm diag}\nolimits}
\def\tr{\mathop{\rm tr}}

\def\Spin{\mathop{\rm Spin}}
\def\SO{\mathop{\rm SO}}
\def\SU{\mathop{\rm SU}}
\def\U{\mathop{\rm U}}
\def\Sp{\mathop{\rm Sp}}
\def\SL{\mathop{\rm SL}}

\def\change#1#2{{\color{blue}#1}{\color{red} [#2]}\color{black}\hbox{}}

\begin{titlepage}
%
 
 \vskip 1cm
 \begin{center}
  
  {\large \bf Direct computation of period polynomials \\
and classification of K3-fibred Calabi--Yau threefolds}
 
 \vskip 1.2cm
  
 Yuichi Enoki, Yotaro Sato and Taizan Watari
  
 \vskip 0.4cm
  
  {\it 
    Kavli Institute for the Physics and Mathematics of the Universe (WPI), 
    University of Tokyo, Kashiwa-no-ha 5-1-5, 277-8583, Japan
   }
 
 \vskip 1.5cm
    
 \abstract{One can assign to four-dimensional $\mathcal{N}=2$ supersymmetric Heterotic string vacua a set of classification invariants including a lattice $\Lambda_S$ and vector-valued modular forms. Some of the classification invariants are constrained by the condition that the Coulomb branch monodromy matrices should be integer-valued. We computed numerically the period polynomials of meromorphic cusp forms for some rank-1 $\Lambda_S$; we then computed the monodromy matrices and extracted general patterns of the constraints on the invariants. The constraints we got imply that a large fraction of the Heterotic string vacua we studied satisfy the necessary conditions for a non-linear sigma model interpretation in the dual Type IIA description. Our computation can also be used to identify diffeomorphism classes of real six-dimensional manifolds that cannot be realized by K3-fibred Calabi--Yau threefolds.

} 
\end{center}
\end{titlepage}
 
\tableofcontents


\section{Introduction}
\label{sec:Intro}

In this article, we report a little progress in the classification 
of Heterotic--Type IIA dual vacua with $\SO(3,1)$ Lorentz 
symmetry and ${\cal N}=2$ supersymmetry. The best known 
example will be the Heterotic compactifications over 
K3$\times T^2$ with 24 instantons distributed within 
$E_8 \times E_8$, which are dual to the Type IIA compactifications  
over Calabi--Yau threefolds that have elliptic fibrations over 
Hirzebruch surfaces \cite{KV, MV-1, MV-2}. They are known to be 
very special cases of the duality correspondence 
(the cases with $\Lambda_S=U$ (see below)), however. For most of 
the branches of the Heterotic--Type IIA dual vacuum moduli space, 
10D supergravity approximation is not available in the Heterotic 
string description; already the ST-model treated in the 
original Ref. \cite{KV} was such an example. Although 
it is known that a Type IIA compactification over a 
Calabi--Yau threefold $M$ has a Heterotic dual when 
there is a K3-fibration $\pi: M \rightarrow \P^1$ \cite{Ferrara:1995yx, 
KLM, Vafa:1995gm},     
there remains a question how to identify a Heterotic--Type IIA 
dual vacuum that does not even have a 10D supergravity 
approximation in the Type IIA description. The classification 
problem requires an approach that does not rely on 
construction of geometries and vector bundles. 

To classify the branches of the Heterotic--IIA dual moduli space, 
we may assign invariants to the branches. 
The invariants that have been around already for a while 
are $(\widetilde{\Lambda}_S, \Lambda_T)$ and $\Phi = \{\Phi_\gamma \}$; 
the first one $(\widetilde{\Lambda}_S, \Lambda_T)$ is a pair of 
even lattices with signature $(2,\rho)$ and $(2,20-\rho)$, respectively, 
both of which are embedded primitively into the even self-dual 
lattice ${\rm II}_{4,20}$ in a way orthogonal to each other; 
the second one $\Phi = \{ \Phi_\gamma \; | \; \gamma \in G_S \}$, 
where $G_S := \widetilde{\Lambda}_S^\vee/\widetilde{\Lambda}_S$,  
is related to the Witten index on the Heterotic string worldsheet
and also to the number of points in the base $\P^1$ where the complex 
structure of the 
K3 fiber hits Noether--Lefschetz divisors in the Type IIA 
description \cite{MP, KMPS}. Those two invariants are not able to distinguish 
multiple Calabi--Yau threefolds in the Type IIA description, however 
(e.g., \cite{MV-2, HM, KKRS, BW-16}); Ref. \cite{EW19} therefore introduced 
one more invariant $\Psi = \{ \Psi_\gamma \; | \; \gamma \in G_S\}$ by 
generalizing what was discussed in \cite{HM} (see also references 
in \cite{EW19}).  
There are certain consistency conditions on the invariants $\Phi$ 
and $\Psi$ derived from string theory; they should be vector-valued modular 
forms of certain weights, and all of their Fourier coefficients have to 
be integers, to name a few; see \cite{EW19} for more. 
One may then change the perspective; think of the set of all the choices of 
the invariants $(\widetilde{\Lambda}_S, \Lambda_T)$, $\Phi$ and $\Psi$ 
 satisfying those understood conditions as the set of hypothetical 
vacuum branches/invariants, regardless of whether reliable 
constructions of the corresponding string vacua are already 
known in the literature. It was found \cite{ESW20} that the 
hypothetical $\Phi$ and $\Psi$ are not necessarily realized 
by truly consistent string vacua, because some of those 
hypothetical $\Phi$ and $\Psi$ lead to 4D particle spectra that have 
the Witten SU(2) anomaly. So, some consistency conditions on those 
invariants $(\widetilde{\Lambda}_S, \Lambda_T)$, $\Phi$ and $\Psi$ 
must be missing in the discussion of \cite{EW19}. 

Two of the present authors (Y.S. and T.W.) found in a discussion 
with Y. Tachikawa the following facts on the hypothetical vacuum branches 
with a non-vanishing Witten SU(2) anomaly: a perturbative duality 
transformation of a closed path in the Coulomb branch moduli space 
cannot be realized by an integer-valued symplectic transformation 
on the electric and magnetic charges; this observation is recorded as the 
appendix \ref{sec:Witten} of this article. 
Motivated by this observation, Ref. \cite{EW21} developed a 
method\footnote{
The algorithm in \cite{EW21} has been established only for the cases 
of $\widetilde{\Lambda}_S= U[-1]\oplus U$ and 
$\widetilde{\Lambda}_S = U[-1]\oplus \Lambda_S$ with 
$\rho := {\rm rank}(\Lambda_S)=1$. There may be a room for 
generalization in the choices of $\widetilde{\Lambda}_S$ in the future.  
}%
---by exploiting the observations 
in \cite{deWit, AFGNT, KapLusThs, AP}---of 
computing perturbative monodromy matrices directly from the hypothetical 
invariants 
$\widetilde{\Lambda}_S$, $\Phi$ and $\Psi$, without constructing a Type IIB 
mirror geometry. The method was then applied \cite{EW21} to the 
cases with $\widetilde{\Lambda}_S = U[-1] \oplus U$ and 
$U[-1] \oplus \vev{+2}$, and the condition that the monodromy 
matrices should be integer valued was translated into the form of extra 
conditions on the invariants $\Phi$ and $\Psi$.  

In this article, we apply the same method to the cases with 
\begin{align}
\widetilde{\Lambda}_S = U[-1]\oplus \vev{+2k}, \qquad \quad  
k=2,3,4,5,6, 11, 
\label{eq:cases-we-worked}
\end{align}
and obtain additional consistency conditions on $\Phi$ and $\Psi$ 
that were not appreciated in \cite{EW19}. 
The analysis also reveals that almost all the 
branches of the Heterotic--Type IIA dual vacua characterized 
by the lattice $\widetilde{\Lambda}_S$ in (\ref{eq:cases-we-worked}) 
satisfy a set of necessary conditions for their Type IIA description
to have a phase of non-linear sigma model (Calabi--Yau compactification). 
It is tempting to speculate that this remains true for all the dual 
vacua with $\widetilde{\Lambda}_S = U[-1] \oplus \Lambda_S$ when $\Lambda_S$ 
is some even lattice\footnote{
In the Type IIA description, if we know for sure that a vacuum branch 
in question has an interpretation as a Calabi--Yau compactification, 
then the lattice $U[-1]$ is interpreted as $H^0(K3;\Z) \oplus H^4(K3;\Z)$ 
of the fiber K3, and $\Lambda_S$ the lattice polarization of the K3-fibration 
$\pi: M \rightarrow \P^1$. 
} %
 with signature $(1,\rho-1)$. Our analysis also serves the purpose 
of narrowing down the set of diffeomorphism classes of real six-dimensional 
manifolds that can be realized as Calabi--Yau threefolds; this is because 
an algorithm is known how to read out the Chern classes and divisor 
intersection rings of the Type IIA Calabi--Yau threefolds from the 
invariants $\Lambda_S$, $\Phi$ and $\Psi$ (see \cite{HM, EW19} and 
references therein). 

The review parts for this article would be much the same as that 
of the preceding article \cite{EW21}, so we decided not to write 
a review from scratch for this article, but to treat the review part 
of Ref. \cite{EW21} also as that of this article. Only 
sections \ref{ssec:to-do-list}, \ref{ssec:MMS} and \ref{ssec:PQ} 
are included as a minimum reminder of, and addenda to, the review 
materials in \cite{EW21}. Section \ref{ssec:vac-id-grp} intends to 
remind ourselves of what we know for sure (and also what we are not 
sure about) on what is the perturbative duality transformation group 
denoted by $[\Gamma_S]$; as we will demand that an appropriate lift of 
the group $[\Gamma_S]$ should have a monodromy representation in the 
integer-valued symplectic transformation group, we cannot avoid asking what the group $[\Gamma_S]$
is. In section \ref{ssec:set-f*} and the appendix \ref{sec:grafting}, 
we will explain extra ideas that are necessary in computing 
the monodromy matrices for the cases with $\widetilde{\Lambda}_S$ 
in (\ref{eq:cases-we-worked}).  
Analysis is presented in sections \ref{sec:PP} 
and \ref{sec:classification}, and lessons are extracted in 
section \ref{sec:discussion}. For busy readers, it may be an option 
to skip sections \ref{ssec:MMS}--\ref{sec:classification} and 
jump to section \ref{sec:discussion}.

\section{The Set-up and the Methods}

\subsection{$\rho =1$ Heterotic--Type IIA Dual Vacua and What to Do}
\label{ssec:to-do-list}

In this article, we work on a class of Heterotic--Type IIA dual 
hypothetical vacua. Review materials in section 2 and 3.1 of \cite{EW21} 
by two of the present authors should be regarded also as a part of 
this article;\footnote{
Whenever we refer to an equation in \cite{EW21}, its equation number 
in JHEP is used. 
} %
 the same notations are maintained also in this article, 
and we often choose not to repeat their explanations here.
It is the class of hypothetical dual vacua characterized by 
$\rho =1$ that we address in this article. Then the lattices 
$\widetilde{\Lambda}_S$ and $\Lambda_T$ have the form of 
\begin{align}
  \widetilde{\Lambda}_S = U[-1] \oplus \vev{+2k}, \qquad 
  \Lambda_T = \vev{-2k} \oplus E_8^{\oplus 2} \oplus U^{\oplus 2}
\end{align}
for $k=2,3,\cdots $; the $k=1$ case has already been studied 
in \cite{EW21}.

Here, we write down what we should do, as a summary of the review 
materials in \cite{EW21}. First, one identifies the set of 
hypothetical classification invariants for branches of 
the Heterotic--Type IIA dual moduli space with the lattices 
$\widetilde{\Lambda}_S$ and $\Lambda_T$. The classification 
invariants consist of $\{ n_\gamma \; | \; \gamma \in G_S\}$, 
$\{ m_\gamma \; | \; \gamma \in G_S \}$, some more integers\footnote{
The rational number $\nu_{|\gamma|} = \nu_\gamma$ is the representative 
of $(\gamma, \gamma)/2 \in \Q/\Z$ of $\gamma \in G_S$ 
chosen in the range $[0, 1)$. 
} %
denoted by $d_\gamma(\nu_{|\gamma|})$ (or simply $d_\gamma$), and 
\begin{align}
b_{a=1} +24\Z \in \R /24\Z, \qquad a_{ab=11} + \Z \in \R/\Z . 
  \label{eq:BPSmass-inv}
\end{align}
A formal definition is given to the invariants $\{ m_\gamma \}$ and 
$d_\gamma$'s in \cite{EW19} in the language of Heterotic string 
worldsheet SCFT. We refer to the classification invariants 
$\{ n_\gamma \}$, $\{ m_\gamma \}$, and $d_\gamma$'s as the BPS classification 
invariants in this article. They are quantized (see \cite{EW19}), 
\begin{align}
n_\gamma \in \Z, \qquad m_\gamma, d_\gamma \in 12\Z, 
   \label{eq:BPS-classifctn-inv}
\end{align}
satisfy  
$n_\gamma = n_{-\gamma}$, $m_\gamma = m_{-\gamma}$, $d_\gamma = d_{-\gamma}$ 
(due to the CPT symmetry of the worldsheet SCFT), and are further subject 
to some linear relations and inequalities that can be worked out for 
each choice of a pair $(\widetilde{\Lambda}_S, \Lambda_T)$.
Interested readers might refer to \cite{EW19}, but it is straightforward 
to apply the argument in \cite{EW19}, so we will just state the result 
of such an analysis in the form of (\ref{eq:BPSpara-k=2}), 
(\ref{eq:BPSpara-k=3}) etc. in section \ref{sec:classification}. 
The invariants $b_{1}+24\Z$ and 
$a_{11}+\Z$, on the other hand, control the masses of 4D BPS particles; 
more comments on the invariants $b_1 + 24\Z$ and $a_{11}+\Z$ 
are found in the ``fourth'' paragraph below. 
 
The second task is to build a cusp meromorphic\footnote{
\label{fn:cusp}
In this article, we understand that there is justification 
for the behavior $f_*(t) \sim e^{2\pi i t}$, being 
exponentially small at $t \simeq i\infty$, but that we do not have 
an argument for how $f_*(t)$ behaves at other cusp points
(the boundary points added to $\Gamma_0(k)\backslash {\cal H}$
for its compactification).
Although the present authors are gaining confidence that 
$f_*$ vanishes exponentially at all the cusps of $\Gamma_0(k)$
\cite{EKSW}, we think of $f_*$ as a (meromorphic) ``{\it cusp}'' form 
in the sense that it vanishes exponentially at the one cusp 
$\Gamma_0(k)\cdot (+i\infty)$ of $\Gamma_0(k)$ in this article 
(strictly speaking, it is not appropriate to use the jargon ``cusp'' 
for such a thing).  This subtlety is relevant only at 
footnote \ref{fn:cusp-k=2} in this article. 
} %
 modular form $f_*(t)$ 
of weight-6 for $\Gamma_0(k)$ from the data $\{ n_\gamma \}$. 
The modular form $f_*$ is uniquely determined from $\{ n_\gamma \}$
as explained in (3.12), (2.45), (2.12) and (2.2) in \cite{EW21}. 
That is enough as a definition, but we will provide more practical information 
in section \ref{ssec:set-f*} and the appendix \ref{sec:grafting}
so that we can compute $f_*$ explicitly. 

Thirdly, we choose a set of generators $\{ g_i \}$ of the vacuum 
identification group $[\Gamma_S]$ (see section 2.2 of \cite{EW21}), 
and compute the period polynomials $P_{\tilde{g}_i}(t;f_*)$. As explained 
in \cite[\S3.1.3]{EW21}, there will be relations among the generators 
$\{g_i\}$, but we do not have to pay attention to those relations. 
In section \ref{ssec:vac-id-grp} in this article, we will make a brief 
comment on what the vacuum identification group $[\Gamma_S]$ will be . 

In the meantime (as the fourth task), one can determine the 4D effective 
Lagrangian parameters 
$d_{111}$, $(c_2)_1$ and $\chi_{\rm match}$ in terms of the 
BPS classification invariants $\{n_\gamma\}$, $\{ m_\gamma \}$, and $d_\gamma$.
The procedure of the computation is explained in \cite{EW19}, especially 
in section 3.2.2 and the appendix B.3 there. In the case-by-case studies 
presented in this article, we did follow this procedure of computation, 
and only the results are shown explicitly in the form of 
(\ref{eq:d111-k=2}--\ref{eq:chiMatch-k=2}), 
(\ref{eq:d111-k=3}--\ref{eq:chiMatch-k=3}) etc. 
 
The fifth step is to combine the result of the period polynomials
$P_{\tilde{g}_i}(t;f_*)$ and the parameters $b_1$, $a_{11}$, $d_{111}$, 
$\chi_{\rm match}$ in the 4D effective prepotential 
(see \cite[(2.9)]{EW21} for conventions) to write down the 
monodromy matrices $M_{\tilde{g}_i}$ (see \cite[(3.8) and (3.18)]{EW21}
for how to determine $M_{\tilde{g}_i}$). For a theoretically sensible branch 
of the Heterotic--IIA dual vacuum moduli space, all the matrix entries 
of $M_{\tilde{g}_i}$ should be integers. So, this condition imposes further
constraints among the BPS classification invariants, $b_1$ and $a_{11}$
(after rewrite $d_{111}$ and $\chi_{\rm match}$ in terms of the BPS 
classification invariants). The invariant $a_{11}+\Z$ is always determined 
uniquely in terms of the BPS classification invariants (as explained 
in \cite{EW21}; see also section \ref{ssec:PQ} here). 
In all the cases we have studied explicitly in this article 
(and also those in \cite{EW21}), $b_{1}+24\Z$ is also uniquely determined 
by the BPS classification invariants as we will see in section \ref{ssec:NLSM}. 
Further constraints among the BPS classification invariants are also 
obtained for some $k$ (i.e., for some lattice pairs 
$\widetilde{\Lambda}_S$ and $\Lambda_T$),
so the range of theoretically possible values of those invariants 
is reduced (i.e., some of the hypothetical branches are eliminated).  
One of those kinds of constraints turn out to be the solution to the 
issue raised in \cite{ESW20}, as we will explain in the 
appendix \ref{sec:Witten} of this article. 

Finally, we may test for the hypothetical vacuum branches with integral 
monodromy matrices whether they may contain a phase interpreted as 
a Calabi--Yau target non-linear sigma model (NLSM) in the Type IIA language;
see \cite[(2.14)--(2.17)]{EW21}. 
To be more explicit, we test two necessary conditions for such a phase
\cite{open-Z-basis}: one is 
\begin{align}
  (c_2)_{a=1} \equiv b_{a=1} \qquad {\rm mod~}+24\Z
 \label{eq:cond-geom-phase-1}
\end{align}
and the other 
\begin{align}
 a_{11} + \frac{d_{111}}{2} \in \Z. 
 \label{eq:cond-geom-phase-2}
\end{align}
The latter condition is always satisfied (see (\ref{eq:cond-PQ-2})
in section \ref{ssec:PQ}).
We will see that the former condition is also 
satisfied in all the cases we study in this article (see
section \ref{ssec:NLSM}), although we do not assume a priori that
a hypothetical branch of vacuum moduli space in question is given by
a Type IIA Calabi--Yau compactification.  
 
\subsection{Massless Matter Singularities}
\label{ssec:MMS}

This section \ref{ssec:MMS} elaborates a little more on 
what has been explained in \cite[\S2.2, \S3.1.1]{EW21}.  
That is partially just for the purpose of setting notations
for later use in this article. 

A 4D state with a purely electric charge (see \cite[\S2.1, \S2.2]{EW21} 
for notations) 
\[
v = (w_0, w_4,w_2) = w_0 e^0 + w_4 e^{\sharp} + w_2 e^1
     = -w_4 e_0 -w_0 e_\sharp + \frac{w_2}{2k} e_1  \in \widetilde{\Lambda}_S^\vee
\]
becomes massless at $X(v) := \{ t \in {\cal H} \; | \; (v, \mho(t)) = 0\}$, 
or equivalently, at $t=t_v$ (with ${\rm Im}(t_v)>0$) satisfying 
\begin{align}
  w_4 k (t_v)^2+w_2 t_v + w_0 = 0.
\end{align}
Such a 4D state\footnote{
We abuse notations and think of $\gamma$ both as an element 
$\gamma \in G_S = \widetilde{\Lambda}_S^\vee/\widetilde{\Lambda}_S$
and as a subset (=a coset) $\gamma \subset \widetilde{\Lambda}_S^\vee$.
} %
 with $v \in \gamma$ exists in the Heterotic string 
compactifications considered in this article if and only if
\cite{AFGNT}
\begin{align}
  -2 \leq (v,v) < 0.
\end{align}
So, for a given $\gamma$, there is a unique rational number 
$\nu_{|\gamma|} -1 \in [-1,0)$ satisfying $(v,v) = 2(\nu_{|\gamma|}-1)$; 
then $D_{|\gamma|}:= 4k(\nu_{|\gamma|} -1)=  2k(v,v)$ is the integer 
within the range $[-4k,0)$ determined uniquely by $\gamma$.  
To summarize, we may introduce a set of purely electric charges for 
a negative integer $D$, 
\begin{align}
  {\rm e.ch}_{\gamma,D} := \left\{ v \in \gamma \; | \; 2k (v,v)=D, \;\;
    w_0, w_4>0  \right\}, 
\end{align}
and use it to label the set of all the massless matter 
singularities\footnote{
Both of the references \cite{AFGNT} and \cite{HM} use the 
word ``enhanced 
symmetry points (ESPs)'' for $X(v)$, but both papers primarily 
worked on the cases where the lattice $\Lambda_S$ is self-dual. 
For a general lattice $\Lambda_S$, however, emergence of massless 
matter fields in the 4D field theory is not necessarily accompanied 
by enhancement of symmetry. So, we avoid using the word ESP here.
 
Reference \cite{HM} also refers to the work of \cite{Bor1} with the word 
``rational quadratic divisor (RQD)''. This notion includes all 
that are contained in ${\cal P}_{\gamma,D}$ with a general integer $D$.
So, massless matter singularities are a special class of RQDs; 
those with $D=D_{|\gamma|} \in [-4k,0)$. 

We do not mind using any one of the terms 
massless matter singularity/divisor/point, 
because the moduli space $D(\widetilde{\Lambda}_S)$ is 1-dimension when 
$\rho =1$. 
}
$\cup_{\gamma \in G_S} {\cal P}_{\gamma,D_{|\gamma|}}$, where 
\begin{align}
  {\cal P}_{\gamma,D_{|\gamma|}} :=
     \left\{ X(v) \; | \; v \in {\rm e.ch}_{\gamma,D_{|\gamma|}} \right\}
 =  \left\{ \left. 
       t_v := \frac{-w_2 + i \sqrt{|D_{|\gamma|}|}}{2k w_4}  \in {\cal H}
   \; \right| \; v \in {\rm e.ch}_{\gamma,D_{|\gamma|}} \right\}. 
\end{align}
Some of those codimension-1 massless matter singularities in ${\cal H} \subset 
D(\widetilde{\Lambda}_S)$ are counted multiple times, as it may happen that 
multiple mutually parallel charges in different $\gamma$'s may still 
satisfy the condition $-2\leq (v,v)<0$ (although the condition $w_0,w_4>0$
eliminates the obvious double counting $t_v = t_{-v}$).  

There are infinitely many charges in ${\rm e.ch}_{\gamma,D_{|\gamma|}}$, 
and infinitely many massless matter singularities in ${\cal H}$. 
The group $\Gamma_0(k)$ acts on the charges ${\rm e.ch}_{\gamma,D_{|\gamma|}}$, 
and also on the massless matter singularities. It is known that 
they organize themselves into only a finite number of $\Gamma_0(k)$ orbits. 
A complete classification of the orbits is found 
in \cite[\S I.1 (p.505), Prop.]{GKZ-II}. 

\subsection{The Vacuum Identification Group $[\Gamma_S]$}
\label{ssec:vac-id-grp}

The Heterotic string worldsheet SCFT varies\footnote{
The moduli of compactification also include 
the hypermultiplet moduli governing the SCFT in the 
$(c,\tilde{c}) = (22-\rho,0) + (0,6)$ sector, and also 
the axi-dilaton complex scalar. 
} %
 over the period 
domain $D(\widetilde{\Lambda}_S)$. There must be some group 
$[\Gamma_S]$ acting on $D(\widetilde{\Lambda}_S)$ and also on 
the set of electric charges $\Lambda_{\rm el} = \widetilde{\Lambda}_S^\vee$ 
in a way elements of $[\Gamma_S]$ are duality transformations 
identifying equivalent SCFTs on the Heterotic string worldsheet 
(see \cite[\S2.2]{EW21} for more review). 
Because we will demand for theoretical 
consistency that the monodromy matrices $M_{\tilde{g}}$ is $\Z$-valued 
for all the elements $g \in [\Gamma_S]$, we need to begin with what 
we are confident about the choice of the group $[\Gamma_S]$. 
In this section \ref{ssec:vac-id-grp}, we add a few discussions 
on top of what is reviewed in \cite[\S3.1.1]{EW21}. 

To (re)start, remember that the isometry group of the lattice 
$\widetilde{\Lambda}_S = U[-1] \oplus \Lambda_S$ (with $\rho =1$)
is of the form of ${\rm Isom}(\widetilde{\Lambda}_S) \cong 
\Gamma_0(k)_+ \rtimes 
(\Z_2\vev{-{\rm id}_U}\times \Z_2\vev{-{\rm id}_{\Lambda_S}})$; here,  
$-{\rm id}_U$ [resp. $-{\rm id}_{\Lambda_S}$] multiply $(-1)$ [resp. $1$] 
on $U[-1]$ and $1$ [resp. $(-1)$] on $\Lambda_S$. 
A $3\times 3$ matrix of the form 
\begin{align}
 & g = \left[ \begin{array}{cc|c} d^2/k_1 & c^2/k_2 & 2cd \\
      k_2 b^2 & k_1 a^2 & 2kab \\ \hline 
      bd/k_1 & ac/k_2 & (ad+bc) \end{array} \right] \longleftrightarrow 
    \pm \left( \begin{array}{cc} k_1 a & b \\ k_1 c & d \end{array} \right)
  \label{eq:convrt-3x3-and-2x2} \\
 & \qquad \qquad  ad-bc=1, \quad k_2|c, \quad k_1|d, 
   \quad {\rm gcd}(k_1,k_2)=1, \quad k_1k_2=k  \nonumber 
\end{align}
acts on the rank-2 $U[-1]$ and rank-1 $\Lambda_S$ 
as above as an isometry; the group of such $3\times 3$ matrices 
is denoted by $\Gamma_0(k)_+$ in this article (as well as in \cite{AP, EW21}). 
The group $\Gamma_0(k)_+$ contains a subgroup $\Gamma_0(k)/\{ \pm 1\}$, 
which consists of those with $k_1 = 1$ and $k_2 = k$. For a little more 
explanation, see \cite[\S3.1.1]{EW21}, or \cite[Thm. 7.1]{Dolgachev} 
and \cite{AP}.  

The group of vacuum identification $[\Gamma_S]$ is a subgroup 
of ${\rm Isom}'(\widetilde{\Lambda}_S)$, where
${\rm Isom}'(\widetilde{\Lambda}_S) \cong \Gamma_0(k)_+ \times
 \Z_2\vev{-{\rm id}}$ is the subgroup of ${\rm Isom}(\widetilde{\Lambda}_S)$
that maps one connected component (${\rm Im}(t)>0$ or ${\rm Im}(t)<0$) 
of $D(\widetilde{\Lambda}_S)$ to itself, and hence is possibly 
relevant to monodromy on the moduli space. Here, $-{\rm id}$ 
is $(-{\rm id}_U) \circ (-{\rm id}_{\Lambda_S})$, represented by 
the $-1_{3\times 3}$ matrix on $\widetilde{\Lambda}_S$. We know for sure 
that the group\footnote{
We have slightly changed the definition of $\Gamma_S$ 
from \cite[(2.22)]{EW21}. 
} %
\begin{align}
 \Gamma_S := {\rm Ker}\left[ {\rm Isom}'(\widetilde{\Lambda}_S) \longrightarrow 
    {\rm Isom}(G_S,q_s) \right] 
\end{align}
is contained in $[\Gamma_S]$; for $g$ in the subgroup $\Gamma_S$, 
the duality between a pair of Heterotic string SCFTs is purely 
within the lattice SCFT with the central charge $(c,\tilde{c}) 
= (\rho, 3)$; all the other sectors in the SCFTs just have to be 
tensored after the duality transformation. 

First, we claim (and verify in the following) that the group 
$\Gamma_S \subset {\rm Isom}'(\widetilde{\Lambda}_S)$ is of the form 
\begin{align}
 \Gamma_S = (\Gamma_0(k)/\{ \pm 1\})\rtimes \Z_2\vev{w^{(k)}}, 
    \label{eq:Gamma_S}
\end{align}
where the element $w^{(k)}$ of order two is the $3 \times 3$ matrix 
\begin{align}
 w^{(k)} := \left[ \begin{array}{cc|c} & -1 & \\ -1 & & \\ \hline
              & & {\rm id}_{\Lambda_S} \end{array} \right]  
     \in {\rm Isom}'(\widetilde{\Lambda}_S). 
   \label{eq:def-w-k-Fricke}
\end{align}

To see this, we begin with a statement\footnote{
Both $\Z_n$ and $\Z/(n\Z)$ denote the cyclic group of $n$ elements. 
} %
 (to be verified below) 
that the group $\Gamma_0(k)_+$ acts on the group $G_S = \Z/(2k\Z)$ 
only through its quotient $\Gamma_0(k)_+ \rightarrow (\Z_2)^s$ 
(see \cite[(3.3)]{EW21}); moreover, the element $\epsilon = (\epsilon_i)$ 
in $(\Z_2)^s = \{ (\epsilon_i)_{i=1,\cdots, s} \; |
 \; \epsilon_i \in \{ 0,1\}\cong \Z_2  \}$ 
acts on the abelian group $G_S = \oplus_j \Z/(p_j^{m_j'}\Z)$ as the 
multiplication of $\oplus_i ((-1)^{\epsilon_i})$; here, 
$k = \prod_i p_i^{m_i}$ and $2k = \prod_j p_j^{m_j' }$ are the prime 
factorizations; when $k$ is odd, $\epsilon = (\epsilon_i) \in (\Z_2)^s$ acts 
trivially on the $\Z_2$ factor for $p_j=2$ in
 $G_S \cong \oplus_j \Z/(p_j^{m'_j}\Z)$.   
This statement is verified by the following observations. 
The action on $G_S$ is trivial on the subgroup $\Gamma_0(k)/\{ \pm 1\}$
because the (3, 3) entry of the matrix 
in (\ref{eq:convrt-3x3-and-2x2})=\cite[(3.5)]{EW21} is 
$ad+bc = ad-bc+2bc=1+2bc \equiv 1$ mod $2k$ for $g \in \Gamma_0(k)/\{ \pm 1\}$. 
Next, the action of $g \in \Gamma_0(k)_+$ in the $\Gamma_0(k)/\{ \pm 1\}$ orbit 
$(\epsilon_i) \in (\Z_2)^s$ is determined by evaluating the multiplication 
of $x:= (ad+bc)$ of a representative $g$ on $G_S \cong \oplus_j\Z/(p_j^{m'_j}\Z)$. 
The relation $ad-bc=1$ along with $k_1|d$ and $k_2|c$ implies that 
\begin{itemize}
        \item $x \equiv -1 \pmod {p_i^{m'_i}}$ if $p_i^{m_i} \mid k_1$, 
        \item $x \equiv +1 \pmod {p_i^{m'_i}}$ if $p_i^{m_i} \mid k_2$, and
        \item for an odd $k$, $x \equiv +1 \pmod 2$. 
\end{itemize}
Because 
\begin{align}
p_i^{m_i} | k_1 \Longleftrightarrow \epsilon_i = 1, \quad {\rm and} \quad 
p_i^{m_i} | k_2 \Longleftrightarrow \epsilon_i = 0, 
  \label{eq:AL-Z2-dict}
\end{align}
the statement at the beginning of this paragraph is verified now. 

The action of ${\rm Isom}'(\widetilde{\Lambda}_S)$ on $G_S$ therefore 
factors through the group $(\Z_2)^s \rtimes \Z_2\vev{-{\rm id}}$
(see \cite[(3.6)]{EW21}). 
Having seen how the group $(\Z_2)^s$ acts on $G_S$, now it is obvious 
that the subgroup $\Gamma_S \subset {\rm Isom}'(\widetilde{\Lambda}_S)$
is $(\Gamma_0(k)/\{ \pm 1\})\rtimes \Z_2\vev{w^{(k)}}$; the element 
$w^{(k)}$ corresponds to $((\epsilon_i =1), -{\rm id})$. 
The generator $w^{(k)}$ acts\footnote{
$w^{(k)}$ acts on the electric charge $v=(w_0,w_4,w_2) \in
 \widetilde{\Lambda}_S^\vee$ from the right, 
$v \mapsto v \cdot w^{(k)}$, and on the moduli space coordinate $t$
from the left, $\mho(t) \longmapsto w^{(k)} \cdot \mho(t) =_{\rm proj} 
\mho(t^{w^{(k)}})$.
} %
 on the connected component 
${\cal H} \subset D(\widetilde{\Lambda}_S)$ as a map $t \longmapsto -1/(kt)$. 
We will abuse notations and terminology and use $w^{(k)}$ to denote 
both the actions on the set of charges and the moduli space, and also 
use the term Fricke involution to refer to both. It will be useful 
later to use the group 
\begin{align}
  {\rm Isom}'(\widetilde{\Lambda}_S) / \; (\Gamma_0(k)/\{ \pm 1\} )
   \cong (\Z_2)^s \times \Z_2\vev{-{\rm id}} 
   \label{eq:qot-grp-fint}
\end{align}
in describing how large the groups $\Gamma_S$ and $[\Gamma_S]$ can be 
within ${\rm Isom}'(\widetilde{\Lambda}_S)$. We have seen that 
$\Gamma_S$ corresponds to $\{ ((0^s),0), \; ((1^s),1) \}$ in this group 
(\ref{eq:qot-grp-fint}). 

Second, there are cases where we can argue that some other Atkin--Lehner 
involutions (elements of $(\Z_2)^s$, possibly accompanied by $-{\rm id}$) 
are a part of $[\Gamma_S]$. That is when there is 
a W-boson whose charge $v_* \in \gamma_* \in G_S$ satisfies\footnote{
The BPS invariant $n_{\gamma_*}$ has the form of 
$n_{\gamma_*} = n^H_{\gamma_*} -2 n^V_{\gamma_*}$; $n^V_{\gamma_*}=+1$ when 
the vertex operator for the 4D W-boson is in the 
Heterotic string worldsheet SCFT, and $n^V_{\gamma_*}=0$ otherwise \cite{HM}. 
Although $n^V_{\gamma*}$ and $n^H_{\gamma*}$ vary as hypermultiplet moduli 
vary, but the combination $n_{\gamma*}$ does not within a branch. 
} %
the conditions (cf \cite[\S2]{EW19})  
\begin{align}
 (v_*, v_*) = -\frac{2}{\ell_{KM}}, \quad \ell_{KM} v_* =0 \in G_S 
 \label{eq:reflection_conditions}
\end{align}
for a positive integer $\ell_{KM}$, and $n^V_{\gamma_*} = 1$; 
at the subvariety $X(v_*) = \left\{ t \in {\cal H} \; | \; 
(v_*, \mho(t))=0 \right\}$, there is a pair of Heterotic string 
vertex operators with charge $\pm v_*$ that forms a part of 
the level-$\ell_{KM}$ SU(2) current algebra. In this case, 
we have the reflection $R_{v_*}$ on the lattice $\Ltil^\vee$:
\begin{align}
    R_{v_*}: \widetilde{\Lambda}_S^\vee \ni v 
       \longmapsto v + (v, \ell_{KM} v_*) v_*.
\end{align}
Notice that the conditions \eqref{eq:reflection_conditions} imply that 
$R_{v_*}(v_*) = -v_*$ and $R_{v_*}(v) \in \Ltil^\vee$ for $v \in \Ltil^\vee$.
It is an involution (i.e., $(R_{v_*})^2 = {\rm id} \in 
{\rm Isom}(\widetilde{\Lambda}_S)$) on the charges, and also acts 
on the complex-codimension-1 neighbor of the massless SU(2) locus 
$X(v_*)$ as the Weyl reflection. 

We argue that this $R_{v_*}$ is in $[\Gamma_S]$ as follows, using the 
language of Heterotic string worldsheet CFT.\footnote{
In the language of the 4D field theory, the argument is simple. 
$R_{v_*}$ generates a finite subgroup in the SU(2) symmetry of 
the SU(2) gauge theory, and the gauge symmetry maps one field 
theory description to another that is equivalent in physics. 
} %
 At a vacuum on $X(v_*)$, 
one may use the zero-mode part of the SU(2) current algebra to 
infer a global SU(2) symmetry of the CFT. The reflection $R_{v_*}$ is 
an element within the global SU(2) symmetry group. As the vacuum 
moduli move away from $X(v_*)$, theoretical consistency (such as 
the modular invariance) of a CFT is maintained when the 
spectrum and the OPE change only in the $(c,\tilde{c})=(\rho,3)$
sector. 
In such SCFTs parametrized by $D(\widetilde{\Lambda}_S)$, $R_{v_*}$
remains to map one consistent SCFT to another that is equivalent 
to the the original SCFT. 

The reflection symmetry $R_{v_*} \in [\Gamma_S]\subset 
{\rm Isom}'(\widetilde{\Lambda}_S)$ can be classified modulo 
$\Gamma_0(k)/\{\pm 1\}$, as an element of (\ref{eq:qot-grp-fint}), 
just like we have done 
for $w^{(k)} \in \Gamma_S$ already. From this information, 
the action of $R_{v_*}$ on $G_S$ is easily extracted.   
Note, first, that $-R_{v_*} \in \Gamma_0(k)_+ \subset 
{\rm Isom}'(\widetilde{\Lambda}_S)$ (rather than $R_{v_*}$ 
is).\footnote{
\label{fn:reflectn-outof-Gamma0kP}
To see this, compare the $3\times 3$ matrix representation 
of the elements of $\Gamma_0(k)_+$ (\ref{eq:convrt-3x3-and-2x2}) 
and of $R_{v_*}$. The (1, 2) entry of any element 
of $\Gamma_0(k)_+$---$c^2/k_2$---is positive. On the other hand, 
the (1, 2) entry of the reflection $R_{v*}$ is $-\ell_{KM} n_0n_0 =
 -\ell_{KM} w_4w_4 < 0$. So, $-R_{v_*} \in \Gamma_0(k)_+$.  
} %
Now the remaining question is to identify the element 
$(\epsilon_i \in \{0,1\}) \in (\Z_2)^s$, or equivalently the pair $(k_1,k_2)$, 
corresponding to $-R_{v_*}$.
Now we claim that $v_*$ is associated in this way with $k_1 = 4k/\ell_{KM}$ 
when $w_{*2}$ is odd, and $k_1 = k/\ell_{KM}$ when $w_{*2}$ is even---(*). 
In particular, W-boson charges $v_* \in \widetilde{\Lambda}_S$
 ($\gamma = 0 \in G_S$) are always associated with $k_1=k$
(like the Fricke involution). 

To derive (*), note first that the 
conditions \eqref{eq:reflection_conditions} can be rewritten as 
 \begin{align}
     4kw_{*0} w_{*4} - w_{*2}^2 = \frac{4k}{\ell_{KM}},
     \qquad
    m := \frac{w_{*2}\ell_{KM}}{2k} \in \Z, 
\end{align}
and also that 
\begin{align}
 R_{v_*}: t \longmapsto
      \frac{(R_{v_*}\cdot \mho(t),e^1)}{(R_{v_*}\cdot \mho(t),e^0)}
   = - \frac{w_{*2}t + 2w_{*0}}{2kw_{*4}t+w_{*2}}. 
  \label{eq:reflection-map}
\end{align}
\label{pg:AL-k1k2}
%
%
{\bf When $w_{*2}$ is odd, } it follows that $4k/\ell_{KM}$ is an odd integer, 
so $\ell_{KM}$ must be divisible by 4. From the relation 
$2 w_{*2} = (4k/\ell_{KM}) m$, we find that $m$ is even. Then the equation
\begin{align}
    \ell_{KM} w_{*0} w_{*4} - \frac{4k}{\ell_{KM}} \left(\frac{m}{2}\right)^2 = 1
\end{align}
implies $\gcd(\ell_{KM}, 4k/\ell_{KM}) = 1$. So, the integers $(\ell_{KM}/4)$ 
and $(4k/\ell_{KM})$ are mutually prime whose product is $k$. 
The map (\ref{eq:reflection-map}) image is seen as 
\begin{align}
  \frac{-(4k/\ell_{KM})(m/2) t - 2w_{*0}}{k(2w_{*4})t+(4k/\ell_{KM})(m/2)} =: 
   \frac{k_1a t + b}{k_1 (k_2c') t + (k_1d')}, \qquad 
          a(k_1d') - b (k_2c') = 1,  
\end{align}
with $k_1 = (4k/\ell_{KM})$ and $k_2 = \ell_{KM}/4$. 
{\bf When $w_{*2}$ is even, } we see that $k/\ell_{KM} \in \Z$. Then the equation
\begin{align}
    \ell_{KM} w_{*0} w_{*4} - \frac{k}{\ell_{KM}} m^2 = 1
\end{align}
implies that the integers $\ell_{KM}$ and $k/\ell_{KM}$ are mutually prime 
and are 
multiplied to be $k$. The map (\ref{eq:reflection-map}) image is regarded as 
\begin{align}
  \frac{-(k/\ell_{KM})m t - w_{*0}}{k(w_{*4})t+(k/\ell_{KM})m} =: 
   \frac{k_1a t + b}{k_1(k_2c') t + (k_1 d')}, \qquad 
       a(k_1d') - b (k_2c') = 1,  
\end{align}
with $k_1=k/\ell_{KM}$ and $k_2 = \ell_{KM}$.   
\begin{table}[tbp]
\begin{center}
\begin{tabular}{c|c|ccccccccccccc}
 \hline 
       $k$ & any & 2 & 3 & 5 & 6 & 6 & 7 & 10 & 10 & 11 & 12 & 13 & 14 & 14 \\
 \hline
$[w_{2*}]$  & 0 & 2 & 3 & $\pm4$ & $\pm 4$ & 6 & 7 & $\pm 6$ & 10 & 11 &
      $\pm 6$ & $\pm 10$ & $\pm 7$ & 14 \\
$\ell_{KM}$ & 1 & 2 & 4 & 5 & 3 & 2 & 4 & 10 & 2 & 4 & 
      4 & 13 & 8 & 2 \\
 \hline
     $k_1$ & $k$ & 1 & 3 & 1 & 2 & 3 & 7 & 1 & 5 & 11 & 
      3 & 1 & 7 & 7 \\
     $k_2$ & 1 & 2 & 1 & 5 & 3 & 2 & 1 & 10 & 2 & 1 & 
      4 & 13 & 2 & 2 \\
$\epsilon$ & $(1^s)$ & 0 & 1 & 0 & (1,0) & (0,1) & 1 & (0,0) & (0,1) & 1 & 
      (0,1) & 0 & (0,1) & (0,1) \\
  \hline 
\end{tabular}
\caption{\label{tab:v*-list}
The list of vacuum branches (with small $k$) 
where an $\SU(2)$ gauge group may be enhanced in the moduli space. 
The top three rows are the same as \cite[Tab. 1]{EW19}, and 
the information in the bottom three rows are determined by following 
the statement (*) and the dictionary (\ref{eq:AL-Z2-dict}) in the 
main text. The last column corresponds to the degree-28 example 
discussed in \cite{ESW20}; the second to last column refers to 
the case $n^V_{\gamma = [7]} = +1$ (without an SU(2) doublet matter) 
whereas the last column to the case $n^V_{\gamma = [14]} = +1$; 
the W-bosons of those two kinds should not coexist.
}
\end{center}
\end{table}


To summarize, we have seen so far that $((\epsilon_i=1), -{\rm id}) \in 
(\Z_2)^s \times \Z_2\vev{-{\rm id}}$ represented by $w^{(k)} \in \Gamma_S$
is guaranteed to be in $[\Gamma_S]$, and besides that, $[\Gamma_S]$
contains $((\epsilon_i), -{\rm id})$ represented by $R_{v_*}$ when 
there is a W-boson with the charge $v_*$. For all other elements 
in the group $(\Z_2)^s \times \Z_2\vev{-{\rm id}}$, we do not have 
a general argument whether they are within $[\Gamma_S]$ or not.
One might have to look into a more detailed information of SCFTs,
%
%
%
%
for example, to determine how large fraction of the group 
$(\Z_2)^s \times \Z_2\vev{-{\rm id}}$ corresponds to the vacuum 
identification group $[\Gamma_S]$. An alternative is an experimental 
approach; we can study how different choices of $[\Gamma_S]$ 
affect theoretically possible choices of the classification invariants. 
This alternative is what we will do in sections \ref{ssec:PP-AL} 
and \ref{sssec:level-6} in this article. 


\subsection{Integrality Conditions from the Peccei--Quinn Symmetry}
\label{ssec:PQ}

For all the $\rho =1$ lattices, the group $\Gamma_0(k)/\{ \pm 1\} 
\subset [\Gamma_S]$ contains the $T$-transformation, 
$g_\infty: t^{a=1} \mapsto t^{a=1}+1$. In the context of string theory, 
it is called the Peccei--Quinn symmetry. The integrality condition of 
each component of the matrix $M_{\tilde{g}_\infty}$ is read out immediately 
(cf \cite{ESW20}); that is \cite[(3.28)]{EW21}, or to write down here again, 
\begin{align}
 d_{111} \in \Z, \qquad 2d_{111}+b_1 \in 12\Z,
  \label{eq:cond-PQ-1}
\end{align}
and 
\begin{align}
 a_{11} \in \frac{d_{111}}{2} + \Z.
  \label{eq:cond-PQ-2}
\end{align}
As promised earlier (``the fifth step'' in section \ref{ssec:to-do-list}), 
the invariant $a_{11}+\Z \in \R/\Z$ is therefore completely fixed 
by the consistency 
condition (\ref{eq:cond-PQ-2}). Furthermore, the condition 
(\ref{eq:cond-PQ-2}) guarantees that one of the necessary conditions 
(\ref{eq:cond-geom-phase-2}) for a non-linear-sigma-model (NLSM) 
interpretation always holds true in this class of compactifications. 
 
This section \ref{ssec:PQ} does not add anything new beyond 
\cite{EW21}, but we included this because we will refer to  
eq. (\ref{eq:cond-PQ-1}) more than a few times in this article.

\subsection{Period Polynomials of a Meromorphic Cuspform}
\label{ssec:set-f*}

Section 3.1.3 of \cite{EW21} has already explained how the 
period polynomial $P_{\tilde{g}}(t;f_*)$ is defined for a weight-6 
meromorphic cusp form $f_*$ for $\Gamma_0(k)$, and how 
the result is used to compute the monodromy matrices $M_{\tilde{g}_i}$.  
We still have to add a few more things here in order to be able to study the 
$\rho =1$ cases with various values of $k=2,3,\cdots$. 

\subsubsection{Numerical Evaluation of the Period Polynomials}
\label{sssec:Pade}

In the case of $k=1$ studied in \cite{EW21}, we may choose a set 
of generators of the group $\Gamma_0(1) = {\rm SL}(2,\Z)$ that consists 
of the two elements: the $T$-transformation and the $S$-transformation. 
The period polynomial for $T$ for a cuspform is always zero; 
that has already been used implicitly in section \ref{ssec:PQ}.  
So, practically it is enough to compute the period polynomial of just 
one element (the $S$-transformation) in $[\Gamma_S]$ to make sure 
that all the monodromy matrices $M_{\tilde{g}_i}$ are $\Z$-valued 
for all $g \in [\Gamma_S]$. That is not true, in general, for a general 
$k$. 

We may use SAGE \cite{SAGE}, for example, to find out a set of generators 
of $\Gamma_0(k)$. It is enough to compute the period 
polynomials for 
such generators, along with those of the Fricke involution $w^{(k)}$, 
and also $R_{v_*}$ where there is $v_* \in \widetilde{\Lambda}_S^\vee$
satisfying the conditions (\ref{eq:reflection_conditions}). 

The period polynomial $P_{\tilde{g}}(t, f_*)$ is defined for $g= \pm 
\left(\begin{array}{cc}
    a & b \\
    c & d
\end{array}\right) \in \Gamma_0(k)_+$ and a meromorphic 
function $f_*$ on the upper half plane ${\cal H}$ satisfying 
$f_*|[g]_6 = f_*$ and vanishing quickly at $t \simeq i\infty$ and 
$t=g^{-1}(+i\infty)$. It is given\footnote{
We use a notation for a general element of $\Gamma_0(k)_+$ 
in this paragraph; $a$ and $c$ in the $2 \times 2$ matrix here 
correspond to $k_1a$ and $k_1c$ 
elsewhere in this article, and also in \cite[(3.4)]{EW21}. 
This is to simplify some of the expressions appearing in this paragraph.
}\raisebox{4pt}{,}\footnote{
\label{fn:on(-id)}
Although it was not stated clearly or emphasized in \cite{EW21}, 
when $g \in [\Gamma_S] \subset {\rm Isom}'(\widetilde{\Lambda}_S)$
but $g \nin \Gamma_0(k)_+$, a monodromy matrix $M_{\tilde{g}}$ should 
be determined as follows. First, when $g \nin \Gamma_0(k)_+$, then 
$(-{\rm id})\cdot g$ is in $\Gamma_0(k)_+$. Choose a path 
$\gamma_{\tilde{g}}$ in the moduli space $D(\widetilde{\Lambda}_S)$
connecting the base point $t_0$ and $t_0^g = t_0^{(-{\rm id})\cdot g}$, 
and compute the matrix $M_{-\tilde{g}} := M((-{\rm id}_{3\times 3})\cdot g, 
\Lambda_{\tilde{g}})$ by 
following the prescription in \cite{EW21}. The monodromy 
matrix $M_{\tilde{g}}$ for the path $\gamma_{\tilde{g}}$ is then 
$(-1)_{6 \times 6} M_{-\tilde{g}}$. 

The prescription above is fine essentially because the 
monodromy matrix is determined by the analytic continuation 
along paths in the moduli space; the analytic continuation 
along a path from $t_0$ to $t_0^{g} = t_0^{-g}$ can be computed 
for an element $(-{\rm id}_{3\times 3})\cdot g \in \Gamma_0(k)_+$, so 
$M_{\tilde{g}} =_{\rm proj} M_{-\tilde{g}} = - M_{-\tilde{g}}$. 
So, it is enough to be able to compute period polynomials 
for elements in $\Gamma_0(k)_+$. 
} %
 by an integral from $g^{-1}(+i\infty)$ to $+i\infty$, 
as we choose $+i\infty := \lim_{t_{0,{\rm Im}} \rightarrow +\infty}(it_{0,{\rm Im}})$ 
as the base point of the Eichler integral. The integration 
from a cusp $(\sigma = -d/c)$ to another cusp $\sigma = +i\infty$ 
may be split into two segments, 
\begin{align}
 P_{\tilde{g}}& (t, f_*) = \frac{(2 \pi i)^5}{4!}
       \int_{g^{-1}(i\infty)}^{i\infty} d\sigma f_*(\sigma) (t-\sigma)^4
           \label{eq:PP-split-not-yet} \\
  & = \frac{(2 \pi i)^5}{4!} \int_{z}^{i\infty} d\sigma f_*(\sigma) (t-\sigma)^4
    - \frac{(2 \pi i)^5}{4!} \int_{g(z)}^{i\infty} d\sigma f_*(\sigma)
                  \frac{((-c\sigma+a)t-(d \sigma-b))^4}{(ad-bc)^2}.
   \label{eq:PP-split-to-2}
\end{align}
We choose $z = (-d + i\sqrt{ad-bc} )/c$
(which means that $\Im z = \Im g(z) = \sqrt{ad-bc}/c =: y_{\mathrm{min}}$), 
so that the integration contour can remain entirely within the region 
${\rm Im}(\sigma)\geq y_{\mathrm{min}}$ in the upper half plane ${\cal H}$. 
To be specific, we fix the (topology of the) integration contour 
as follows: in the first [resp. second] term of (\ref{eq:PP-split-to-2}), 
the contour is a half line parallel to the imaginary axis from $z$ to 
$z+i\infty$ [resp. $g(z)$ to $g(z) + i\infty$], followed by 
the segment from $z+i \infty$ to $+i\infty$ [resp. from 
$g(z)+i\infty$ to $+i\infty$]; the additional segment yields zero 
contribution to the period polynomial, however. 
If there are some poles of $f_*$ on those lines, we shift $z \to z + \epsilon$ 
by a small positive constant $\epsilon$.
This prescription specifies the integration contour 
$g^{-1}(\tilde{\gamma}_{\tilde{g}})$ 
in (\ref{eq:PP-split-not-yet})=\cite[(3.17)]{EW21}, and the element 
$\tilde{g}$ of the perturbative duality group after the 
lift (cf \cite[\S2.2]{EW21}).
%
%

{\bf A (Variation of the) Pad\'e Approximation}

The prescription above allows us to evaluate the period polynomials 
of a meromorphic cusp form $f_*(t)$, even when we know $f_*(t)$ 
only as a power series\footnote{
An analytic (all order) expression for $f_*$ is known 
in the $k=1$ case (\cite{KapLusThs} and \cite[(3.46)]{EW21}), and also 
in the $k=2$ case with $n_1=n_2=0$ \cite{AP}, 
quoted as (\ref{eq:f*-AP95-k=2-n12=0}) in this article.  
} %
in $q := e^{2\pi i t}$. As we will explain in 
section \ref{sssec:build-f*}, it is possible in all the $\rho=1$ cases 
(i.e., $k=1,2,\cdots$) to determine $f_*$ as such a power series for 
arbitrary choice of the BPS classification invariants $\{n_\gamma \}$ 
without an obstruction (indicated e.g. by (\ref{eq:lin-rltn-k=11})) 
up to any power, limited only by the computational resources. 

The modular form $f_*(t)$ is meromorphic, i.e., has a pole in the 
interior of the upper halfplane ${\cal H}$. Massless matter singularities 
$X(v)$ characterized by $(v,\mho(t))=0$ give rise to the logarithmic 
singularity in the prepotential, and a pole of order three in $f_*$
\cite{KapLusThs}. The power series expansion of $f_*$ truncated 
at a finite order becomes 
useless at $|q|\simeq |q_{\rm pole}|$, where $q_{\rm pole}$ is the value of $q$
at a massless matter singularity. As a way out of this issue, we 
use a variation of the Pad\'e approximation. 

The idea is to replace the power series truncated at the $q^n$ term, 
denoted by $[f_*(t)]_n$, by the following rational function of $q$, 
\begin{align}
   [f_*(t)]_n \rightarrow f_*^{\rm Pade}(t) :=
   \frac{[[f_*(q)]_n f_*^{\mathrm{denm}}(q)]_n}{[f_*^{\mathrm{denm}}(q)]_{n}};
  \label{eq:Pade-general}
\end{align}
suppose $f_*^{\mathrm{denm}}(q)$ has no poles but some zeros of order 3 
(or more) and reproduces the poles of $f_*$ at least in the region 
$\Im t > y_{\mathrm{min}} \delta_y $ with some $0 < \delta_y < 1$;
set the numerator to be the power series 
$[f_*(t)]_n f_*^{\mathrm{denm}}(q)$ truncated at $q^n$; then the 
singularity of $f_*(t)$ is captured properly by the denominator 
$[f_*^{\mathrm{denm}}(q)]_{n}$ of $f_*^{\rm Pade}$ in that region, the 
power series expansion of $f_*^{\rm Pade}$ agrees with that of $[f_*]_n$, and 
yet $f_*^{\rm Pade}(t)$ has a better-controlled analytic continuation 
in the region with ${\rm Im}(t)$ smaller than the largest imaginary 
part of the poles.  One simple choice\footnote{
In a widely used version of the Pad\'e approximation, the coefficients 
of both the denominator polynomial and the numerator polynomial are 
treated as variables in fitting the original truncated power series 
on the left hand side. Here, we know where the poles are, so only 
the coefficients in the numerator are treated as the variables in fitting 
the original truncated power series $[f_*(t)]_n$.
} %
 for $f_*^{\mathrm{denm}}(q)$ is 
\begin{align}
    f_*^{\mathrm{denm}}(q) = \prod_j (q-e^{2\pi i t_j})^3,
   \label{eq:poles-4-denm}
\end{align}
where $t_j$'s are the finite number of massless matter singularities 
 with ${\rm Im}(t_j) > y_{\rm min} \delta_y$ modulo the relation 
$t_j \sim t_j' =t_j+1$. 

\subsubsection{How to Compute $f_*$ in Practice}
\label{sssec:build-f*}

Once the weight-(21/2) holomorphic vector-valued modular form 
$\Phi = \{ \Phi_\gamma(\tau) \}$ is   
determined for the classification invariants $\{ n_\gamma \}$, 
then it is straightforward to determine $f_*(t)$ as a power 
series in $q = e^{2\pi i t}$ \cite{HM} (see \cite{EW19} for other 
references):\footnote{
In this article, elements $\gamma \in G_S = \Z/(2k\Z) \cong 
\{ [i/(2k) e_1] \; | \; i=1-k,\cdots, -1,0,1,\cdots, k\}$ may also 
be denoted by $[i]$, or sometimes $i$. For example, $n_i$ stands for 
$n_{[i]} = n_{[e_1i/(2k)]}$. 
} %
\begin{align}
 f_* (t) = \frac{1}{(2\pi i)^3} \sum_{d=1}^\infty 
   d^5 c_{[d]}(d^2/(2k))  {\rm Li}_{-2}(q^d)
\end{align}
in any $\rho =1$ cases with $k=1,2,\cdots, \in \N$. 
This power series truncated at finite order can be used 
as the input $[f_*(t)]_n$ in (\ref{eq:Pade-general}). 

It is only in the $k=1,2,3$ cases, however, that independent generators 
of the vector space of $\Phi$'s are known explicitly in terms 
of the Eisenstein series and theta series \cite{KKRS, MP}. 

Rademacher expansion can be used (see \cite{Dijkgraaf:2000fq,
 deBoer:2006vg, Manschot:2007ha} and references therein) 
to determine the higher order 
terms of the vector-valued modular form\footnote{
As a reminder, $F=\Phi/\eta^{24}$, $F_\gamma = \Phi_\gamma/\eta^{24}$, 
so it is of weight-$(-3/2)$, and has a pole at the cusp. 
} %
\begin{align}
 F_\gamma = n_\gamma e^{2\pi i \tau (\nu_{|\gamma|}-1)} + \sum_{n=0,1,\cdots}^\infty 
   c_\gamma(\nu_{|\gamma|}+n)e^{2\pi i \tau (\nu_{|\gamma|}+n)}
\end{align}
for a given\footnote{
We mean by $\{ n_\gamma \}$ that satisfy appropriate linear 
relations (e.g. see \cite{EW19}) such that the obstruction 
for the Rademacher expansion (e.g., \cite{Manschot:2007ha}) 
vanishes.
} %
 $\{ n_\gamma \}$. This method is applicable not just for 
the $\rho=1$ lattices with arbitrary large $k$, but also 
for lattices with any $\rho=1,2,\cdots$. In this method, 
we obtain the leading contribution to $c_\gamma(\nu_{|\gamma|}+n)$ 
simultaneously for all $n$, while we need to add up many terms to 
determine those coefficients with good precision\footnote{
\label{fn:Rademacher}
The Rademacher expansion does not come to our mind as the first 
option in computing $c_\gamma(\nu)$'s of $\Phi$ for the following reason. 
Remember, first, that there are two purposes for us to compute the 
Fourier coefficients of the vector-valued modular form $\Phi$ for 
a given set $\{ n_\gamma \}$, and $\Psi$ for a given set $\{n_\gamma \}$, 
$\{m_\gamma \}$ and $d_\gamma$'s. One purpose has been explained in the 
main text (the Fourier coefficients of $\Phi$ determine the coefficients 
of the power series $f_*$; we need to evaluate the period polynomial 
with $f_*$ in the integrand). For this purpose, we may only need 
{\it approximate} values of the coefficients of $f_*$ for many terms. 
The other purpose is that we need to use the {\it precise} 
Fourier coefficients of $\Phi$ and $\Psi$ at the $e^{2\pi i \tau \nu}$ terms 
with $\nu \leq 2$, in order to determine the 4D effective Lagrangian 
parameters $d_{111}$, $(c_2)_1$ and $\chi_{\rm match}$ (as 
in (\ref{eq:d111-k=2}--\ref{eq:chiMatch-k=2})). Without obtaining those 
Fourier coefficients as rational numbers, we cannot argue for which 
$\{ n_\gamma \}$, $\{ m_\gamma\}$ and $d_\gamma$'s the Lagrangian parameter 
$d_{111}$ is an integer. 
It might have been an option to use the Rademacher expansion for 
the first purpose, but the authors just did not try. 
} %
 even for small $n$. 

Instead, we used a method of Refs. \cite{EW19, ESW20} 
(explained shortly) that is applicable 
to vector-valued modular forms for the $\rho=1$ cases with any $k$. 
To determine all the Fourier coefficients in the expansion 
$\Phi_\gamma = \sum_{\nu} NL_{\nu, \gamma} e^{2\pi i \tau \nu}$ with 
$\nu \leq N_{{\rm cut}\Phi/\Psi}$, one may first generate weight-$w$ index-$k$ 
holomorphic Jacobi forms to list up weight-$(w-1/2)$ holomorphic 
vector-valued modular forms $\varphi = \{ \varphi_\gamma \}$ 
for the group ${\rm Mp}(2,\Z)$ in the dual of the Weil 
representation associated with the lattice $\widetilde{\Lambda}_S$ 
(see \cite{DMZ}). 
The combination $\varphi \cdot \Phi$ must be equal to a 
scalar-valued modular form of weight $10+w$, which must be an element 
of the weight-$(10+w)$ subspace of the ring $\C [E_4, E_6]$ generated
by the Eisenstein series $E_4$ and $E_6$. For smaller $w$'s starting 
with $w=4, 6, \cdots$, one may obtain non-trivial constraints
 on $NL_{\nu,\gamma}$'s by comparing the coefficients 
of the $e^{2\pi i \tau n}$ terms ($n \leq N_{{\rm cut}\Phi/\Psi}$) of 
$\varphi \cdot \Phi$ and the element in $\C[E_4,E_6]$; for 
$10+w=12N_{{\rm cut}\Phi/\Psi}$ and $(10+w)\geq 12N_{{\rm cut}\Phi/\Psi}+4$, 
however, no constraint is obtained on $NL_{\nu,\gamma}$'s with 
$\nu \leq N_{{\rm cut}\Phi/\Psi}$ because higher order terms ($NL_{\nu,\gamma}$'s 
with $\nu > N_{{\rm cut}\Phi/\Psi}$) are necessary in identifying 
$(\varphi \cdot \Phi)$ within the space of weight-$(10+w)$ 
scalar-valued modular forms.
For all the cases of $\widetilde{\Lambda}_S=U[-1]\oplus \vev{+2k}$ 
that we have studied explicitly, this method determines all of $NL_{\nu,\gamma}$
with $\nu \leq N_{{\rm cut}\Phi/\Psi}$ completely in terms of the leading 
coefficients $NL_{\nu_{|\gamma|}, \gamma}$, well before exploiting the constraints 
obtained from $\varphi$'s with the maximum $w = 12N_{{\rm cut}\Phi/\Psi}-8$. 

This method allows us to determine $c_\gamma(\nu)$'s in terms 
of $\{ n_\gamma \}$ for all $\nu \leq N_{{\rm cut}\Phi/\Psi}-1$. 
We call this method in \cite{EW19, ESW20} the brute force order-by-order 
method for the $\rho=1$ cases. 
Now, one can compute all the coefficients\footnote{
Note that the notation $q$ 
is reserved for $e^{2\pi i t}$, where $t$ is the Coulomb branch moduli 
parameter (i.e., Narain moduli, complexified K\"{a}hler moduli), not 
for $e^{2\pi i \tau}$ in this article; $\tau$ is the complex structure 
parameter of the $g=1$ worldsheet in the Heterotic description.  
In Ref. \cite{EW21}, $q$ was used in both contexts. 
} %
 of the $q^n=e^{2\pi i t n}$ terms of $f_*(t)$ 
for $n \leq \sqrt{2k(N_{{\rm cut}\Phi/\Psi}-1)}$.

\vspace{5mm} 

{\bf Graft method}

One may further determine the coefficients of $q^n = e^{2\pi i t n}$ in $f_*(t)$
for even larger $n$, as we explain below and also in the 
appendix \ref{sec:grafting}, 
by using the fact that $f_*(t)$ is a weight-6 meromorphic 
modular form for $\Gamma_0(k)$ (and is even under the Fricke involution 
$w^{(k)}$).  
When the integration contour extends to small ${\rm Im}(\sigma)$ region
(when $y_{\rm min}$ is small), this is necessary 
in order for us to judge with confidence whether a numerically 
computed coefficient of a period polynomial should be interpreted 
as a rational number or not. 

Here is the idea. We know that $f_*$ is meromorphic, because the massless 
matter singularities in section \ref{ssec:MMS} give rise to the logarithmic 
singularity in the 2nd derivative of the prepotential, and a pole of 
order 3 in $f_*(t)$ \cite{KapLusThs}. So, suppose that we manage to find a weight-$w_D$ 
holomorphic modular form ${\rm Denm}f_*(t)$ for $\Gamma_0(k)$ that has 
a zero at least of order 3 at all the massless matter singularities 
($k$ and $\{n_\gamma \}$ are fixed).
Then the product $f_* \times {\rm Denm}f_*$ is a holomorphic scalar-valued 
modular form of weight-$(6+w_D)$ for $\Gamma_0(k)$. 
Now, the vector space of weight-$(6+w_D)$ modular forms 
for $\Gamma_0(k)$ is of finite dimensions, and one can use SAGE \cite{SAGE}, 
for example, to compute its basis elements as a power series of 
$q=e^{2\pi i t}$ up to the $q^{N_{{\rm cut}f*}}$ term for very large $N_{{\rm cut}f*}$. 
So, it is enough to set $N_{{\rm cut}\Phi/\Psi}$ large enough to determine the 
coefficients of $f_*$ so that $f_* \times {\rm Denm}f_*$ can be chosen 
uniquely\footnote{
The required $N_{{\rm cut}\Phi/\Psi}$ is smaller, when we fit 
$f_* \times {\rm Denm}f_*$ within the vector space of weight-$(6+w_D)$ 
Fricke-even [resp. odd] modular forms, provided ${\rm Denm}f_*$ is 
Fricke-even [resp. odd]. 
} %
 from the vector space of weight-$(6+w_D)$ modular forms 
for $\Gamma_0(k)$. We call this the graft method/construction.\footnote{
Oxford English Dictionary describes the noun {\it graft} as follows: 
a piece cut from a living plant and fixed in a cut made 
in another plant, to form a new growth; the process or result of doing this.
} %

The appendix \ref{sec:grafting} will explain an idea on how to 
find ${\rm Denm}f_*$, 
and also illustrate the graft method with an example. We will also 
see how the known analytic expression for $f_*$ in the case of 
$k=2$ and $n_1=n_2=0$ \cite{AP} fits into the 
graft construction of $f_*$ here. 

The graft method allows us to identify $f_*$ for a given $k$ 
and $\{n_\gamma \}$ in the form of (\ref{eq:f*-as-a-ratio-of-MF}),
but what we can handle in practice is 
\begin{align}
   \frac{1}{(2\pi i)^3} \frac{[{\rm Numr}f_*]_n}{[{\rm Denm}f_*]_n}
  \label{eq:f*-as-a-ratio-of-MF-trunc}
\end{align}
instead of (\ref{eq:f*-as-a-ratio-of-MF}). It is one way to use 
this rational function of $q$ for $f_*$ in the integrand of 
(\ref{eq:PP-split-to-2}), and it is another to compute $[f_*]_n$
from (\ref{eq:f*-as-a-ratio-of-MF}) and apply the procedure 
described in (\ref{eq:Pade-general}). 
For the cases with small $k$, we used the latter---sometimes even without
resorting to the graft method---with the choice (\ref{eq:poles-4-denm}) 
for the denominator; for the cases with larger $k$, we used the former.



\section{Period Polynomials Numerically Evaluated}
\label{sec:PP}

The period polynomials can be evaluated by doing the contour integrals 
in the upper half plane numerically. The results are collected 
in this section \ref{sec:PP}; we will use the results in the next section 
to constrain the BPS classification invariants and run the test 
for a geometric phase interpretation. 

The period polynomials for cuspforms have been a subject of interest 
in their own right in number theory; they are regarded as a generalization 
of the period integrals of the modular curves.  
The results here are for a class of meromorphic weight-$6$ cuspforms $f_*$ 
for $\Gamma_0(k)$ for $k=2,3,\cdots, 6$ and $11$, constructed from 
vector-valued modular forms $\Phi$ of weight 21/2 (equivalently,  
$\Phi/\eta^{24}$ of weight $-3/2$).  

For any $k=2,3,\cdots$, the group $[\Gamma_S]$ contains an 
element $w^{(k)}$, which acts on the upper half plane as the Fricke 
involution. The period polynomials for this $w^{(k)}$ are recorded in 
section \ref{ssec:PP-Fricke}. Those for the generators of the subgroup 
$\Gamma_0(k)/\{ \pm 1\}$ of $[\Gamma_S]$ are found in 
section \ref{ssec:PP-Gamma0k}. 
In section \ref{ssec:PP-AL}, we will record the numerical evaluation 
of the period polynomial for an element of 
${\rm Isom}'(\widetilde{\Lambda}_S)$ that acts as Atkin--Lehner involutions 
(other than the Fricke involution); the numerical result will be fed into 
the discussion in section \ref{sssec:level-6} on when such an isometry can 
be in the vacuum identification group $[\Gamma_S]$. 

\subsection{For the Fricke Involution}
\label{ssec:PP-Fricke}

The period polynomial for the Fricke involution $w^{(k)}$ can be 
determined by evaluating the integral (\ref{eq:PP-split-to-2}) numerically. 
We have computed $P_{\widetilde{w}^{(k)}}(t;f_*)$ numerically\footnote{
Here is a reminder. There are multiple topologically different 
choices of the integration contour in the period 
polynomial (\ref{eq:PP-split-not-yet}) for 
$w^{(k)} \in {\rm Isom}'(\widetilde{\Lambda}_S)$; our choice of the 
topology of the contour has already been described below 
(\ref{eq:PP-split-not-yet}, \ref{eq:PP-split-to-2}). Such a choice also 
identifies the lift $\widetilde{w}^{(k)}$ uniquely modulo $+\Z D$ 
for $w^{(k)} \in {\rm Isom}'(\widetilde{\Lambda}_S)$; see 
\cite[\S 2.2]{EW21} for more background materials. 
This comment will not be repeated in the rest of this article, but 
is applied to all the other period polynomials $P_{\tilde{g}}$ for 
$g \in [\Gamma_S]$. 
} %
for $k=2,3,4,5,6$ and $k=11$ with arbitrary $\{ n_\gamma \}$;
in the case of $k=11$, however, $\{ n_{\gamma} \}$ should be 
subject to the linear constraint (\ref{eq:lin-rltn-k=11}) for 
a vector-valued modular form $\Phi$ to exist.

It turns out that the numerical results are fitted very well by 
a universal formula 
\begin{align}
 P_{\tilde{w}^{(k)}}(t,f_*) \simeq \left[\frac{-n_0}{4}\right](kt^2+1)^2
   + \frac{(kt^3+t)}{6k} [d'_{\rm pp}]  
   + \frac{\zeta(3)}{(2\pi i)^3} \frac{[\chi_{\rm  pp}]}{2} (k^2 t^4-1) ; 
   \label{eq:PP-Fricke-formula-gen-1}
\end{align}
this includes observations that (i) when the polynomial with numerical 
coefficients on the left hand side is split into the one with purely 
imaginary coefficients and the one with real coefficients, 
the former is numerically almost proportional to $(k^2t^4-1)$; 
$\chi_{\rm pp}$ is meant to be the $\{ n_\gamma \}$-dependent 
(but $t$-independent) coefficient such that the numerically evaluated 
period polynomial is reproduced. Another observation is that (ii) 
the $t^3$ and $t^1$ 
part of the real coefficient polynomial turns out to be numerically 
almost proportional to $(kt^3+t)$; $d'_{\rm pp}$ is meant to be the 
$\{ n_\gamma \}$-dependent (but $t$-independent) coefficient that 
reproduces the numerically evaluated period polynomial. 

Moreover, we have confirmed that the $\{ n_\gamma \}$-dependent fit 
coefficients $\chi_{\rm pp}$ and $d'_{\rm pp}$ of the period polynomial 
can be set to the following combinations 
\begin{align}
  \chi_{\rm pp} = \chi_{\rm match}, \qquad   d'_{\rm pp} = d'_{\rm match}, 
  \label{eq:PP-Fricke-formula-gen-2}
\end{align}
that appear in the 4D effective Lagrangian.
%
Here,  
\begin{align}
  d'_{\rm match} := d'_{111} := d_{111} - \frac{k}{4}(c_2)_1, 
\end{align}
is the combination \cite[\S3.1.3]{EW19} that 
is independent of $\Psi$ (i.e., independent of $\{ m_\gamma \}$ 
and $d_\gamma$'s). 
The 4D Lagrangian parameter $\chi_{\rm match}$ and the combination 
$d'_{\rm match}$ are computed (following the prescription in \cite{EW19}, 
independent of numerical evaluations of $P_{\tilde{w}^{(k)}}(t,f_*)$), 
and are presented in section \ref{sec:classification}; 
see (\ref{eq:d111-k=2}--\ref{eq:chiMatch-k=2}) for $k=2$, 
(\ref{eq:d111-k=3}--\ref{eq:chiMatch-k=3}) for $k=3$, 
(\ref{eq:d111-k=4}--\ref{eq:chiMatch-k=4}) for $k=4$, 
(\ref{eq:d111-k=5}--\ref{eq:chiMatch-k=5}) for $k=5$, 
(\ref{eq:d111-k=6}--\ref{eq:chiMatch-k=6}) for $k=6$
and (\ref{eq:d111-k=11}--\ref{eq:chiMatch-k=11}) for $k=11$, in particular.  
The fitting formula (\ref{eq:PP-Fricke-formula-gen-1},
 \ref{eq:PP-Fricke-formula-gen-2}) 
also covers the $k=1$ case studied already in \cite{EW21}. 

Here is a detail of how good the universal fitting 
formula (\ref{eq:PP-Fricke-formula-gen-1},
 \ref{eq:PP-Fricke-formula-gen-2}) is when 
we think of the fitting coefficients in $[ \cdots ]$
on the right hand side literally as $-n_0/4$, $\chi_{\rm match}$
and $d'_{\rm match}$ with coefficients in $\Q$. 
In the case $k=6$ with $n_1=n_5=0$, we used $n=N_{{\rm cut}f*}=100$
in (\ref{eq:Pade-general}, \ref{eq:f*-as-a-ratio-of-MF-trunc}) 
for the integrand on the left hand side; in the case $k=6$ with 
$(n_1,n_5) \neq (0,0)$ and in the case $k=11$, we used 
$n=N_{{\rm cut}f*}=300$ on the left hand side. The coefficients 
obtained in the numerical integrals in this way do not differ 
from the rational coefficients in $[ \cdots ]$ more than 
${\cal O}(1) \times 10^{-6}$; some terms on the right hand side 
have zero coefficients ($t^1$, $t^2$ and $t^3$ terms with 
pure imaginary coefficients in (\ref{eq:PP-Fricke-formula-gen-1}))
and the corresponding terms on the left hand side turn out not 
to be more than ${\cal O}(1) \times 10^{-6}$. In the case 
of $k \leq 5$, we have also confirmed that the formula 
(\ref{eq:PP-Fricke-formula-gen-1}, \ref{eq:PP-Fricke-formula-gen-2})
fits just as well as in the cases $k=6,11$ when we use even 
smaller $n=N_{{\rm cut}f*}$ 
in (\ref{eq:Pade-general}, \ref{eq:f*-as-a-ratio-of-MF-trunc}) 
in the integrand on the left hand side. We also computed $P_{\tilde{w}^{(k)}}$
for $k=7$ and $22$ for some choices of $\{ n_\gamma \}$ and obtained 
the results consistent with (\ref{eq:PP-Fricke-formula-gen-1}, 
\ref{eq:PP-Fricke-formula-gen-2}). 

It is highly non-trivial that $\chi_{\rm pp}$ and $d'_{\rm pp}$ for the 
period polynomials and $\chi_{\rm match}$ and $d'_{\rm match}$ for the 
4D effective Lagrangian are likely to be identical, until one finds 
a mathematical proof of equivalence between them.  As of now, 
they are both determined from $\{n_\gamma \}$, but in completely 
independent ways, as remarked already in \cite[\S5]{EW21}.  
An analytical evaluation of the period polynomials (rather than numerical) 
is under way \cite{EKSW}, which is meant as a step toward a better 
understanding of their equivalence. 

\subsection{For Generators of $\Gamma_0(k)$}
\label{ssec:PP-Gamma0k}

{\bf The cases with $k=2,3,4$:} the 
group $\Gamma_S = (\Gamma_0(k)/\{\pm 1\}) \rtimes \Z_2\vev{w^{(k)}}$ 
is generated\footnote{
Throughout this article, 
we say that a set of elements $\{ g_a, g_b, \cdots \}$ {\it generates a group}
when any element of $G$ can be expressed as a product of a finite number 
of integral (possibly negative) powers of elements of 
$\{ g_a, g_b, \cdots, \}$.  
} %
 by $g_\infty$ and $w^{(k)}$ for $k=2,3,4$ (a little more details are 
written below). So, we 
have already computed all the period integrals in order to impose 
the integrality conditions of the monodromy matrices 
for the duality transformations in $\Gamma_S$. Let us still 
leave the numerical results of the period polynomials for 
a few elements in $\Gamma_0(k)/\{ \pm 1\}$ with $k=2,3,4$, because 
the period polynomials themselves are objects of interest 
in mathematics. 

In the case $k=2$, think of $g_3 \in {\rm Isom}'(\widetilde{\Lambda}_S)$ 
that corresponds to $ \pm \left(\begin{array}{cc} 1 & -1 \\ 2 & -1 \end{array} \right) \in \Gamma_0(k)/\{ \pm 1\}$. We have confirmed numerically that 
\begin{align}
    \nonumber
    P_{\tilde{g}_3}(t,f_*) &\; \simeq  
 -\left[(n_0/2)+n_1+(n_2/2)\right](1-4t+6t^2-4t^3)
+[n_2](-t^2/2 + t^3-t^4)  \nonumber \\
 & \;    +\frac{\zeta(3)}{(2\pi i)^3}\frac{[\chi_{\rm pp}]}{2}((2t-1)^4-1) .
  \label{eq:pp-k=2-g3}
\end{align}
The pure imaginary part of the numerically evaluated period polynomial 
turns out to be numerically almost proportional to $((2t-1)^4-1)$, 
and the same coefficient $\chi_{\rm pp}$ as for $\widetilde{w}^{(k)}$ 
to be used for the fitting here. We will comment at the end of 
this section \ref{ssec:PP-Gamma0k} on how close the numerically 
evaluated period polynomials are to the 
formulae (\ref{eq:pp-k=2-g3}, \ref{eq:pp-k=4-g2},
 \ref{eq:pp-k=5-g2}, \ref{eq:pp-k=5-g3}, \ref{eq:pp-k=6-g2},
 \ref{eq:pp-k=11-g1}, \ref{eq:pp-k=11-g2}, \ref{eq:PP-k=6-Weyl4-full})
when the coefficients in $[ \cdots ]$ are regarded literally 
as rational numbers. A forthcoming paper \cite{EKSW} 
will provide analytic derivation of this pure imaginary part of the 
period polynomial.  All those comments on the purely imaginary 
part of the period polynomial (\ref{eq:pp-k=2-g3}) apply also to 
all the other period polynomials 
(\ref{eq:pp-k=4-g2}, \ref{eq:pp-k=5-g2}, \ref{eq:pp-k=5-g3}, 
\ref{eq:pp-k=6-g2}) in this article, and we will not repeat 
them each time in the rest of this article. 

While the group $\Gamma_0(2)/\{ \pm 1\}$ is generated by 
$\{ g_\infty, g_3\}$, it is enough \cite{AP} to 
choose $\{ g_\infty, w^{(2)} \}$ as a set of generators of the group 
$\Gamma_S$ for $k=2$. This is because 
\begin{align}
 g_3 = g_{\infty} w^{(2)} g_\infty w^{(2)} \in {\rm Isom}'(\widetilde{\Lambda}_S). 
 \label{eq:relatn-k=2-g3-generated}
\end{align}
The situation is similar in the case $k=3$. 

Let us give one more example, which is in the case $k=4$; think of 
$g_2 \in {\rm Isom}'(\Lambda_S)$ that corresponds to 
$\pm \left(\begin{array}{cc} -1 & 1 \\ -4 & 3\end{array} \right)$ in 
$\Gamma_0(k)/\{ \pm 1\}$. It turns out numerically that 
\begin{align}
 P_{\tilde{g}_2}& \; (t,f_*) \simeq 
     t^4[24n_0 + 36n_1-40n_2 -10n_3-40n_4 ] \nonumber \\
& \!\! + t^2\left[64n_0+78n_1-174n_2-\frac{87}{2}n_3 -174n_4 \right]
  + \left[ \frac{9}{2}n_0+3n_1-22n_2- \frac{11}{2} n_3 -22n_4  \right]
      \nonumber \\
& \!\! + t^3 [-64n_0-88n_1+136n_2+34n_3+136n_4]
       + t [-28n_0-28n_1+100n_2+25n_3+100n_4]    \nonumber \\
& \!\!  + \frac{\zeta(3)}{(2\pi i)^3}\frac{[\chi_{pp}]}{2}((4t-3)^4-1).   
         \label{eq:pp-k=4-g2} 
\end{align}
While the group $\Gamma_0(4)/\{ \pm 1\}$ is generated by 
$\{ g_\infty, g_2\}$, it is enough to choose $\{ g_\infty, w^{(4)}\}$
as a set of generators of $\Gamma_S$, because 
\begin{align}
 g_2 = w^{(4)}\cdot g_\infty^{-1} \cdot w^{(4)}\cdot g_\infty^{-1}
        \in {\rm Isom}'(\widetilde{\Lambda}_S).  
   \label{eq:relatn-k=4-g2-generated}
\end{align}

For both $g_3$ in the $k=2$ case and $g_2$ in the $k=4$ case, 
we have confirmed that their period polynomials shown above satisfy 
the relations that follow\footnote{
See \cite[(3.5), (3.6) and (3.9)]{AFGNT}, \cite[(4.4), (4.6)]{AP} 
and \cite[(3.19), (3.21))]{EW21}.
} %
from the generator relations (\ref{eq:relatn-k=2-g3-generated}, 
\ref{eq:relatn-k=4-g2-generated}) above. This only serves as 
a check of our numerical calculation.\footnote{
Reference \cite[\S4]{AP} imposed the generator 
relations \cite[(4.4), (4.6)]{AP} and found a set of polynomials 
$\{ P_{\tilde{g}_i}(t) \}_{g_i \in [\Gamma_S]}$---\cite[(4.16)]{AP}---satisfying
\cite[(4.4)+(4.6)]{AP}, instead of computing $\{ P_{\tilde{g}_i}(t;f_*) \}$
from a general $f_*$. The solution \cite[(4.16)]{AP} corresponds to 
$n_1=n_2=\ell_{\rm aux}=0$ in this article. 
} %

\vspace{5mm}

{\bf In the cases with $k \geq 5$, } the whole group $\Gamma_S = 
(\Gamma_0(k)/\{ \pm 1\}) \rtimes \Z_2\vev{w^{(k)}}$ is not necessarily 
generated by $g_\infty$ and $w^{(k)}$. {\bf In the case $k=5$,} one 
can use SAGE \cite{SAGE} to see that the group $\Gamma_0(k)/\{ \pm \}$ can be 
generated by\footnote{
From here on, we abuse the notation and use the same symbol for 
both an element in $\Gamma_0(k)/\{ \pm 1\} \subset {\rm SL}(2;\Z)/\{ \pm 1\}$ 
(i.e., a mod-($\pm 1$) $2\times 2$ matrix)
and one in $\Gamma_0(k)/\{ \pm 1\} \subset {\rm Isom}(\widetilde{\Lambda}_S)$
(i.e., a $3\times 3$ matrix). The correspondence between them is given 
by \cite[(3.4) and (3.5)]{EW21} = (\ref{eq:convrt-3x3-and-2x2}).
} %
\begin{align}
 g_2 = \pm \left( \begin{array}{cc} 2 & -1 \\ 5 & -2 \end{array} \right),
     \qquad 
 g_3 = \pm \left( \begin{array}{cc} 3 & -2 \\ 5 & -3 \end{array} \right)
\end{align}
along with $g_\infty$. Because there is a relation 
\begin{align}
 g_\infty^{-1} \cdot g_3 \cdot g_\infty = w^{(5)} \cdot g_2 \cdot w^{(5)} 
    \in  {\rm Isom}'(\widetilde{\Lambda}_S), 
  \label{eq:relatn-k=5-g3-generated}
\end{align}
we just have to compute the period polynomial of either $\tilde{g}_2$ 
or $\tilde{g}_3$, not the one for both. We still did, out of 
mathematical interest, and record the numerical result below. 
For $g=g_2$, 
\begin{align}
 P_{\tilde{g}_2}(t, f_*)
 & \; \simeq 
    \left[ 25n_0 + 140 n_1 - 10 n_2 + 60 n_3 - 30 n_4 - 200 n_5 \right]
        \; \; t^4/4      \nonumber \\
 & \quad + \left[ -60n_0 + 54 n_1 - 140 n_2 + 6 n_3 - 44 n_4 - 270 n_5\right]
         \; \; t^2/4    \nonumber \\ 
 & \quad + \left[ -5n_0 - 2 n_1 - 10 n_2 - 2 n_3 - 2 n_4 - 10 n_5 \right]/4
              \nonumber \\
 & \quad + \left[ 30n_0 - 157 n_1 + 118 n_2 - 53 n_3 + 59 n_4 + 385 n_5 \right]
          \; \;  t^3/4   \nonumber \\
 & \quad + \left[ 30n_0 - n_1 + 62 n_2 + 7 n_3 + 15 n_4 + 85 n_5 \right]
          \; \; t/4     \nonumber \\
 & \quad  + \frac{\zeta(3)}{(2\pi i)^3}\frac{[\chi_{pp}]}{2} ((5t-2)^4-1).
   \label{eq:pp-k=5-g2}
\end{align}
For $g_3$, 
\begin{align}
 P_{\tilde{g}_3}(t,f_*) & \; \simeq
    - [25n_0 + 140 n_1 + 190 n_2 + 60 n_3 + 20 n_4 - 200 n_5] \; t^4/4
      \nonumber \\
 & \quad - [180n_0 + 423 n_1 + 602 n_2 + 207 n_3 + 65 n_4 - 315 n_5] \; t^2/4
     \nonumber  \\
 & \quad  + [-20n_0 - 34 n_1 - 52 n_2 - 18 n_3 - 6 n_4 + 10 n_5] \; /4
     \nonumber \\
 & \quad -  [-130n_0 - 403 n_1 - 558 n_2 - 187 n_3 - 59 n_4 + 415 n_5] \; t^3/4
      \nonumber  \\
 & \quad - [-100n_0 - 196 n_1 - 288 n_2 - 100 n_3 - 32 n_4 + 100 n_5] \; t/4
      \nonumber  \\
 & \quad + \frac{\zeta(3)}{(2\pi i)^3} \frac{[\chi_{pp}]}{2}((5t-3)^4-1). 
    \label{eq:pp-k=5-g3}
\end{align}

{\bf In the case $k=6$,} the group $\Gamma_0(6)/\{ \pm 1\}$ can be 
generated \cite{SAGE} by two elements 
\begin{align}
 g_1 = \pm \left( \begin{array}{cc} 5 & -1 \\ 6 & -1 \end{array} \right),
     \qquad \qquad 
 g_2 = \pm \left( \begin{array}{cc} 7 & -3 \\ 12 & -5 \end{array} \right)
\end{align}
along with $g_\infty$. 
Because $g_1 = g_\infty \cdot w^{(6)} \cdot g_\infty \cdot w^{(6)}$ in 
${\rm Isom}'(\widetilde{\Lambda}_S)$, it is enough to choose 
$\{ g_\infty, g_2, w^{(6)} \}$ as a set of generators of $\Gamma_S$. 
We found numerically that 
\begin{align}
P_{\tilde{g}_2}(t,f_*) & \; \simeq
  t^4 \; \; [-624 n_0 + 242 n_1 - 1040 n_2 - 780 n_3 - 208 n_4
       - 3038 n_5 + 750 n_6 ]   \nonumber \\
& \quad + 
 t^2 \; \; [-732 n_0 + (67/2)n_1 - 1220 n_2 - 915 n_3 - 244 n_4 
       - (7091/2) n_5 + 735 n_6 ] \nonumber \\
& \quad + \; \; 
  [-24 n_0  -(11/2) n_1 - 40 n_2 - 30 n_3 - 8 n_4
     - (231/2) n_5+ 20 n_6]  \nonumber \\
& \quad + 
 t^3 \; \; [1104 n_0 - 226 n_1 + 1840 n_2 + 1380 n_3 + 368 n_4 
        + 5362 n_5 - 1212 n_6 ] \nonumber \\
& \quad + 
 t \; \; [216 n_0  + 21 n_1 + 360 n_2 + 270 n_3 + 72 n_4 
        + 1043 n_5 - 198 n_6 ]  \nonumber \\
& \quad + \frac{\zeta(3)}{(2\pi i)^3} \frac{[\chi_{pp}]}{2}
   ((12t-5)^4-1).
   \label{eq:pp-k=6-g2}
\end{align}

{\bf In the case $k=11$,} the group $\Gamma_0(11)/\{ \pm 1\}$ is generated 
by  \cite{SAGE}
\begin{align}
  g_1 = \pm \left( \begin{array}{cc} 7 & -2 \\ 11 & -3 
            \end{array} \right), \qquad 
  g_2 = \pm \left( \begin{array}{cc} 8 & -3 \\ 11 & -4 
            \end{array} \right) 
\end{align}
along with $g_\infty$. Their period polynomials (for $n_7=0$) 
turned out to be approximately 
\begin{align}
P_{\tilde{g}_1}(t;f_*) \simeq & \;
  t^4  \; \; \left[
        -393.25n_0  -555.5n_1 + 269.5n_4  -522.5n_5 -5.5x_6+126.5n_{11}
   \right.  \nonumber \\
& \qquad \quad  
   \left. -935_* n_2 -561_* n_3 -698.5_* n_8 +1276_* n_{10}
   \right]   \nonumber  \\
+ & \; t^2 \; \;
    \left[-255.75n_0  -332.75 n_1 +26.5n_4 -228n_5 -17.75x_6+47.875n_{11}
    \right.  \nonumber \\
& \qquad \quad 
 \left. -583.5 n_2 -332.5 n_3 -361.5_* n_8 +414n_{10} \right]  \nonumber \\
+ & \; 
 t^0 \; \; \left[ -4.5 n_0  -5.5n_1 -n_4 - 3n_5 -0.5x_6+0.5n_{11}
   \right.  \nonumber  \\
& \qquad \quad  
   \left. -10.5 n_2  -5.5 n_3 -5 n_8 +2.5 n_{10}
   \right]  \nonumber \\ 
+ & \;
 t^3 \; \; \left[
    525.25n_0 +710.25 n_1 -176.5n_4 + 559.5n_5 +23.75x_6-126.875n_{11}
    \right.  \nonumber \\
& \qquad \quad 
   \left. + 1210.5_* n_2 +712_* n_3 +824.5_* n_8 -1209_* n_{10}
   \right]  \nonumber  \\
+ & \; t^1 \; \; \left[ 55 n_0 +69n_1 +4n_4 + 42 n_5 + 5x_6-8n_{11}
    \right.  \nonumber \\
& \qquad \quad 
\left. + 126_* n_2 +69 n_3 +70 n_8 -58 n_{10}
\right]  \nonumber \\
+ & \; \frac{\zeta(3)}{(2\pi i)^3} \frac{[\chi_{\rm pp}]_*}{2}
      ((11t-3)^4-1) , 
  \label{eq:pp-k=11-g1}
\end{align}
and 
\begin{align}
P_{\tilde{g}_2}(t;f_*) \simeq & \;   
t^4 \; \; \left[ -635.25 n_0 -1424.5n_1 -401.5n_4 - 660n_5 -77_* x_6 + 77n_{11}
   \right.  \nonumber \\
& \quad 
  \left. -2260.5n_2 -1078_{**}n_3  +929.5_{**}n_8 + 137.5_{**}n_{10}
 \right]  \nonumber \\
+ & \; t^2 \; \; \left[
     -610.5 n_0 -1250n_1 -440n_4 -538.5 n_5 - 84.5_* x_6 + 52.75n_{11}
   \right.  \nonumber \\
& \qquad \quad 
   \left. -2064_*n_2 -986.5_*n_3 +723_{**}n_8 -105_*n_{10} 
\right]  \nonumber \\
+ & \; t^0 \; \;
 \left[ -16.25 n_0 -30.5n_1 -12.5n_4 - 12.5n_5 -2.5x_6 + n_{11} 
   \right.  \nonumber \\
& \quad 
  \left. -53_*n_2 -25_*n_3 +15.5_*n_8 -8_*n_{10} 
  \right]  \nonumber \\
+ & \; t^3 \; \;
 \left[ 1020.25 n_0 +2182.25n_1 + 697.5n_4 + 970.5 n_5 +132.75_*x_6 
  -103.875 n_{11} 
   \right.  \nonumber \\
& \qquad \quad 
  \left. + 3526.5_* n_2 +1688_{**} n_3 -1337.5_{**} n_8 -9_* n_{10} 
  \right]  \nonumber \\
+ & \; t^1 \; \;
  \left[ 162.25 n_0 +318.25n_1 +121.5n_4 -133.5 n_5 +23.75x_6 -11.875n_{11}
  \right.  \nonumber \\
& \qquad \quad 
  \left. +538.5_*n_2  +256_{**}n_3 -173.5_* n_8 + 55_*n_{10} 
  \right]  \nonumber \\
+ & \; \frac{\zeta(3)}{(2\pi i)^3} \frac{[\chi_{\rm pp}]_*}{2}
      ((11t-4)^4-1)
  \label{eq:pp-k=11-g2}, 
\end{align}
respectively. Here, $x_6 := n_6 -9n_2 -n_3+3n_8-4n_{10}$. 

\vspace{5mm}

{\bf Notes on the precision:} 
we have used the power-series cut-off at $n=N_{{\rm cut}f*}=300$ in setting 
the integrand in the case $k=11$ (\ref{eq:pp-k=11-g1}, \ref{eq:pp-k=11-g2}),  
and also in the case $k=6$ (\ref{eq:pp-k=6-g2}, \ref{eq:PP-k=6-Weyl4-full}) 
when $(n_1,n_5)\neq (0,0)$; the cut-off at $n=N_{{\rm cut}f*}=100$ was  
used in the case $k=6$ when $(n_1,n_5)=(0,0)$. Large fraction of 
the coefficients of the period polynomials 
(\ref{eq:pp-k=6-g2}, \ref{eq:PP-k=6-Weyl4-full}, \ref{eq:pp-k=11-g1}, 
\ref{eq:pp-k=11-g2}) evaluated numerically 
are confirmed to agree with the rational coefficients in $[ \cdots ]$ 
on the right hand side\footnote{
Use $\chi_{\rm match}$ for $\chi_{\rm pp}$. Read $0.25$ and $0.875$ 
in (\ref{eq:pp-k=11-g1}, \ref{eq:pp-k=11-g2}) as $1/4$ and $7/8$, 
respectively.  
} %
 within the error ${\cal O}(1) \times 10^{-4}$; the coefficients 
in (\ref{eq:pp-k=11-g1}, \ref{eq:pp-k=11-g2}) with ${}_*$ and ${}_{**}$
agree within the error of ${\cal O}(1)\times 10^{-3}$ and 
${\cal O}(1)\times 10^{-2}$, respectively. 

It tends to be difficult to evaluate the period polynomials reliably 
in the method explained in section \ref{ssec:set-f*} in certain 
situations. That is when the integration 
contour in (\ref{eq:PP-split-to-2}) extends closer to the boundary 
of the upper complex half plane (i.e., when $y_{\rm min}$ is small), 
and also when we need to use the denominator factor with large weight $w_D$
(e.g., in the $k=11$ case, we used denominators with 
$w_D =18$ when any one of $n_{2,3,8,10}$ is non-zero (see the appendix 
\ref{sec:grafting} for more explanations)). So, we do not 
regard the disagreement by ${\cal O}(1) \times 10^{-2}$ or $10^{-3}$
as a sign that the period polynomials have non-rational coefficients.  
In the case $k=11$ with $n_7 \neq 0$, we could not obtain reliable 
numerical integrals\footnote{
When $n_7 \neq 0$, we will have to use 
${\rm Denm}f_ * =(H^{(11)}_{|[7]|})^3(H^{(11)}_{[11]})^2$ 
when $(n_7, n_{11})\propto (1,3)$ and $n_{\gamma \neq 7,11} = 0$, 
or ${\rm Denm}f_* = (H^{(11)}_{|[7]|})^3(H^{(11)}_{|[9]|})^3$ 
when $(n_7,n_9) \propto (1,-3)$ and $n_{\gamma \neq 7,9}=0$. 
This means that $w_D=28$ or 30. Moreover, the integration 
contour extends to $y_{\rm min} = 1/11$. In these situations, 
numerical integrals using the ``NIntegrate'' function 
in Mathematica return 
the coefficients that vary more than ${\cal O}(1)$ for 
a small change in cut-off parameters that should not affect 
the true value of the coefficients. On the other hand, we have 
managed to compute the period polynomial for $w^{(k)}$ (not for 
$g_1$ and $g_2$) with $k=11$ in (\ref{eq:PP-Fricke-formula-gen-1}), 
due to 
the fact that the integration contour of (\ref{eq:PP-split-to-2})
extends only to $y_{\rm min} = 1/\sqrt{11}$ then. 
} %
 for the generators $g_1$ and $g_2$, 
so the fitting formulae (\ref{eq:pp-k=11-g1}, \ref{eq:pp-k=11-g2})
should be read as those only for $\{ n_\gamma \}$ with $n_7=0$. 
 
In the cases with $k\leq 5$, our numerical evaluation of the 
period polynomials agrees with the formulae 
(\ref{eq:pp-k=2-g3}, \ref{eq:pp-k=4-g2}, 
\ref{eq:pp-k=5-g2}, \ref{eq:pp-k=5-g3}) with much smaller 
discrepancy, even when the power-series cut-off $n=N_{{\rm cut}f*}$
in (\ref{eq:Pade-general}, \ref{eq:f*-as-a-ratio-of-MF-trunc}) 
is set below 50.

%

\subsection{For Non-trivial Atkin--Lehner Involutions}
\label{ssec:PP-AL}

For all of the levels $k=2,3,4,5$, $k$ contains powers of just one prime 
number; $s=1$. So, there is no non-trivial Atkin--Lehner involution 
$\epsilon \in (\Z_2)^s$ other than the Fricke involution 
$\epsilon = (1,\cdots, 1) = (1)$. 
The level $k=6$ is the first example ($k=10$ next) where $k$ 
contains multiple prime factors.  

Primarily for mathematical interest, and also as input data for 
the experimental approach of the monodromy study (announced at 
the end of section \ref{ssec:vac-id-grp} and carried out 
in section \ref{sssec:level-6}), we compute the period polynomial 
for an element in $\Gamma_0(k)_+$ that corresponds to a non-trivial 
Atkin--Lehner involution other than the Fricke involution. 

Now, set $k=6$. $G_S \cong \Z_{12} \cong \Z_4 \times \Z_3$ 
and the quotient group of $\Gamma_0(k)_+$ by $\Gamma_0(k)/\{ \pm 1\}$
is $(\Z_2)^s \cong \Z_2 \times \Z_2$; the first factor ($\Z_4$ and $\Z_2$)
and the second factor ($\Z_3$ and $\Z_2$) are for $p=2$ and $p=3$ 
of the prime decomposition ($12 = 2^2\cdot 3$ and $6 = 2\cdot 3$), 
respectively.
As a representative of $\epsilon = (1,0) \in (\Z_2)^{s}|_{s=2}$
and $\epsilon = (0,1)$, one may choose 
\begin{align}
\pm  \left( \begin{array}{cc} 2 \cdot 1 & -1 \\ 2 \cdot 3 & -2
 \end{array} \right)_{k_1=2,k_2=3} \Leftrightarrow
  - R_{v_*(4)} = 
\left[ \begin{array}{cc|c} 
   2 & 3 & -12 \\
   3 & 2 & -12 \\
  \hline 
   1 & 1 & - 5 \end{array} \right]
\end{align}
and 
\begin{align}
  \pm \left( \begin{array}{cc} 3 \cdot 1 & -2 \\ 3 \cdot 2 & -3
 \end{array} \right)_{k_1=3, k_2=2} \Leftrightarrow 
  -R_{v_*(6)} = \left[ \begin{array}{cc|c}
     3 & 2 & -12 \\
     8 & 3 & -24 \\
  \hline 
     2 & 1 & -7 \end{array} \right], 
\end{align}
respectively. The representatives above are chosen 
so that they are multiplied by $(-{\rm id})$ to be 
Weyl reflection symmetries associated with the 
W-boson charge vectors $v_{*(4)}=(-1,-1,4) \in \widetilde{\Lambda}_S^\vee$
and $v_{*(6)} = (-2,-1,6)$, respectively; 
see section \ref{ssec:vac-id-grp} and Table \ref{tab:v*-list}. 


The lattice isometry $R_{v*(4)} \in {\rm Isom}'(\widetilde{\Lambda}_S)$ maps 
\begin{align*}
 G_S \cong \Z/(12\Z) \ni [1], [5] & \; \longmapsto
        [5], [1] \in \Z/(12\Z), \\
  [2], [4] & \; \longmapsto -[2],-[4] , \\
  [3], -[3] & \; \longmapsto [3], -[3],  
\end{align*}
respectively. So, the lattice isometry $R_{v*(4)} $ has a chance to be 
a duality transformation of a Heterotic string vacuum, only 
when $n_1 = n_5$. 
So, we are motivated to compute the period polynomial only for the case 
$n_1 = n_5 =: \bar{n}$. In this case, it turns out numerically that 
\begin{align}
P_{R_{v_*(4)}}(t,f_*) & \simeq  t^4 \; \; 
 [15 n_0 + 22 n_2 + 12 n_3 - 10 n_4 - 6 n_6 -42\bar{n} ]
    \nonumber \\
& \quad +  t^2 \; \; 
 [3 n_0 + 2 n_2 - 3 n_3 - 12 n_4 - (15/2) n_6 -75\bar{n} ]
           \nonumber \\
& \quad +  \; \; \; \left[ -\frac{1}{4} n_0 - \frac{1}{2} (n_2+n_3+n_4)
    - \frac{1}{4} n_6 - 3\bar{n} \right] \nonumber \\
& \quad +  t^3 \; \; [-13 n_0 - (50/3) n_2 - 5 n_3 + (53/3) n_4
    + (23/2) n_6 +103 \bar{n} ] \nonumber \\
& \quad +  t \; \; \left[ \frac{1}{2} n_0 + \frac{5}{3} n_2
 + \frac{5}{2} n_3 + \frac{23}{6} n_4 + \frac{9}{4} n_6
     +\frac{49}{2}\bar{n} \right]    \nonumber \\
& \quad + \frac{\zeta(3)}{(2\pi i)^3}
     \left( \frac{(6t-2)^4}{2^2}-1\right)
     \frac{[\chi_{\rm pp}|_{n_1 = n_5 = : \bar{n}}]}{2} . 
       \label{eq:PP-k=6-Weyl4-full}
\end{align}

The period polynomial for $R_{v_*(6)}$ can be computed from the one 
for $R_{v*(4)}$ above, and those for $w^{(k)}$ and $\Gamma_0(k)/\{ \pm 1\}$. 
So we will not record the result for $R_{v_*(6)}$. 

\section{Classification of Vacuum Branches}
\label{sec:classification}

In this section \ref{sec:classification}, we will impose the condition 
that all the monodromy matrices have integer entries. 
Although the numerical evaluation of the period polynomials 
was only with limited precision, now we regard the period 
polynomials to be equal to the fitting formulae in the previous section 
with rational coefficients in $[ \cdots ]$ in the rest of this article. 

\subsection{Non-linear-sigma-model Interpretation}
\label{ssec:NLSM}

The condition that the monodromy matrix for $g_\infty$ (Peccei--Quinn 
symmetry) should be integer valued determines the invariant 
$a_{11}+\Z \in \R/\Z$ uniquely in terms of other classification 
invariants (see section \ref{ssec:PQ}). 
The same condition for the Fricke involution $w^{(k)}$ also 
determines the invariant $b_{1}+24\Z \in \R/24\Z$ uniquely. 
To see this, we just have to note that (see \cite[(3.18), (3.36)]{EW21})
\begin{align}
 \left[Q|[w^{(k)}]_{-4} -Q\right](t) =
    \left( \frac{b_1}{24} - \frac{d_{111}}{6k}\right) (kt^3 +t)
      - \frac{\zeta(3)}{(2\pi i)^3}\frac{\chi_{\rm match}}{2} (k^2 t^4-1);
\end{align}
combining this with the period polynomial (\ref{eq:PP-Fricke-formula-gen-1}), 
we find that 
\begin{align}
 \Lambda_{\tilde{w}^{(k)}} = \left( \begin{array}{cc|c}
      1 & 1 & (b_1-(c_2)_1)/24 \\ 1 & 1 & (b_1-(c_2)_1)/24 \\ \hline 
      (b_1-(c_2)_1)/24 & (b_1-(c_2)_1)/24 & 0 \end{array} \right)
    + \Z \widetilde{C}.
\end{align}
The condition that this matrix is $\Z$-valued is equivalent to 
\begin{align}
  b_1 +24\Z = (c_2)_1 + 24\Z. 
  \label{eq:cond-Fricke-inZ}
\end{align}
The invariant $b_1+24\Z$ is determined uniquely as promised. 

The result (\ref{eq:cond-Fricke-inZ}) also implies that 
one of the necessary conditions for a NLSM 
interpretation, (\ref{eq:cond-geom-phase-1}), is always 
guaranteed\footnote{
Here, we do not assume geometric phases, so $b_1$ is not equal to $(c_2)_1$ 
a priori. However, the integrality of $\Lambda_{\tilde{w}^{(k)}}$ 
exactly implies this necessary condition for a geometric phase. 
} %
 in all the branches of the Heterotic--Type IIA dual ${\cal N}=2$ $\rho=1$ 
moduli spaces with the values of $k$ that we have studied numerically 
($k=2,3,4,5,6,11$). As explained in \cite{EW21}, 
having (\ref{eq:cond-geom-phase-1}) guaranteed also implies that 
we can read the conditions (\ref{eq:cond-PQ-1}) as 
\begin{align}
 d_{abc} \in \Z, \qquad  2d_{aaa}+(c_2)_a \in 12\Z, \qquad 
  d_{aab}+d_{abb} \in 2\Z
\label{eq:cond-Wall}
\end{align}
of Wall's Theorem \cite{Wall} for existence of a diffeomorphism class 
of real six-dimensional manifolds whose intersection ring and the 
Pontryagin class have the property designated by $d_{111}$ and $(c_2)_1$. 

The result above indicates that the same conclusion obtained 
in the case $k=1$ (and also $\Lambda_S = U$) in \cite{EW21} 
actually holds for a much larger class of the branches of the 
moduli space. Our numerical study even hints that all the 
$\rho =1$ branches have a phase interpreted as a NLSM 
in its Type IIA description. 
 
In the following sections, we impose the integrality condition 
of the monodromy matrices $M_{\tilde{g}_i}$ for generators $g_i$ 
of the vacuum identification group $[\Gamma_S]$ other than $w^{(k)}$. 
There, we may eliminate the invariants $a_{11}+\Z$ and $b_{11}+24\Z$ from 
the analysis; the $+\Z$ and $+24\Z$ ambiguity correspond to 
a choice of a basis of magnetic charges of the $(\rho+2)$ U(1) vector 
fields in the 4D effective field theory, so we set 
\begin{align}
 a_{11} = \frac{d_{111}}{2}, \qquad \qquad 
 b_1 = (c_2)_1 
 \label{eq:eliminate-aNb}
\end{align}
and forget the conditions (\ref{eq:cond-PQ-2}, \ref{eq:cond-Fricke-inZ}) 
in the rest of the analysis. The integrality condition of some of 
$M_{\tilde{g}_i}$ may still involve $d_{111}$ and $(c_2)_1$ 
(e.g., (\ref{eq:cond-PQ-1}) for $M_{\tilde{g}_\infty}$), but such conditions 
can be read as those on the BPS classification invariants 
$\{ n_\gamma \}$, $\{ m_\gamma \}$ and $d_\gamma$'s, because $d_{111}$ and 
$(c_2)_1$ are determined by the BPS classification invariants
(e.g., (\ref{eq:d111-k=2}--\ref{eq:chiMatch-k=2})). 
We will find that certain values of the BPS classification invariants 
are not realized by branches the vacuum moduli space, and we also learn 
that certain diffeomorphism classes of real six-dimensional manifolds 
do not fit for Type IIA compactifications. 

\subsection{BPS Invariants with Integral Monodromy Matrices}
 
\subsubsection{Level-2 (Degree-4: Quartic-K3 fibrations)}
\label{sssec:level-2}


First, let us list up the BPS classification invariants for $\rho =1$ 
degree-$(2k=4)$ branches of the vacuum moduli space by 
following \cite{EW19}. They are\footnote{
The invariants $m_0$ and $m_1$ are equal to 0 as explained in \cite{EW19}.
} %
\begin{align}
& \left\{ n_0 = -2, \; n_1 \in \Z_{\geq 0}, \; n_2 \in \Z_{\geq -2} \; | \; 
    \chi_{\rm match} \leq 4 \right\}, \nonumber \\
& m_{2} \in 12\Z, \qquad    d_0(0) \in 12 \Z_{\geq 0}; 
 \label{eq:BPSpara-k=2}
\end{align}
because the invariant $d_0(0)/24$ has an interpretation as the 1-loop beta 
function of a probe gauge group in the language of the 4D effective 
field theory, we will use 
\begin{align}
 b_{\cal R} := \frac{d_0(0)}{24} \in 2^{-1} \Z_{\geq 0}
\end{align}
instead of $d_0(0)$ whenever $d_0(0)$ is in the list of the BPS 
classification invariants in the rest of this article. 
For $\chi_{\rm match}$ in the inequality above, see the next paragraph. 

The 4D effective Lagrangian parameters are determined 
in terms of the BPS classification invariants as follows:
\begin{align}
  d_{111}& \; =-4n_0 -7 n_{1} -n_{2} + 3\ell_{\rm aux} + 12 \delta n,
      \label{eq:d111-k=2} \\
      (c_2)_1& \; =-28n_0 -34n_{1}-4n_{2} + 6\ell_{\rm aux} + 24 \delta n,
       \label{eq:c2-k=2} \\
  & d'_{\rm match} = 10n_0+10 n_{1}+n_{2}, \\
\chi_{\rm match} & \; =84n_0+128n_{1}+14n_{2}.  \label{eq:chiMatch-k=2}
\end{align}
where $\ell_{\rm aux} := [-d_0(0)/24 + m_{2/4}/12]$. This was 
described as the fourth task in section \ref{ssec:to-do-list}; 
the second and third tasks had been done and the results recorded 
in the previous section.  

Let us demand that all the monodromy matrices of the duality transformations 
in $\Gamma_S$ be $\Z$-valued and derive constraints on the BPS classification 
invariants (\ref{eq:BPSpara-k=2}). In the case $k=2$, $g_\infty$ and 
$w^{(2)}$ are enough to generate $\Gamma_S$, which means that 
we just have to impose the condition (\ref{eq:cond-PQ-1}) along with\footnote{
This reminder---stated already at the end of section \ref{ssec:NLSM}---will 
no longer be repeated in the rest of this article.
} %
(\ref{eq:eliminate-aNb}) on the BPS classification 
invariants (\ref{eq:BPSpara-k=2}). 
The first one in (\ref{eq:cond-PQ-1}) and 
the second one are translated to 
\begin{align}
b_{\cal R} \in \Z_{\geq 0}, 
  \label{eq:b-is-Z}
\end{align}
and\footnote{
It thus follows as a consequence automatically that the probe gauge group 
must be free of Witten SU(2) anomaly. This was observed already 
in \cite{ESW20} for a few other choices of $\Lambda_S$. 
More examples are found later in this article ($k=5,6$). In fact, by applying 
the discussion in the appendix \ref{sec:Witten} to the symmetry-enhanced 
branch, we can conclude that $b_{\cal R} \in \Z$ immediately for any $\Lambda_S$.
} %
\begin{align}
 n_{2} \in 2\Z, 
  \label{eq:cond-k=2-n2-even}
\end{align}
respectively. A list of Calabi--Yau threefolds that belong to 
the case $k=2$ \cite[Table 1]{KKRS} already provided a hint 
for the pattern $n_2\in 2\Z$; now we have a theory for the 
empirical pattern.

For most of branches of vacua with $k=2$, the conservative estimate 
of the vacuum identification group $[\Gamma_S]$ is $\Gamma_S$, so 
we have already implemented all the conditions for the monodromy matrices 
of $[\Gamma_S]=\Gamma_S$ to be integer valued. 
In a branch where $n_2^V=1$ for some hypermultiplet moduli vacuum 
expectation value (vev), however, a massless SU(2) vector field 
emerges at a massless matter singularity for a charge 
$v_* \in \gamma=[2] \in \Z_4 = G_S$ 
satisfying (\ref{eq:reflection_conditions}), and the reflection 
symmetry $R_{v_*}$ should be within $[\Gamma_S]$. The group $[\Gamma_S]$
cast into the group $\Z_2 \times \Z_2$ (see (\ref{eq:qot-grp-fint}))
consists not only of $(0,0)$, $(1,1)$, but also $(0,1)$ represented 
by $R_{v_*}$ (read out $\epsilon =0$ in the $k=2$ column of 
Table \ref{tab:v*-list}, and footnote \ref{fn:reflectn-outof-Gamma0kP}), 
and hence also of $(1,0)$. So, $[\Gamma_S] = 
{\rm Isom}'(\widetilde{\Lambda}_S)$. 
Having $[\Gamma_S]$ strictly larger than $\Gamma_S$ in the $k=2$ case, however,
does not yield more conditions on the invariants $\{ n_\gamma \}$.  
We have seen\footnote{
\label{fn:prm-pwr-level-think-just-GammaS}
With this observation in mind, one can also argue as follows 
when the level $k$ contains just one prime factor (i.e., $k=2,3,4,5,7,\cdots$).
First, $(\Z_2)^s = \Z_2$. Second, suppose that we have identified 
all the constraints so that the monodromy matrices of $\Gamma_S$ 
are integer valued. Now, if there is $g \in [\Gamma_S]$ that is cast 
into $(0,1) \in (\Z_2)^s \times \Z_2$, then the monodromy matrix 
$M_{\tilde{g}}$ is integer valued without imposing an extra constraint, 
regardless of whether there is a tune in hypermultiplet moduli so that 
$n_{\gamma*}^V=1$. There must also be elements of $[\Gamma_S]$ then 
that are cast into $(1,0) \in (\Z_2)^s \times \Z_2$, but their 
monodromy matrices are also integer valued without an extra constraint.
} %
 in footnote \ref{fn:on(-id)}
that the matrix $M_{\tilde{g}}$ for the duality transformation $g \in [\Gamma_S]$
(with $g \nin \Gamma_0(k)_+$) is integer valued if and only if the 
$3 \times 3$ matrix $\Lambda_{-\tilde{g}}$ calculated formally for the 
monodromy of the symplectic section along a path from $t_0$ to $t_0^{-g}$ 
for $-g \in \Gamma_0(k)_+$. In the present case, $-g$ for $g = R_{v_*}$ is 
classified as $(0,0)$ in the group $(\Z_2)^s|_{s=1} \times \Z_2$; 
the constraints (\ref{eq:b-is-Z}) and (\ref{eq:cond-k=2-n2-even}) 
therefore guarantee that the $3\times 3$ matrix $\Lambda_{-\tilde{R}_{v*}}$
is integer valued, and also that the matrix $M_{\widetilde{R}_{v*}}$ is.

To conclude, all the possible range of the BPS classification 
invariants in a vacuum branch with $k=2$ 
are as follows: (\ref{eq:BPSpara-k=2}) with the 
constraints (\ref{eq:b-is-Z}, \ref{eq:cond-k=2-n2-even}).

All those vacuum branches pass the test for Type IIA NLSM interpretations, 
as remarked already in section \ref{ssec:NLSM}.  
So, we can derive a list that covers all the diffeomorphism classes 
of Calabi--Yau threefolds that have a lattice-$\vev{+4}$-polarized 
regular K3-fibration, by exploiting the BPS classification 
invariants in the range derived above.
Noting that (i) the combination of (\ref{eq:BPSpara-k=2}) 
and (\ref{eq:b-is-Z}) allows $\ell_{\rm aux}$ to be scanned 
freely within $\Z$, and that (ii) $\delta n \in \Z$ 
corresponds to $\Delta \ell_{\rm aux} \in 4\Z$ in (\ref{eq:d111-k=2},
 \ref{eq:c2-k=2}), we find that there are at most four choices of 
$\chi_{\rm match}$, $d_{111}$ and $(c_2)_1$ for a given $\{ n_\gamma \}$ 
that cannot be reinterpreted by a change of basis of $H^2(M;\Z)$.  
So, as a whole, there can be at most 40 diffeomorphism classes,\footnote{
\label{fn:caution} 
We use the expression ``there can be at most ...'' here, because 
our study only imposes the theoretical consistency conditions 
that we are aware of, and are technically not too difficult to implement. 
We will not repeat this reminder in the rest of this section, but 
this reminder applies to all the cases with other $\Lambda_S = \vev{+2k}$. 
} %
 listed in Table \ref{tab:CY-k=2}.
\begin{table}[tbp]
\begin{center}
\begin{tabular}{|c||ccccccc|ccc|}
  \hline 
$(n_1,n_2)$ & (0,$-2$) & (0,0) & (0,2) & (0,4) & (0,6) & (0,8) & (0,10) &
    (1,$-2$) & (1,0) & (1,2) \\
\hline 
$\chi$ & -196 & -168 & -140 & -112 & -84 & -56 & -28 & -68 & -40 & -12 \\
$d_{111}+12\delta n$ & 10$_{+}$ & 8$_+$ & 6$_+$ & 4$_+$ & 2$_+$ & 0$_+$ & -2$_+$ 
  & 3$_+$ & 1$_+$ & -1$_+$    \\
$(c_2)_1+24\delta n$ & 64$_{+}$ & 56$_+$ & 48$_+$ & 40$_+$ & 32$_+$ & 24$_+$ & 16$_+$ & 30$_+$ & 22$_+$ & 14$_+$ \\
  \hline
\end{tabular}
\caption{\label{tab:CY-k=2}
the list of possible values of the topological invariants of a Calabi--Yau threefold $M$ with a regular $\vev{+4}$-polarized K3-fibration. 
$\chi = \chi_{\rm Euler}(M)$ in the 2nd row. The 3rd and 4th rows are $H^3$ 
and $\vev{H, c_2(TM)}$, respectively, where $H$ is a generator of $H_4(M;\Z)$ 
along with the K3 fiber class $D_s$; their values change as we replace $H$ 
by $H + (\delta n)D_s$ for arbitrary $\delta n \in \Z$. We should read 
an entry $X_+$ in the 3rd row [resp. $Y_+$ in the 4th row] as the 
set of four different values $d_{111}+12 \delta n \in \{X+12\delta n, 
X+3+12\delta n, X+6+12\delta n, X+9+12\delta n\}$ [resp. 
$(c_2)_1+24\delta n \in \{
Y+24\delta n, Y+6+24\delta n, Y+12+24\delta n, Y+18+24\delta n\}$]. 
There are ten columns here, and each has four distinct values 
of $\ell_{\rm aux}$, so there can be forty diffeomorphism classes of 
such Calabi--Yau threefolds $M$. }
\end{center}
\end{table}

Table 1 of \cite{KKRS} includes a list of explicit constructions 
of Calabi--Yau threefolds that belong to the case $k=2$ in the 
form of complete intersections of toric varieties obtained by 
scanning the toric data within a certain range; the list contains 
ten entries corresponding to $(n_1,n_2) = (0, -2\sim 6)$. 
Logically one of the following three must be right: (a) Calabi--Yau 
threefolds do exist for the remaining 30 diffeomorphism classes, 
(b) some of the 30 classes are not realized by a Calabi--Yau threefold, 
yet there are consistent string vacua 
that appear to have some observables consistent with an interpretation 
as a compactification over such a real six-dimensional manifold (that does 
not actually exist), and (c) we still fail to implement some 
consistency conditions in string theory, and those consistency 
conditions rule out some of the remaining 30 hypothetical choices of 
the classification invariants.  Search in a larger database of 
Calabi--Yau threefolds will give a hint on whether (a) is right, 
or (b, c) is right, but we do not try to do so by ourselves 
within this article, due to our limited experience.  A related 
comment is also found at the end of section \ref{sssec:level-3}. 

\subsubsection{Level-3 (Degree-6)}
\label{sssec:level-3}

First, we begin with listing up the BPS classification invariants 
in a $\rho=1$ $k=3$ branches of the Heterotic--IIA dual vacua moduli space. 
They are 
\begin{align}
& \left\{ n_0 = -2, \; n_{1}, n_{2} \in \Z_{\geq 0}, \; n_{3} \in \Z_{\geq -2} \; 
    | \; \chi_{\rm match} \leq 4 \right\}, \nonumber \\
& m_{2}, m_{3} \in 12\Z, \qquad b_{\cal R} \in 2^{-1} \Z_{\geq 0}. 
 \label{eq:BPSpara-k=3}
\end{align}
The following parameters of the 4D effective Lagrangian are also computed by 
following the procedure in \cite{EW19}:
\begin{align}
 d_{111} & \; =-8n_0 -11n_{1} -8n_{2} -n_{3}/2 -3 \ell_{\rm aux} + 18\delta n,
     \label{eq:d111-k=3} \\
     (c_2)_1 & \; = -34n_0-38n_{1}-20n_{2}-n_{3} -4 \ell_{\rm aux} + 24 \delta n,
       \label{eq:c2-k=3} \\
   & d'_{\rm match} = \frac{35}{2}n_0 + \frac{35}{2}n_{1}+ 7n_{2}
       + \frac{1}{4}n_{3},  \\
 \chi_{\rm match} & \; = 74n_0 + 108n_{1} + 54n_{2} + 2n_{3}, 
     \label{eq:chiMatch-k=3}
\end{align}
where $\ell_{\rm aux}$ is the following combination 
\begin{align}
    \ell_{\rm aux} := 2b_{\cal R} -m_{2}/2 - m_{3}/12 
  \label{eq:BPSpara-k=3-aux}
\end{align}
guaranteed to be an integer. 

To make sure that all the monodromy matrices $M_{\tilde{g}}$ are integer 
valued for all $g \in \Gamma_S$, it is enough to impose the condition 
(\ref{eq:cond-PQ-1}) on the BPS classification 
invariants (\ref{eq:BPSpara-k=3}) in the $k=3$ case, just like 
in the $k=2$ case; this is because 
$g_\infty$ and $w^{(3)}$ are enough in generating the group $\Gamma_S$
The first and second conditions in (\ref{eq:cond-PQ-1}) 
are read as 
\begin{align}
  n_{3} \in 2\Z_{\geq -2} 
 \label{eq:cond-PQ-inZ-k=3}
\end{align}
and 
\begin{align}
 \ell_{\rm aux} \in n_{3}-2+6\Z,
 \label{eq:cond-aux-k=3}
\end{align}
respectively. 

When there is a tune in the hypermultiplet moduli such that 
$n_3^V = 1$, we do have a logic that the SU(2) Weyl reflection $R_{v*}$
must be in the vacuum identification group $[\Gamma_S]$. This thought 
does not yield an extra constraint, however, because $R_{v*}$ in the 
case $k=3$ is cast into the element $(1,1)$ in the 
group (\ref{eq:qot-grp-fint}) (read out $\epsilon =1$ in the $k=3$ 
column of Table \ref{tab:v*-list}); this means that $R_{v_*}$ is within 
$(\Gamma_0(k)/\{ \pm 1\}) \cdot w^{(k)}$, and hence $R_{v_*}$ represents 
an equivalence of the $(c,\tilde{c}) = (\rho,3)$ lattice SCFT that 
is trivially extended to the equivalence of the $(c,\tilde{c}) = (22,9)$ 
SCFT on the Heterotic string worldsheet. In particular, the monodromy 
matrix $M_{\tilde{R}_{v_*}}$ is guaranteed to be integer valued, once all 
of the conditions so far---(\ref{eq:cond-PQ-inZ-k=3}) 
and (\ref{eq:cond-aux-k=3})---are satisfied. 

To summarize, in the case of $\Lambda_S = \vev{+6}$, 
the BPS classification 
invariants (\ref{eq:BPSpara-k=3}, \ref{eq:BPSpara-k=3-aux}) 
are further subject to (\ref{eq:cond-PQ-inZ-k=3}, \ref{eq:cond-aux-k=3}). 
A pair of vacuum branches that differ only by 
$\Delta \ell_{\rm aux} \in 6\Z$ just correspond to one 
diffeomorphism class, because $\Delta \ell = +6$ is absorbed 
by $\delta n = +1$. So, $n_{1}, n_{2} \in \Z_{\geq 0}$ and 
$n_{3} \in 2\Z_{\geq -2}$ combined determines a diffeomorphism 
class. Their list is as follows:
\begin{itemize}
\item $(n_1,n_2,n_3,\ell_{\rm aux},\chi) = (0,0,n_3,n_3-2+6\Z, -148+2n_3)$ with 
   $n_3 = -2, 0,2,\cdots, 76$, 
\item $(n_1,n_2,n_3,\ell_{\rm aux},\chi) = (0,1,n_3,n_3-2+6\Z,-94+2n_3)$ with 
   $n_3 = -2, 0, 2, \cdots, 48$, 
\item $(n_1,n_2,n_3,\ell_{\rm aux},\chi) = (0,2,n_3,n_3-2+6\Z,-40+2n_3)$ with 
   $n_3 = -2, 0, 2, \cdots, 22$, 
\item $(n_1,n_2,n_3,\ell_{\rm aux}, \chi) = (1,0,n_3,n_3-2+6\Z,-40+2n_3)$ with 
   $n_3 = -2, 0, 2, \cdots, 22$. 
\end{itemize}
There can be at most 40 + 26 + 13 + 13 = 92 diffeomorphism classes of 
Calabi--Yau 
threefolds that have a lattice-$\vev{+6}$-polarized regular K3-fibration. 
Only seven of them have been listed in Table 1 of \cite{KKRS}. 

Searching for Calabi--Yau threefolds in the $k=3$ case in databases  
requires more effort than to search them in the $k=2$ case. 
Whereas degree-4 K3 surfaces can be constructed as a degree-4 hypersurface 
of $\P^3$, it is not known how to construct a degree-6 K3 surface as 
a {\it hypersurface} of a projective space (see also discussion in 
p. \pageref{page:construct-K3}). So it is essential to look for 
a database that covers constructions involving multiple defining equations 
on an ambient space \cite{Hubsch:1986ny, Green:1986ck, Candelas:1987kf}. 
{\it Complete intersection} constructs a manifold 
$M$ within an ambient space $\P$ as the common zero locus 
$f_1 = f_2 = \cdots = 0$ of multiple holomorphic sections of 
line bundles $L_1$, $L_2$, $\cdots$ on $\P$. {\it Generalized complete 
intersection} constructs a manifold also as the common zero locus 
in $\P$, $f_1=f_2 = \cdots = 0$, but some of $f_i$'s---let's say 
that is $f_1$---are/is allowed to be meromorphic on $\P$ so long as 
$f_1$ is holomorphic when restricted to the common zero locus 
of all other $f_i$'s \cite{Anderson:2015iia, Berglund:2016yqo, 
Garbagnati:2017rtb}. The choice of ambient spaces 
may be scanned within the class of direct products of projective 
spaces (e.g., \cite{Anderson:2017aux}), those of toric varieties, 
and of general toric varieties that have a toric fibration morphism over 
a base toric variety (e.g., \cite{Braun:2011ux} and references therein) 
in which the base curve $\P^1$ is constructed.  

It is natural to ask the same question on the 92-7 missing diffeomorphism 
classes as in the $k=2$ case. Partially due to limited experience of the 
present authors, and partially due to the fact that such database building 
is still in the growing stage than in the matured phase, the authors 
think of it beyond the scope of this article to look for the 92-7 
remaining diffeomorphism classes in the databases available already.

\subsubsection{Level-4 (Degree-8)}
\label{ssec:level-4}

First, let us list up the BPS classification invariants for the 
$\rho =1$ $k=4$ moduli spaces. They are 
\begin{align}
& \left\{  n_0 = -2, \; n_{1}, n_{2}, n_{3}, n_4 \in \Z_{\geq 0}, \; 
     | \; \chi_{\rm match}\leq 4\right\},  \nonumber \\
& m_{3}, \; d_1(1/16) \in 12\Z, \qquad  b_{\cal R} \in 2^{-1} \Z_{\geq 0}.
\label{eq:BPSpara-k=4}
\end{align}
The 4D effective Lagrangian parameters are determined as follows, 
by following the procedure outlined in \cite{EW19}:
\begin{align}
  d_{111} & \; = -16n_0-24 n_1-8 n_2-n_3 + \ell_{\rm aux} + 24 \delta n,
      \label{eq:d111-k=4} \\
 (c_2)_1 & \; = -40n_0-54 n_1-20 n_2-4 n_3 - 12n_4
         + \ell_{\rm aux} + 24 \delta n ,   \label{eq:c2-k=4} \\
 & d'_{\rm match} = 24n_0 + 30n_{1}+12n_{2}+3n_{3}+12n_4, \\
 \chi_{\rm match} & \; = 56n_0 + 112 n_{1} + 56 n_{2} + 16 n_{3} +70 n_4, 
    \label{eq:chiMatch-k=4}
\end{align}
where
\begin{align}
 \ell_{\rm aux} := [-4 d_0-d_1+m_3]/12. 
\end{align}
Here, as well as in sections \ref{ssec:PP-Fricke} and \ref{ssec:PP-Gamma0k}, 
we did all the computation including the cases where the invariant 
$n_4$ is not necessarily zero (Ref. \cite{EW19} was concerned only 
about the $n_4=0$ cases, where a NLSM interpretation is available in the 
Type IIA description). 

The first condition in (\ref{eq:cond-PQ-1}) is satisfied automatically 
when imposed on the BPS classification invariants  (\ref{eq:BPSpara-k=4}), 
while the second condition in (\ref{eq:cond-PQ-1}) is translated to 
\begin{align}
 \ell_{\rm aux} -2(n_{1}+n_{3}) \in 4\Z.  
 \label{eq:cond-Fricke-on-BPS-k=4}
\end{align}
This is all the constraint on the invariants (\ref{eq:BPSpara-k=4})
when the monodromy matrices are required to be integer valued for 
all the duality transformations in $\Gamma_S$, because $\Gamma_S$ 
is generated by $g_\infty$ and $w^{(4)}$ 
(see (\ref{eq:relatn-k=4-g2-generated})). There cannot be 
a W-boson charge $v_*$ for an SU(2) gauge group enhancement
(see Table \ref{tab:v*-list}) in vacuum branches with $k=4$; 
the authors do not find other reasonings that hints at 
$[\Gamma_S]$ strictly larger than $\Gamma_S$ for cases with $k=4$. 

In a different perspective, one may also think of a case where the 
group $[\Gamma_S]$ is the maximal ${\rm Isom}'(\widetilde{\Lambda}_S)$
for whatever reasons, and study its consequences. In the case with 
$k=4$, the argument in footnote \ref{fn:prm-pwr-level-think-just-GammaS}
applies. So, the monodromy matrices of all the elements in $[\Gamma_S] = {\rm Isom}'(\widetilde{\Lambda}_S)$ are guaranteed to be integer valued, 
when one imposes all the conditions---(\ref{eq:cond-Fricke-on-BPS-k=4})---for 
the monodromy matrices of $\Gamma_S$ to be integer valued. This argument 
holds true for all the levels $k$ that is a prime power ($s=1$).  

That makes it ready for us to list up all the diffeomorphism 
classes of Calabi--Yau threefolds with a $\vev{+8}$-polarized 
regular K3-fibration. For a given $\{ n_\gamma \}$, the value of 
$\ell_{\rm aux}$ is quantized by $+4\Z$ whereas $\delta n \in \Z$
is equivalent to $\Delta \ell_{\rm aux} \in 24\Z$. 
This means that there are 6 different diffeomorphism classes for 
a given $\{ n_\gamma \}$. The possible combinations of $\{ n_\gamma \}$
can be worked out just like we did in Table \ref{tab:CY-k=2} 
for the $k=2$ cases, and also at the end of section \ref{sssec:level-3}
for the $k=3$ cases (see also a comment at the beginning 
of section \ref{ssec:mod-diffeo-summary} on the cases with $n_4>0$). 

\subsubsection{Level-5 (Degree-10)}
\label{sssec:level-5}

First, the BPS classification invariants of $\rho=1$ and $k=5$ 
branches are listed up, by following \cite{EW19}:
\begin{align}
& \left\{ n_0 = -2, \;  n_{1}, n_{2}, n_{3}, n_{5} \in \Z_{\geq 0},
  \; n_{4} \in \Z_{\geq -2} \; | \; \chi_{\rm match} \leq 4 \right\}, \nonumber \\
& m_{3}, \; m_{4} \in 12\Z, \quad 
 b_{\cal R} \in 2^{-1}\Z_{\geq 0}. 
  \label{eq:BPSpara-k=5}
\end{align}
The 4D effective Lagrangian parameters are computed as follows, 
by following the procedure in \cite{EW19}:
\begin{align}
 d_{111} & \; = 30 \delta n + 5 \ell_{\rm aux}
 -20n_0 -19 n_1-24 n_2-11 n_3-2 n_4-15 n_5,  \label{eq:d111-k=5} \\
 (c_2)_1 & \; = 24 \delta n + 4 \ell_{\rm aux} 
 -44n_0-46 n_1-36 n_2-14 n_3-2 n_4-18 n_5,  \\
 & d'_{\rm match} = 35n_0+\frac{77}{2}n_1+21n_2+\frac{13}{2}n_3
         +\frac{1}{2}n_4+\frac{15}{2}n_5, \\
 \chi_{\rm match} & \; = 60n_0 +88 n_1+66n_2+24n_3+2n_4+32n_5, 
    \label{eq:chiMatch-k=5}   
\end{align}
where
\begin{align}
 \ell_{\rm aux} := [-((3 d_0)/2)+6 m_3+3 m_4]/12. 
\end{align}

To make sure that all the monodromy matrices are integer valued 
for all $g \in [\Gamma_S]$, let us first impose the 
condition (\ref{eq:cond-PQ-1}), which is for $g_\infty \in \Gamma_S$. 
The two conditions in (\ref{eq:cond-PQ-1}) are translated into 
\begin{align}
 b_{\cal R} \in \Z_{\geq 0}, \qquad \ell_{\rm aux}-3n_{4} \in 6\Z.
  \label{eq:b-is-Z-k=5}
\end{align}
Thus, $\ell_{\rm aux}$ is quantized by $+6\Z$, while $\delta n \in \Z$ 
corresponds to $\Delta \ell_{\rm aux} \in 6\Z$. 
So, there can be just one diffeomorphism class for a given $\{n_\gamma\}$, 
for now. 

In the case of $k=5$, we need either $g_2$ or $g_3$ 
along with $g_\infty$ and $w^{(k)}$ in generating the group $\Gamma_S$, 
and just one of $g_2$ or $g_3$ is enough because of the relation 
(\ref{eq:relatn-k=5-g3-generated}). The off-diagonal $3\times 3$ part 
$\Lambda_{\tilde{g}_2}$ of the monodromy matrix $M_{\tilde{g}_2}$ is 
computed by using $\left(Q|[g_2]_{-4}-Q\right)(t)$ and the period 
polynomial $P_{\tilde{g}_2}(t;f_*)$ in (\ref{eq:pp-k=5-g2}); see \cite{EW21}
and the notation (\ref{eq:def-|opr});  
the matrix is written down in terms of the BPS classification invariants 
for the choice (\ref{eq:eliminate-aNb}):  
\begin{align*}
 (\Lambda_{\tilde{g}_2})_{00} & \; = -56 \delta n
    - \frac{28}{3}(\ell_{\rm aux}-3n_4) 
    + (107/2)n_0 + 55n_1 + 51n_2 + 23n_3 - 25n_4 + 27n_5,  \\
 (\Lambda_{\tilde{g}_2})_{01} & \; = 262 \delta n
    + \frac{131}{3}(\ell_{\rm aux}-3n_4)
    -257n_0 -265n_1 -248n_2 -111n_3 + 116n_4 -129 n_5,   \\
 (\Lambda_{\tilde{g}_2})_{\#\#} & \; = -64\delta n
    - \frac{32}{3}(\ell_{\rm aux}-3n_4)
    + (133/2)n_0 +69n_1 + 65 n_2 + 29n_3 -28n_4 +33n_5,           \\
 (\Lambda_{\tilde{g}_2})_{\# 1} & \; = 278 \delta n
    + \frac{139}{3}(\ell_{\rm aux}-3n_4)
    - 283n_0 -293n_1 -276n_2 -123 n_3 +122n_4 -141n_5 , 
\end{align*}
and finally,
\begin{align*}
  (\Lambda_{\tilde{g}_2})_{11} & \; +2k(\Lambda_{\tilde{g}_2})_{0\#} \\
  & \; = -1800 \delta n -300 \ell_{\rm aux}
   +1800n_0 + 1860 n_1 +1748 n_2 +780n_3 + 107 n_4 + 900 n_5. 
\end{align*}
In light of the constraint (\ref{eq:b-is-Z-k=5}), we see that 
the condition that the monodromy matrix
$M_{\tilde{g}_2} = M(g_2,\Lambda_{\tilde{g}_2})$ should be integer valued
does not yield an extra condition.

The monodromy matrix
$M(g_3,\Lambda_{\tilde{g}_3})$ is automatically integer valued; we can
compute it either by using the relation (\ref{eq:relatn-k=5-g3-generated})
(while paying attention to the monodromy around the massless matter 
singularities (cf \cite{AFGNT})), 
or by following the same procedure as above for the period polynomial
(\ref{eq:pp-k=5-g2}). We do not do that, however, because we have already
obtained all the constraints on the classification invariants from the
integrality of the monodromy matrices. 

It is true that the perturbative duality transformation group 
$[\Gamma_S]$ can be larger than $\Gamma_S$ in a $k=5$ branch with 
a hypermultiplet tune so that $n_4^V=1$ (see Table \ref{tab:v*-list}). 
The SU(2) Weyl reflection $R_{v*}$ is not within $\Gamma_S$. 
Its monodromy matrix is guaranteed to be integer valued, however, once the 
conditions---(\ref{eq:b-is-Z-k=5})---are imposed so that the monodromy 
matrices for those in $\Gamma_S$ are all integer valued.  
To see this, we just have to repeat the same logic as in the $k=2,4$ cases. 
So, for all the vacuum branches with $k=5$, we have obtained all the 
constraints for the integrality of the monodromy matrices of $\Gamma_S$ 
and $R_{v_*}$. 

Now one can find a list of diffeomorphism classes that still have a chance 
to be realized by Calabi--Yau threefolds with a $\vev{+10}$-lattice-polarized 
regular K3-fibration.
We have already seen below (\ref{eq:b-is-Z-k=5}) that there is
just one diffeomorphism class for a given choice of the invariants
$\{ n_\gamma\}$. So, one just has to count the number of choices of
$\{ n_\gamma \}$ satisfying the constraint in (\ref{eq:BPSpara-k=5}).
There can be at most  
\begin{align}
1 \times & \left[ (65+53+41+29+17+5)
+(49+37+25+13+1)
+(33+21+9) \right. \nonumber \\
& \left. +17+5+1+20+8+16+4+21+9+5 \right]
\end{align}
diffeomorphism classes; each number within $[ \quad ]$
corresponds to the number of different choices of $n_4$ 
for a given $\{ n_{1,2,5,3} \}$ (see Table \ref{tab:k=5-all-n} and 
read out (the 5th row + 3)). 
\begin{table}[tbp]
  \begin{center}
    \begin{tabular}{c||c|c|c|c|c|c|c|c|c|c|c|c|c|c}
$n_1$ & 0          & 0         & 0   & 0&0& 0 & 0 & 0 & 0 & 0 & 1 & 1 & 1 \\
$n_2$ & 0          & 0         & 0   & 0&0& 0 & 1 & 1 & 1 & 1 & 0 & 0 & 0 \\
$n_5$ & 0          & 1         & 2   & 3&3& 4 & 0 & 0 & 1 & 1 & 0 & 0 & 1 \\
$n_3$ & 0$\cdots$5 & 0$\cdots$4 & 0.1.2 & 0&1& 0 & 1 & 2 & 0 & 1 & 0 & 1 & 0 \\
$n_4\leq$
    & 62$\cdots$2 & 46$\cdots$(-2) & 30.18.6 &14&2& -2& 17& 5 & 13& 1 & 18& 6 & 2 \\
 \hline 
$\leq \chi$ & -124$\cdots$4 & -92$\cdots$-4 & -60.-36.-12 &-28&-4&4 &-34&-10&-26& -2& -36 & -12 & -4
     \end{tabular}
\caption{\label{tab:k=5-all-n} All the possible choices of
$\{ n_\gamma \}$ in a Calabi--Yau threefold with a
$\vev{+10}$-polarized regular K3 fibration are listed here.
Each one of the ten columns on the right is for a specific 
$\{ n_1, n_2,n_5,n_3\}$ in the upper four rows. The possible value 
of $n_4$ is $\{ -2, -1,0, \cdots, \}$ whose upper bound is written 
in the 5th row of the column. The corresponding value of 
$\chi_{\rm match} = \chi(M)$ is (the lower bound)$+2(n_4+2)$, where 
the lower bound is written in the 6th row. Each one of the three 
remaining columns on the left is for a specific $\{ n_1, n_2, n_5\}$
where $n_3$ may take value in the range $0,1,\cdots$ specified 
in the 4th row. The upper bounds on $n_4$ for those $n_3$'s 
are written in the 5th row. The lower bounds on $\chi(M)$ 
for those $\{ n_{1,2,5,3} \}$ are in the 6th row. 
} 
    \end{center}
\end{table}

Because a K3 surface with Picard number 1 and degree more than 8
cannot be constructed as a complete intersection in a projective space 
(cf p. \pageref{page:construct-K3}), 
we cannot hope to construct a Calabi--Yau threefold and
a $\vev{+2k}$-polarized K3 fibration for large $k$ easily 
by using a toric ambient space and a toric projection morphism. 
The study above has managed to avoid the limitation of 
construction-based approach and obtained a complete list of those threefolds. 

\subsubsection{Level-6 (Degree-12)}
\label{sssec:level-6}

The BPS classification invariants of the $\rho = 1$ $k=6$ vacuum 
branches are as follows: 
\begin{align}
 \left\{ n_0 = -2, \quad n_{1,2,3,5} \in \Z, \quad n_{4,6} \in \Z_{\geq -2} 
    \; | \; \chi_{\rm match} \leq 4 \right\}, \nonumber \\
 m_{3,4}, \; d_1(1/24) \in 12\Z, \quad b_{\cal R} \in 2^{-1}\Z_{\geq 0}.
   \label{eq:BPSpara-k=6}
\end{align}
The 4D effective Lagrangian parameters for those branches 
turn out to be 
\begin{align}
d_{111} & \; =   3\ell_{\rm aux} +36\delta n
    -27n_0 -46 n_1-26 n_2-9 n_3-n_4+n_5, \label{eq:d111-k=6} \\
(c_2)_1 & \; = 2\ell_{\rm aux} +24\delta n
    -46n_0 -68 n_1-36 n_2-14 n_3-2 n_4-18 n_5-2 n_6, \\
  & \; d'_{\rm match} = 42n_0 +56n_1 +28n_2 +12n_3 +2n_4 +28n_5 +3n_6, \\
 \chi_{\rm match} & \; =46n_0+96n_1+60n_2+32n_3+6n_4+96n_5+10n_6, 
   \label{eq:chiMatch-k=6}
\end{align}
where 
\begin{align}
 \ell_{\rm aux} := [-5 d_0/2- d_1+7 m_3+ m_4- m_6]/12.  
\end{align}

The condition (\ref{eq:cond-PQ-1}) for the monodromy matrix of 
$g_\infty$ to be integer valued is translated to 
\begin{align}
  b_{\cal R} \in \Z_{\geq 0}, \qquad 
  6n'_6 := 4\ell_{\rm aux}-2n_0-2n_1-2n_2-4n_3-2n_4-2n_5-n_6 \in 6\Z. 
  \label{eq:b-is-Z-k=6}
\end{align}
So, $\ell_{\rm aux}$ is always integer valued. One may also read out 
that an integer $\ell_{\rm aux}$ exists for a given $\{ n_\gamma \}$ 
if and only if
\begin{align}
 n_6 \in 2\Z_{\geq -2};
   \label{eq:n6-is-even-k=6}         
\end{align}
the value of $\ell_{\rm aux}$ for a given $\{ n_\gamma \}$ is quantized 
by $+3\Z$. On the other hand, $\delta n = 1$ corresponds to 
$\delta \ell_{\rm aux} = 12$. So, there are four distinct diffeomorphism 
classes of Calabi--Yau threefolds with a $\vev{+12}$-polarized regular 
K3 fibration for a given $\{ n_\gamma \}$ satisfying 
(\ref{eq:BPSpara-k=6}, \ref{eq:n6-is-even-k=6}). 

The monodromy matrix $M_{\tilde{g}_2} = M(g_2, \Lambda_{\tilde{g}_2})$
can be calculated by using the period polynomial (\ref{eq:pp-k=6-g2}), 
just like we have done in the cases with $k=1,2,3,4$. We choose to 
write down the $3 \times 3$ part of the monodromy matrix in terms of 
$\{n_{1,2,3,4,5} \}$, $\ell_{\rm aux}$ and $n'_6$ (instead of $n_6$),  
which can take integer values independently from one another.  
The result is this: 
\begin{align*}
 (\Lambda_{\tilde{g}_2})_{00} & \;
    = -615 n'_6 + 145 \ell'_{\rm aux} + 3007 n_0 + 5014 n_1 + 2595 n_2
          + 575 n_3 - 91 n_4 + 59 n_5, \\
 (\Lambda_{\tilde{g}_2})_{01} & \;
    = 2976 n'_6 - 723 \ell'_{\rm aux} - 14313 n_0 - 23885 n_1 - 12358 n_2
         - 2721 n_3 + 445 n_4 - 326 n_5, \\
 (\Lambda_{\tilde{g}_2})_{\sharp \sharp} & \;
    = -586 n'_6 + 155 \ell'_{\rm aux} + 2675 n_0 + 4477 n_1 + 2312 n_2
      + 497 n_3 - 91 n_4 + 80 n_5, \\
 (\Lambda_{\tilde{g}_2})_{\sharp 1} & \;
    = 2904 n'_6 - 747 \ell'_{\rm aux} - 13497 n_0 - 22565 n_1 - 11662 n_2
       - 2529 n_3 + 445 n_4 - 374 n_5, 
\end{align*}
and 
\begin{align*}
&  (\Lambda_{g3})_{11} + 12 (\Lambda_{g3})_{0\sharp} = \\
 &  - 21600 n'_6 + 5403 \ell'_{\rm aux} + 102129 n_0 + 170581 n_1
  + 88214 n_2 + 19281 n_3 - 3269 n_4 + 2590 n_5, 
\end{align*}
where $\ell'_{\rm aux} := \ell_{\rm aux}+12 \delta n$. 
So, the monodromy matrix $M_{\tilde{g}_2}$ is automatically integer valued. 

The group $\Gamma_S$ is generated by $w^{(6)}$, $g_\infty$ and $g_2$. After 
demanding that all of their monodromy matrices are integer valued, now we 
see that (\ref{eq:b-is-Z-k=6}) is the only condition on the invariants 
(\ref{eq:BPSpara-k=5}). 
 
\vspace{5mm}

In a $k=6$ vacuum branch with $n_1 = n_5$, there is a chance 
that $R_{v*(4)}$ is in the group $[\Gamma_S]$ of Het-perturbative 
duality transformations. Using the period 
polynomial (\ref{eq:PP-k=6-Weyl4-full}), let us compute the monodromy 
matrix on the electric and magnetic charges, assuming that $R_{v*(4)}$ 
is a duality transformation.  The procedure of the computation is 
the same as before, except that $n_1 = n_5 =: \bar{n}$. 
\begin{align*}
 (\Lambda_{R_{v*(4)}})_{00} & \; = (33/2)n_0 + 14n_2 + 5n_3 + 24\bar{n}
       -2\ell'_{\rm aux} + n'_6, \\
 (\Lambda_{R_{v*(4)}})_{01} & \; = (-183/2)n_0 -79n_2 -28n_3 - n_4/2
   -138 \bar{n} + (17/2)\ell'_{\rm aux} -2 n'_6 ,   \\
 (\Lambda_{R_{v*(4)}})_{\sharp\sharp} & \; = (45/2)n_0 +20n_2 + 7n_3 +36\bar{n}
    - \ell'_{\rm aux} - n'_6 ,   \\
 (\Lambda_{R_{v*(4)}})_{\sharp 1} & \; = (-207/2)n_0 -91n_2 -32n_3 -n_4/2
     -162\bar{n} +(13/2)\ell'_{\rm aux} +2n'_6 , 
\end{align*}
and 
\begin{align*}
 (\Lambda_{R_{v*(4)}})_{11}+12 (\Lambda_{R_{v*(4)}})_{0\sharp} = 
  702 n_0 +612n_2 +216n_3 + 4n_4 + 1080 \bar{n} -54 \ell'_{\rm aux} . 
\end{align*}
So, noting that $n_0 = -2$, the hypothetical monodromy matrix 
$M_{\widetilde{R}_{v*(4)}}$ of the lattice isometry $R_{v*(4)} \in \Gamma_0(k)_+$ 
representing a non-trivial Atkin--Lehner involution is integer valued 
if and only if 
\begin{align}
  n_4 + \ell_{\rm aux} \equiv 0 \quad {\rm mod} \; 2. 
   \label{eq:cond-k=6-AL}
\end{align}

We have thus learned that the reflection $R_{v_*(4)}$ may be in the 
duality group $[\Gamma_S]$ only when $n_1 = n_5$ and 
$n_4 + \ell_{\rm aux} \equiv 0$ mod 2. In a $k=6$ branch with a 
tune in hypermultiplet moduli so that $n_4^V=1$ or $n_6^V=1$, 
we know that the SU(2) Weyl reflection $R_{v_*(4)}$ 
(or $R_{v*(6)}$, and hence $R_{v_*(4)} \in \Gamma_S R_{v_*(6)}$ again) should 
be in the duality group $[\Gamma_S]$ at least for the tuned hypermultiplet 
moduli, so $n_1 = n_5$ and $n_4 + \ell_{\rm aux} \equiv 0$ (mod 2) must 
be satisfied in the entire branch. 
The authors do not have a field-theory argument for why $n_4 + \ell_{\rm aux}$
has to be even; this constraint on a theory with an SU(2) gauge group 
seems to have arisen from abstract mathematical analysis that uses 
heavily the property of modular forms; the authors wonder if 
this constraint should be regarded as a consequence of 
string theory (than quantum field theory). 

Let us note in passing that the monodromy matrix $M_{\widetilde{R}_{v*}}$
of the Weyl reflection $R_{v_*}$ of a W-boson charge  is integer valued 
guarantees that the 4D SU(2) gauge theory is free of the Witten SU(2) 
anomaly; see the appendix \ref{sec:Witten}. The W-boson charges $v_*$
that appear in the cases of $k=2,3,5,6$ so far are primitive 
elements in $\widetilde{\Lambda}_S^\vee$, so there cannot be SU(2) doublet 
matter fields in the first place. 

\subsubsection{Level-11 (Degree-22)}
\label{sssec:level-11}

The list of BPS classification invariants of the $\rho = 1$ and $k=11$ 
branches turns out to be 
\begin{align}
 \left\{ n_0 = -2, \; n_{1,\cdots, 10} \in \Z_{\geq 0}, \; n_{11} \in \Z_{\geq -2}\;
   | \; 
   {\rm linear~relation~}(\ref{eq:lin-rltn-k=11}) , 
  \chi_{\rm match} \leq 4 \right\}, \nonumber \\
 m_{4,5,6,8,9,10,11} \in  12\Z, \quad b_{\cal R} \in 2^{-1}\Z_{\geq 0}; 
   \label{eq:BPSpara-k=11} \\
 n_9 = 6n_1 -9n_2 -1n_3 +8n_4 -5n_5+ 1n_6 -3n_7+ 3n_8 -4n_{10}+ 1n_{11}.
  \label{eq:lin-rltn-k=11}
\end{align}
So, the spectrum (multiplicity) of light matter multiplets in the 4D
effective field theory is not as general as one expects from general 
${\cal N}=2$ supersymmetric gauge theory \cite{EW19}. The linear constraint 
on the light matter spectrum\footnote{
Due to the relation $n_\gamma = NL_{\nu_{|\gamma|}, \gamma}$, this is also 
regarded as a constraint on the number of points in the base $\P^1$ 
of a $\vev{+22}$-polarized regular K3-fibration above which 
transcendental cycles $\beta_T \in \Lambda_T^\vee$ (where 
${}^\exists w \in \gamma \subset \Lambda_S^\vee$ such that 
$w+\beta_T \in {\rm II}_{3,19}$) with $\beta^2_T = -2\nu_{|\gamma|}$ 
become algebraic \cite{MP, KMPS}.
} %
 (\ref{eq:lin-rltn-k=11}) is understood as obstruction 
to building a vector-valued modular form $\Phi$ in the Rademacher expansion
(i.e., as constraint from the modular invariant UV completion 
in string theory). 

The 4D effective Lagrangian parameters on such a branch are computed 
to be as follows:
\begin{align}
d_{111} & \; = 66\delta n + 11\ell_{\rm aux} - 88n_0 -85 n_1
-17 n_2-63 n_3-112 n_4+16 n_5-15 n_6   \nonumber \\
 & \qquad \qquad   -135 n_7-69 n_8 -68 n_{10}-(25 n_{11})/2,
        \label{eq:d111-k=11}\\
(c_2)_1 & \; = 24\delta n + 4\ell_{\rm aux}
-70n_0 -82 n_1 -38 n_2-42 n_3-52 n_4+4 n_5-6 n_6 \nonumber \\
 & \qquad \qquad -66 n_7-30 n_8 -32 n_{10}-5 n_{11}, \\
 & \; d'_{\rm match} =  \frac{209}{2}n_0 + \frac{281}{2}n_1 + \frac{175}{2}n_2 
   + \frac{105}{2}n_3 + 31 n_4 + 5 n_5 + \frac{3}{2} n_6 \nonumber \\
 & \qquad \qquad + \frac{93}{2}n_7 + \frac{27}{2}n_8 + 20 n_{10}
         + \frac{5}{4}n_{11},   \\
\chi_{\rm match} & \; = 42n_0 + 68 n_1 + 52 n_2 + 40 n_3 + 32 n_4 + 4 n_5 + 2 n_6
  + 64 n_7 + 20 n_8  + 34 n_{10} + 2 n_{11},  
   \label{eq:chiMatch-k=11}
\end{align}
where    
\begin{align}
 \ell_{\rm aux} := 3[- d_0+6 m_{10}+ m_{11}+4 m_8+2 m_9+4 m_4
  +2 m_5+2 m_6]/12. 
  \label{eq:def-ell-aux-k=11}
\end{align}

The condition (\ref{eq:cond-PQ-1}) for the monodromy matrix 
$M_{\tilde{g}_\infty}$ to be integer valued is translated to 
\begin{align}
  n_{11} \in 2\Z, \quad {\rm and} \quad 
  \ell_{\rm aux}  \in 6\Z;   
  \label{eq:n11-and-ell-are-even-k=11}
\end{align}
although we learn from (\ref{eq:def-ell-aux-k=11}) that 
$\ell_{\rm aux}$ is automatically divisible by 3 but not by 6, 
this condition says that the combination $(m_{11}+d_0(0))$ should be 
divisible not just by 12 as in (\ref{eq:BPSpara-k=11}), but by 24.

One can see that the monodromy matrices $M_{\tilde{g}_1}$ and 
$M_{\tilde{g}_2}$ are integer valued automatically at least when $n_7=0$, 
once the conditions (\ref{eq:n11-and-ell-are-even-k=11}) are imposed 
on the classification invariants (\ref{eq:BPSpara-k=11}). To do so, 
we need to compute the $3\times 3$ part $\Lambda_{\tilde{g}_1}$ of the 
matrix $M_{\tilde{g}_1}$ and $\Lambda_{\tilde{g}_2}$ of $M_{\tilde{g}_2}$ from 
the period polynomials (\ref{eq:pp-k=11-g1}) and (\ref{eq:pp-k=11-g2}), 
respectively. It turns out that 
\begin{align*}
 (\Lambda_{\tilde{g}_1})_{00} & \; = -128 \ell'_{\rm aux} + 1186 n_0 + 1208 n_1
  + 329 n_2 + 809 n_3 + 1352 n_4 - 184 n_5 + 176 n_6 \\ 
  & \qquad + 812 n_8 + 833 n_{10} + 297 (n_{11}/2), \\
 (\Lambda_{\tilde{g}_1})_{01} & \; = 2779 (\ell'_{\rm aux}/3) - 8597 n_0 - 8758 n_1
  - 2390 n_2 - 5865 n_3 - 9796 n_4 + 1330 n_5 - 1276 n_6 \\
  & \qquad - 5886 n_8 - 6042 n_{10} - 2151 (n_{11}/2), \\
 (\Lambda_{\tilde{g}_1})_{\sharp\sharp} & \; = -1085 (\ell'_{\rm aux}/6)
   + 1684 n_0 + 1716 n_1 + 469 n_2 + 1149 n_3 + 1918 n_4 - 260 n_5 + 250 n_6 \\
  & \qquad + 1154 n_8  + 1186 n_{10} + 421 (n_{11}/2), \\
 (\Lambda_{\tilde{g}_1})_{\sharp 1} & \; = 3302 (\ell'_{\rm aux}/3) - 10238 n_0
    - 10431 n_1 - 2850 n_2 - 6985 n_3 - 11662 n_4 + 1581 n_5 \\
  & \qquad - 1520 n_6 - 7014 n_8  - 7206 n_{10} - 1280 n_{11}, 
\end{align*}
and 
\begin{align*}
  (\Lambda_{\tilde{g}_1})_{11} + 22 (\Lambda_{\tilde{g}_1})_{0\sharp} & \; = 
 -10043 \ell'_{\rm aux} + 93313 n_0 + 95064 n_1 + 25961 n_2 
 + 63662 n_3
 + 106308 n_4 \\  & \quad - 14419 n_5 + 13853 n_6 + 63909 n_8  + 65634 n_{10}
  + 23339 (n_{11}/2) 
\end{align*}
for $M_{\tilde{g}_1}$; here, $\ell'_{\rm aux} := \ell_{\rm aux} +6\delta n$.  
As for $M_{\tilde{g}_2}$, 
\begin{align*}
 (\Lambda_{\tilde{g}_2})_{00} & \; = -460 \ell'_{\rm aux} + 8511 (n_0/2) + 4311 n_1
  + 1159 n_2 + 2895 n_3 + 4839 n_4 - 665 n_5 \\
  & \qquad + 631 n_6 + 2980 n_8  + 2964 n_{10} + 532 n_{11}, \\
 (\Lambda_{\tilde{g}_2})_{01} & \; = 2511 \ell'_{\rm aux} - 23246 n_0 - 23549 n_1
  - 6336 n_2 - 15817 n_3 - 26430 n_4 + 3627 n_5 \\
  & \qquad - 3448 n_6 - 16282 n_8 - 16198 n_{10} - 2905 n_{11}, \\
 (\Lambda_{\tilde{g}_2})_{\sharp \sharp} & \; = -632 (\ell'_{\rm aux}/3)
   + 3907 (n_0/2) + 1979 n_1 + 533 n_2 + 1330 n_3 + 2221 n_4 - 304 n_5 \\
   & \qquad + 290 n_6 + 1369 n_8 + 1363 n_{10} + 244 n_{11} , \\
 (\Lambda_{\tilde{g}_2})_{\sharp 1} & \; = 1699 \ell'_{\rm aux} - 15746 n_0
      - 15951 n_1 - 4295 n_2 -  10718 n_3 - 17902 n_4 + 2452 n_5 \\ 
    & \qquad - 2337 n_6 - 11033 n_8 - 10982 n_{10} - 1967 n_{11}, 
\end{align*}
and 
\begin{align*}
 (\Lambda_{\tilde{g}_2})_{11} & \; + 22 (\Lambda_{\tilde{g}_2})_{0\sharp} = 
-20548 \ell'_{\rm aux} + 190333 n_0 + 192810 n_1 + 51899 n_2 + 129530 n_3 \\
 &  + 216396 n_4 - 29665 n_5 + 28241 n_6 + 133341 n_8  + 132690 n_{10} 
  + 47561 (n_{11}/2). 
\end{align*}
All of them are automatically integers indeed, after 
using (\ref{eq:n11-and-ell-are-even-k=11}). 
It is true that the numerically evaluated coefficients of 
the period polynomials (\ref{eq:pp-k=11-g1}, \ref{eq:pp-k=11-g2})
are away from their approximate values in $\Q$ by as much as 
${\cal O}(1)\times 10^{-2}$, as stated at the end of 
section \ref{ssec:PP-Gamma0k}. The approximate rational values we chose 
are still highly non-trivial; this is because a change by $+1$ 
in the coefficient of $t^4$ [resp. $t^3$] 
in (\ref{eq:pp-k=11-g1}, \ref{eq:pp-k=11-g2}) would result in 
a change by $+2/11^2$ [resp. $+1/11$] in the 
monodromy matrix elements $(\Lambda_{\tilde{g}})_{\sharp \sharp}$ 
[resp. $(\Lambda_{\tilde{g}})_{\sharp 1}$], and the monodromy matrices 
$M_{\tilde{g}_1}$ and $M_{\tilde{g}_2}$ would not be integer valued. 

So, we conclude that (\ref{eq:n11-and-ell-are-even-k=11}) is 
the necessary and sufficient condition on (\ref{eq:BPSpara-k=11}) 
for the monodromy matrices of the elements in $\Gamma_S$ to be 
integer valued. One also learns from this study that there can be 
just one diffeomorphism class of Calabi--Yau threefolds 
that have a $\vev{+22}$-polarized K3-fibration with a given 
$\{ n_\gamma \}_{n_7=0}$; this is because $\ell_{\rm aux}$ 
is quantized by $+6\Z$, while $\delta n \in \Z$ corresponds 
to $\Delta \ell \in 6\Z$. 

\section{Discussions}
\label{sec:discussion}

A case-by-case study has been presented in 
sections \ref{sec:PP} and \ref{sec:classification}. 
Here, we extract general patterns in those cases and generalize 
by guesswork, from a couple of different perspectives in 
sections \ref{ssec:even-BPS-inv}--\ref{ssec:mod-diffeo-summary}.  

\subsection{SU(2) Halfhyper Multiplets}
\label{ssec:even-BPS-inv}

Some of the constraints we obtained are on the classification invariants 
$\{ m_\gamma \}$ and $d_\gamma(\nu_\gamma)$'s, or in other words, on the 
combination $\ell_{\rm aux}$; see 
 (\ref{eq:b-is-Z}, \ref{eq:cond-aux-k=3}, \ref{eq:cond-Fricke-on-BPS-k=4}, 
\ref{eq:b-is-Z-k=5}, \ref{eq:b-is-Z-k=6}, \ref{eq:cond-k=6-AL},
 \ref{eq:n11-and-ell-are-even-k=11}). There are also constraints 
exclusively on the classification invariants $\{ n_\gamma \}$, which 
directly govern the spectrum of BPS states; see  (\ref{eq:cond-k=2-n2-even}, 
\ref{eq:cond-PQ-inZ-k=3}, \ref{eq:n6-is-even-k=6}, 
\ref{eq:n11-and-ell-are-even-k=11}). To the latter list of constraints, 
we may add \cite[(32)]{ESW20}, which is on the $k=14$ case in the 
$\rho=1$ series. 

All the constraints on $\{n_\gamma \}$ are of the form of $n_\gamma \in 2\Z$ 
for $\gamma = [k] \in \Z_{2k} \cong G_S$ in the $\rho = 1$ series, and 
we have seen that this constraint should be imposed for $k=2,3,6,11$ 
and $k=14$, whereas such constraint is not necessary for $k=1,4,5$; 
in the $k=1$ case, Calabi--Yau threefolds with an odd $n_{k=1}$ are known 
\cite{KKRS}, so Type IIA string vacua with an odd $n_k$ do exist. 
The list of $\Lambda_S \cong \vev{+2k}$ where the condition $n_k \in 2\Z$ 
was derived agrees with the entries of the form 
$(k, [w_{2*}]) = (k, [k])$ in Table \ref{tab:v*-list}. 

Suppose that $v_* \in [k] \subset \widetilde{\Lambda}_S^\vee$ is 
the U(1) charge of an SU(2) W-boson, and $n^V_{[k]} = 1$ (i.e., 
the W-boson exists in the spectrum); it is then not difficult 
to understand what the condition $n_k \in 2\Z$ means in terms of the 
4D ${\cal N}=2$ supersymmetric field theory. The Coulomb branch 
moduli space has a point where the 4D effective field theory has 
an enhanced $\SU(2)$ symmetry and massless W-bosons are in the particle 
spectrum. Whereas the super-Poincare algebra and the 4D CPT transformation 
can be realized on a half-hyper multiplet in a pseudo-real 
representation\footnote{
The SU(2) representations of half-odd isospin 
$j \in \{ 1/2, 3/2, 5/2, \cdots \}$ are pseudo-real, while 
those of integral isospin $j \in \{ 0,1,2,\cdots \}$ are 
strictly real. 
}\raisebox{4pt}{,}\footnote{
The authors thank Y. Tachikawa for pointing this out.
} %
 of the $\SU(2)$ gauge group \cite{Bagger:1983tt, 
Derendinger:1984bu}, the algebra cannot be realized on an odd number 
of ``half-hyper multiplets'' in a strictly real representation of the SU(2). 
On the other hand, an odd $n_{[k]} = n_{[k]}^H-2n^V_{[k]}$ would imply that 
there are an odd number of ``half-hyper multiplets'' in the isospin-1 
representation in the 4D spectrum; so, $n_{[k]}$ must be even.  
It is conceivable that this observation is translated into the language 
of the Heterotic string operator product expansion (OPE); the action of 
the SU(2) algebra and the 4D ${\cal N}=2$ supersymmetry on a state 
will be replaced by the OPE of the state with the vertex operator of the 
W-boson and the spectral flow operators, respectively. 
One might learn a lesson from here that whether the OPE coefficients 
can be constructed properly (consistently with the 2D CPT transformation
of the worldsheet SCFT) can yield consistency conditions besides what 
we learn from the modular invariance of the spectrum/partition 
function (cf \cite{Dymarsky:2022kwb}).  

The constraint $n_{k} \in 2\Z$ obtained in this article, however, 
might say something a little more than the argument above. 
Suppose, for now, that there is a way to tune hypermultiplet moduli in a vacuum 
branch so that $n_k^V = 1$. Then the argument above is applied, 
and we should have $n_k \in 2\Z$. Because the value of $n_k$ does not change 
as hypermultiplet moduli vary, the property $n_k \in 2\Z$ remains to 
hold in the same branch. The analysis in this article derived $n_k \in 2\Z$,
however, without assuming that there is such a tune in the moduli space. 

The same constraint $n_k \in 2\Z$ for those $k$'s should have an 
explanation also in the Type IIA description. Presumably a way to go 
is to exploit\footnote{
As we now know that almost all of the branches of the Heterotic--Type IIA 
dual vacua satisfy a set of necessary conditions for a NLSM interpretation 
in the Type IIA description 
(sections \ref{ssec:NLSM} and \ref{ssec:guess-pattern}), 
we do not lose much by using the formulation relying on a geometric phase 
description. 
} %
 the formulation of \cite{MP, KMPS, D4D2D0}: 
in a Type IIA compactification 
on a Calabi--Yau threefold $M$ that has a $\Lambda_S$-polarized regular 
K3 fibration $\pi: M \rightarrow \P^1$, a section $\sigma_\pi : \P^1 
\rightarrow {\cal M}^{\cal V}_{\Lambda_S}$ is well-defined; here, 
${\cal M}^{\cal V}_{\Lambda_S}$ is the total space of a fibred space 
over the base $\P^1$ with the period domain $D(\Lambda_T)$ as 
the fiber; the vector-valued modular form $\Phi$ is interpreted as 
the generating function of the Noether--Lefschetz numbers obtained 
as the intersection between the curve $\sigma_\pi(\P^1)$ and the 
Noether--Lefschetz divisors ${\cal D}_{\gamma,\nu}$ within the 
total space ${\cal M}^{\cal V}_{\Lambda_S}$. The authors expect that 
the constraint $n_k \in 2\Z$ for those $\Lambda_S = \vev{+2k}$ 
can be derived also from the geometry of $\sigma_\pi(\P^1)$ and the 
Noether--Lefschetz divisors of ${\cal M}^{\cal V}_{\Lambda_S}$. Such a 
derivation would add another supporting evidence for the Heterotic--Type IIA 
duality, but that is beyond the scope of this article.

The constraint $n_k \in 2\Z$ for $\gamma = [k] \in \Z_{2k} \cong G_S$ 
in the cases of $\Lambda_S = \vev{+2k}$ with $k=2,3,6,11,14$ will be
generalized as follows; 
this is a guess than a derivation, however. Now, $\Lambda_S$ is a general 
even lattice of signature $(1,\rho-1)$ that fits within the lattice 
${\rm II}_{3,19} \cong H^2(K3;\Z)$. Suppose that 
\begin{itemize}
\item there is $\gamma_* \in G_S$ such that 
$(\gamma_*, \gamma_*) = -2/\ell_{KM} + 2\Z$, $\ell_{KM} \in \N_{\geq 1}$, and 
$\ell_{KM} \gamma_* = 0 \in G_S$, 
%
%
\item there is a positive integer $m$ (for the integer isospin) such that 
 $m\gamma_* \neq 0 \in G_S$ but $2m\gamma_* = 0 \in G_S$, 
and $m^2 \leq \ell_{KM}$. 
\end{itemize}
The first condition is just for the existence of an SU(2) 
symmetry (\ref{eq:reflection_conditions}), and the second condition 
is for the state with a charge $mv_* \in m\gamma_*$ to generated 
a ``halfhyper multiplet'' of the isospin-$m$ ($m \in \N_{\geq 1}$) 
representation of the SU(2) symmetry. 
We expect that the condition $n_{m\gamma_*} \in 2\Z$ should be imposed 
in such a branch of the moduli space\footnote{
Within the $\rho =1$ series, $k\equiv 2$ or $\equiv 3$ mod 4 (where $m=1$), 
$k \equiv 12$ or $\equiv 14$ mod 16 (where $m=2$), 
$k \equiv 33$ mod 36 (where $m=3$, and 
$k \equiv 60$ mod 64 (where $m=4$). 
There will be more examples with $\rho > 1$. 
} %
 for the monodromy matrices to be integer valued.

\subsection{Constraints on the BPS Classification Invariants}
\label{ssec:guess-pattern}

As we have reviewed in section \ref{ssec:to-do-list}, 
branches of 4D ${\cal N}=2$ Heterotic--Type IIA dual vacua 
have classification invariants $\{ n_\gamma \}_{\gamma \in G_S}$, 
$\{ m_\gamma \}_{\gamma \in G_S}$ and $d_\gamma(\nu_\gamma)$'s associated 
with BPS state counting (\ref{eq:BPS-classifctn-inv}), and 
some parameters of the 4D prepotential $a_{ab}$ and $b_a$ 
with $a, b=1,\cdots, \rho$ (\ref{eq:BPSmass-inv}). By demanding that 
the monodromy matrices of the Peccei--Quinn symmetry 
($M_{\tilde{g}_{\infty(a)}}$ for $a=1,\cdots, \rho$) should be 
integer valued, the values of $a_{ab} + \Z$ are completely 
determined in terms of the BPS classification invariants 
as in (\ref{eq:cond-PQ-2}) when $\rho = 1$, 
or as in \cite[(2.38)]{EW21} for a general $\rho$. 
Moreover, the value $a_{ab} +\Z$ is determined precisely in the 
way one of the necessary conditions (\ref{eq:cond-geom-phase-2}) 
(or \cite[(2.16)]{EW21}) for the Type IIA NLSM interpretation is 
satisfied. 

In this article, we have worked on the $\rho = 1$ vacuum branches 
with several small values of $k$. In all those cases, we found 
by requiring that $w^{(k)} \in \Gamma_S$ have integral monodromy 
matrix $M_{\widetilde{w}^{(k)}}$ that the parameter 
$b_{a=1}+24\Z$ is also determined uniquely in terms of the BPS 
classification invariants as in (\ref{eq:cond-Fricke-inZ}); 
moreover, it is determined precisely in the way one of the other 
necessary conditions (\ref{eq:cond-geom-phase-1}) for the Type IIA 
NLSM interpretation is satisfied. It is a natural guess that the 
relation 
\begin{align}
 b_{a} \equiv (c_2)_a \qquad \qquad ({\rm mod~} 24), \quad a=1,\cdots, \rho 
   \label{eq:cond-Fricke-inZ-genRho}
\end{align}
holds true in all the theoretically consistent 4D ${\cal N}=2$ 
Heterotic string vacua with the lattice $\widetilde{\Lambda}_S$ of 
the form\footnote{
It will be interesting to explore the 4D ${\cal N}=2$ Heterotic--Type IIA 
duality with the lattice $\widetilde{\Lambda}_S$ not in this form. 
For example, such a branch of the moduli space should be there 
on the other side of the Coulomb--Higgs transition at a massless 
matter singularity in the moduli space with $\rho =1$.  
In the Type IIA 
description, already the K3 fiber loses supergravity approximation 
at the massless matter singularity, and the geometry picture is gone 
(because there is no $U[-1] \cong H^0(K3;\Z) \oplus H^4(K3;\Z)$ component
in $\widetilde{\Lambda}_S$)
after the transition. The absence of the $U[-1]$ component also implies 
that there is no lift to the Heterotic string--M-theory duality preserving 
the $\SO(4,1)$ Lorentz symmetry (unless we generalize the treatment 
of M-theory as 11D supergravity). 
} %
$U[-1] \oplus \Lambda_S$. That may be perhaps by demanding that the 
$2(\rho+2) \times 2(\rho+2)$ monodromy matrices of 
the duality transformations\footnote{
For the cases with $\rho >1$, there is still a lattice isometry 
of the form (\ref{eq:def-w-k-Fricke}) in 
$\Gamma_S \subset {\rm Isom}(\widetilde{\Lambda}_S)$ (although it is no
longer appropriate to call it the Fricke involution). 
} %
 in $[\Gamma_S]$ are all integer valued. 
If so, then it will be enough to focus on the BPS classification 
invariants to label the vacuum branches. 

It then follows by combining the relation (\ref{eq:cond-Fricke-inZ-genRho})
with the conditions \cite[(2.37)]{EW21} and (\ref{eq:cond-PQ-1}), 
that Wall's condition (\ref{eq:cond-Wall}) is satisfied, 
and hence there must be a diffeomorphism class of real six-dimensional 
manifold $M$ with $b_1(M)=0$, $\chi(M) = \chi_{\rm match}$, 
the symmetric trilinear form $H^2(M) \times H^2(M) \times H^2(M)
 \rightarrow \Z$ by the cup product given by $\mu = (C_{ab}, d_{abc})$, 
and the linear form $(c_2(TM)\wedge) : H^2(M) \rightarrow \Z$ given by 
$-p_1/2 = (24, (c_2)_a)$ (for more information, see discussion 
around \cite[(2.14) and (2.15)]{EW21} and references therein). 
This strongly hints that all the vacuum branches in this class 
(4D ${\cal N}=2$ dual vacua, with the lattice $\widetilde{\Lambda}_S 
= U \oplus \Lambda_S \subset {\rm II}_{4,20}$) have a phase interpreted 
as Type IIA compactification on a Calabi--Yau threefold $M$ with 
a $\Lambda_S$-polarized regular K3-fibration. There is a caveat 
to that, however, as we will remark at the beginning of 
 section \ref{ssec:mod-diffeo-summary}.

In addition to the speculative 
extrapolation (\ref{eq:cond-Fricke-inZ-genRho}) from the case studies 
in this article, there is one more speculative extrapolation, which 
is less motivated, however. In all the cases we studied, the 
condition (\ref{eq:cond-Fricke-inZ}, \ref{eq:cond-Fricke-inZ-genRho}) 
was for the monodromy matrix of the Fricke involution $w^{(k)}$ to be
integer valued, and all other non-trivial constraints on the 
vacuum classification invariants were for the monodromy matrix  
of the Peccei--Quinn symmetry $g_\infty$ to be integer valued. 
Although the $\Gamma_S$ of duality transformations is not 
necessarily generated by just $g_\infty$ and $w^{(k)}$, all other 
generators did not yield additional constraints on the vacuum 
classification invariants. It is tempting to speculate that 
this pattern persists for all of $\Lambda_S = \vev{+2k}$ in the 
$\rho =1$ series. 
%
%

There is also a chance that the vacuum identification group $[\Gamma_S]$
is larger than $\Gamma_S$. Within the cases we studied in this article, 
extra constraints were obtained from the integral monodromy condition 
only when $\Lambda_S = \vev{+2k}=\vev{+12}$; for $R_{v_*}$ not in $\Gamma_S$ 
to be in $[\Gamma_S]$, we found the constraints $n_1 = n_5$ and 
$n_4 + \ell_{\rm aux} \equiv 0$ (mod 2). Having just one example, without 
a 4D field theory interpretation on the latter constraint, makes it hard 
for us to generalize those constraints even 
in a wild guess.  One may also be interested in whether the possibility 
$\Gamma_S \subsetneq [\Gamma_S]$ is always associated with an enhanced 
non-abelian gauge symmetry in the 4D effective field theory, but 
we do not have enough clues to be able to answer this question.  

Let us also note here that this theoretical consistency condition
of even $n_4 + \ell_{\rm aux}$ restricts the value of the coupling
constants ($d_{111}$ and $(c_2)_1$ through $\ell_{\rm aux}$) for a given 
set of parameters governing the BPS particle spectrum (i.e., $n_4$). 
This is in
contrast with the consistency conditions such as $n_{k} \in 2\Z$
discussed in section \ref{ssec:even-BPS-inv} and the linear
relations among $\{ n_\gamma \}$ (such as (\ref{eq:lin-rltn-k=11})), 
which restricts the spectrum of BPS states.

\subsection{Mod-diffeomorphism Classification of Calabi--Yau Threefolds}
\label{ssec:mod-diffeo-summary}

Certainly we have pointed out that there always exists a real 
six-dimensional manifold $M$ for each of the branches of 
4D ${\cal N}=2$ vacua in our case studies with integral monodromy 
matrices; the manifold $M$ has all the properties that we can check 
easily (that is, $b_1(M)=0$, 
$\chi(M)=\chi_{\rm match}$, $\mu = (C_{ab},d_{abc})$, 
$c_2(TM) = (24, (c_2)_a)$ and (\ref{eq:cond-geom-phase-1},
 \ref{eq:cond-geom-phase-2})) for the branch to be interpreted as 
the Type IIA compactification over $M$. There are branches among them, 
however, which would not be a K3-fibred Calabi--Yau compactification 
in the Type IIA description. 
They are the cases with $\Lambda_S= \vev{+2k}=\vev{+8}$ and $n_4 > 0$, 
where $\gamma_0 = [4/8] \in \Z_{8} \cong G_S$ is non-zero and isotropic 
(i.e., $(\gamma_0 ,\gamma_0) = 0 \in \Q/2\Z$). 
More generally, hypothetical branches of vacua with $n_{\gamma_0} > 0$ 
at a non-zero isotropic $\gamma_0 \in G_S$ with the constraint 
(\ref{eq:cond-Fricke-inZ-genRho}) may survive the test of integral 
monodromy matrices of the vacuum identification group $[\Gamma_S]$. 
If such vacuum were a Calabi--Yau compactification with a 
$\Lambda_S$-polarized regular K3-fibration, then $n_{\gamma_0} > 0$ would 
be interpreted as the number of points in the base $\P^1$ over which 
the fiber K3-surface hits the Noether--Lefschetz locus of a transcendental 
cycle $\beta_T \in \gamma_0 \subset \widetilde{\Lambda}_T^\vee$ with the 
norm $(\beta_T, \beta_T)=0$. In fact, that is not possible, because 
a norm-0 element $\beta_T \in \Lambda_T\otimes \R$ cannot be completely 
orthogonal to the positive definite 2-dimensional subspace 
${\rm Span}_\R \{ {\rm Re}(\Omega^{(2,0)}), \; {\rm Im}(\Omega^{(2,0)}) \}
 \subset \Lambda_T \otimes \R$ of the holomorphic (2,0) form $\Omega^{(2,0)}$ 
of the fiber K3 surface within the signature $(2,20-\rho)$ vector space 
$\Lambda_T \otimes \R$ \cite{Bor2, MP}. 

There will be two logical possibilities one may think of. 
One is that we still fail to capture some of theoretical consistency 
conditions, those hypothetical vacua with $n_{\gamma_0} > 0$ are actually 
in conflict with some consistency conditions, and those vacua do not exist 
in fact. The other is that those vacua do exist, are perfectly 
theoretically consistent, and yet their Type IIA description does not 
have a phase interpreted as a Calabi--Yau target non-linear sigma 
model.\footnote{
Certainly Wall's theorem guarantees that there are real six-dimensional 
manifold $M$ as indicated at the beginning of this 
section \ref{ssec:mod-diffeo-summary}. Due to the reasoning presented 
in the main text, however, such a six-dimensional manifold $M$ will not 
admit a K\"{a}hler structure (a compatible pair of a symplectic structure 
and a complex structure) to be a Calabi--Yau threefold, and is not relevant 
to the Type IIA vacuum in question. 
} %
We cannot tell which is right at this moment. Either way, we will 
take those (hypothetical) vacua with $n_{\gamma_0} > 0$ at a non-zero 
isotropic $\gamma_0 \in G_S$ out of consideration in the rest 
of this section \ref{ssec:mod-diffeo-summary}. 

\vspace{5mm}

Classification of Calabi--Yau threefolds $M$ that admit K3-fibrations 
is different from classification of pairs $(M,\pi)$ 
of Calabi--Yau threefolds $M$ and their K3-fibration maps  
$\pi: M \rightarrow \P^1$. One Calabi--Yau threefold $M$ may admit
multiple different lattice-polarized K3-fibrations, in principle, 
and lots of such examples are known in fact \cite{multi-fiberAsp, multi-fiber}. 
Such study as in section \ref{sec:classification}
of this article is suitable for the latter, because we specify 
the lattice polarization $\Lambda_S$ of the K3 fiber at the beginning.  

One may note, however, that there is no double counting within the 
Calabi--Yau threefolds $M$ that show up in the $\rho=1$ series. 
To see this, let $M$ be a Calabi--Yau threefold,  
$D_s \in H^2(M;\Z)$ the Poincare dual to the fiber K3 class, 
and $H$ one more generator so that $H^2(M;\Z) = {\rm Span}_\Z\{ D_s, H\}$ 
($b^2(M)=2$ because $\rho=1$). 
The intersection of $D_s$ and $H$ satisfy $D_s^2 = 0$ and 
$D_s\cdot H\cdot H=2k$, where $2k$ is the degree of the generic fiber K3.
Now, suppose that this threefold $M$ also admits another K3-fibration. 
That is, there exists a new fiber K3 class $D_s' = m_1 D_s + m_2 H$ with 
$m_2 \neq 0$. $D_s'$ also has to satisfy the condition $D_s'^2=0$. 
This is impossible, however, because we run into a contradiction:
\begin{align}
    0=D_s\cdot D_s'^2=D_s\cdot(m_1 D_s + m_2 H)^2=m_2^2D_s\cdot H^2=m_2^2\cdot 2k.
\end{align}
So, such a threefold $M$ has just one fibration map, where 
$D_s$ is the K3-fiber class.

\vspace{5mm}

Although we wish to be able to classify Calabi--Yau threefolds modulo 
holomorphic isomorphisms, it can be a hard task to claim that a pair
of manifolds are {\it not} isomorphic to each other, when the pair
share the same set of topological invariants. By comparing the
Gopakumar--Vafa / Gromov--Witten invariants, we may test whether
a pair of Calabi--Yau manifolds have a symplectomorphism to each other;
we throw away the complex structures of those manifolds in this test.  
The Fourier coefficients $c_\gamma(\nu)$ of the vector-valued modular form
$\Phi$ we deal with in this study determine the Gopakumar--Vafa
invariants of the vertical curve classes of $M$ (those that are mapped
to points in the base $\P^1$ by the fibration map $\pi: M \rightarrow \P^1$)
\cite{MP, KMPS}. The Gopakumar--Vafa invariants for non-vertical curve
classes, however, are not captured by $c_\gamma(\nu)$; we do know in fact
that there is a pair of Calabi--Yau manifolds not sharing the same
symplectic structure (elliptic fibration $M=M_0$ over the Hirzebruch
surface $F_0$ and the one $M=M_2$ over $F_2$) although they share all
the invariants treated in this article. On the other hand, it is known
(Wall's theorem) that the set of topological invariants $\chi(M)$, $b_2(M)$,
$b_3(M)$, the trilinear symmetric form
$\mu: H^2(M) \times H^2(M) \times H^2(M) \rightarrow H^6(M)$ and $c_2(TM)$
uniquely identifies the diffeomorphism class a real six-dimensional manifold
$M$ with $b_1(M)=0$ belongs to. So, the study in this article can be used
to find out which diffeomorphism classes cannot be realized as a Calabi--Yau
threefold.

We can do so without relying on explicit construction of Calabi--Yau 
threefolds. Certainly explicit constructions are known for rank-1 lattice 
polarization with low degrees. To talk about constructing lattice polarized 
K3 surfaces before threefolds, a degree-2 K3 surface can be 
constructed as $(6) \subset WP^2_{[1:1:1:3]}$,
a degree-4 (quartic) K3 as $(4) \subset \P^3$, 
a degree-6 K3 as $(3) \cdot (2) \subset \P^4$ and 
a degree-8 K3 as $(2) \cdot (2) \cdot (2) \subset \P^5$. 
\label{page:construct-K3}
A little more different toric constructions of those K3 surfaces are
found in Ref. \cite{KKRS}. So, we may hope to construct Calabi--Yau
threefolds with a $\vev{+2k}$-polarized K3-fibration for $k=1,2,3,4$
as complete intersections of a toric ambient space with a toric fibration
morphism; Table 1 of Ref. \cite{KKRS} already did so to some extent. 
Beyond that, even for $k \geq 5$ in the $\rho =1$ series, such
construction approach will be much more difficult task.\footnote{
It is known that a K3 surface with a degree-$2k$ polarization
(at least with $k \in \{5,6,\cdots, 10\}$) can be constructed as 
a subvariety of a Grassmannian rather than a projective space.
To mass-produce Calabi--Yau threefolds that have such a K3 surface
in the fiber, however, it is desirable that the way to bring the Grassmannians 
into the fiber over $\P^1$ can be treated as combinatorial data (as in
the case of toric varieties). To the present authors, such techniques
are not known. 
} %
Even in the $k=2,3,4$ cases, construction-based approaches do not allow
us to conclude that a manifold that has not been constructed does not exist. 
The approach in this article manages to avoid all those problems. 
That is just to repeat what we have already remarked in the previous
section, however. 

Having worked out the modulo-diffeomorphism classification of
Calabi--Yau threefolds with $\rho =1$ for some small $k$'s,
one may extrapolate the results in the way we guessed speculatively
in sections \ref{ssec:even-BPS-inv} and \ref{ssec:guess-pattern}
to all the $\rho =1$ cases (with larger $k$'s).
This would already imply\footnote{
\label{fn:caution2}
This conclusion, however, relies on whether the guess is right, 
and also on the assumption that theoretical consistency in string 
theory has been exploited by the modular transformation of 
$\Phi$'s, $\Psi$'s, integrality of their Fourier coefficients, 
and the integrality of the monodromy matrices of the duality group 
$[\Gamma_S]$ (cf footnote \ref{fn:caution} and a comment 
at the end of section \ref{sssec:level-2}). 
It should also be remembered that we have only checked a set 
of necessary conditions (\ref{eq:cond-geom-phase-1},
 \ref{eq:cond-geom-phase-2}, \ref{eq:cond-Fricke-inZ-genRho}) 
 for a Type IIA NLSM interpretation on hypothetical branches of vacua; 
we do not know for sure whether they are sufficient conditions. 
} %
 that there would be infinitely many
diffeomorphism classes of Calabi--Yau threefolds $M$ with
$b_2(M)=\rho+1=2$; certainly the consistency condition
$n_{k} \in 2\Z$ in section \ref{ssec:even-BPS-inv} should be
added on top of the linear relations on $\{ n_\gamma\}$ in \cite{EW19}
(such as (\ref{eq:lin-rltn-k=11})), but there remain more and more
independent $n_\gamma$'s governing $\Phi$ for larger $k$, as discussed
already in \cite{EW19}. 

It was a highly non-trivial result of our numerical evaluation
that the coefficients of the period polynomials turn out to be  
rational numbers, or rational numbers multiplied by $\zeta(3)/(2\pi i)^3$.
If that were not the case, then we would have concluded that 
there would be only finite number of diffeomorphism classes of 
K3-fibred Calabi--Yau threefolds, as opposed to our argument above. 
It is known that the coefficients of the period polynomial of a cusp form
is, in general, not rational numbers, as remarked already 
in \cite[\S5]{EW21}.  So, $f_*$ obtained form $\Phi/\eta^{24}$ in 
the $\rho =1$ series must somehow belong to a special subclass.  
A work in progress \cite{EKSW} will explain more about this, 
and also provide references in mathematics. 

The authors are aware of Ref. \cite{KanWil} where an inequality 
has been derived on $\chi(M)$, $c_2(TM)$ and the trilinear form
$\mu$ (equivalent to $(C_{ab}, d_{abc})$ when a regular K3-fibration 
exists) on a Calabi--Yau threefold $M$. It is thus a natural question whether those inequalities
give rise to more constraints on the classification than what has
been captured in the study in this article. This inequality in
Ref. \cite{KanWil}, however, is derived by referring to an ample
divisor on $M$. In the way we deal with the information of vacuum
branches in this article, complex structure of the target space
Calabi--Yau threefold $M$ has been thrown away, which makes it
impossible to judge which element of $H^2(M;\Z)$ corresponds to
an ample divisor. Although there are a few more ideas 
in \cite{EW19} and \cite{BW-16} that may (or may not) be used 
in capturing more structures than smooth manifolds, an extra effort
in that direction is beyond the scope of this article. 

\vspace{5mm}

Let us now have a closer look at the results\footnote{
See footnote \ref{fn:caution2}. Despite the issues there, 
at least we can use the result in the way which diffeomorphism 
classes are not realized by Calabi--Yau threefolds. 
} %
 of the modulo-diffeomorphism classifications of Calabi--Yau
threefolds $M$ in the $\rho =1$ series. This is
a continuation of the program of \cite{EW19}.\footnote{
While the computation of $d_{111}$, $(c_2)_1$ and $\chi(M)$
in terms of $\{ n_\gamma \}$, $\{ m_\gamma \}$ and $d_\gamma(\nu)$'s 
in (\ref{eq:d111-k=2}--\ref{eq:chiMatch-k=2}),
(\ref{eq:d111-k=3}--\ref{eq:chiMatch-k=3}),
(\ref{eq:d111-k=4}--\ref{eq:chiMatch-k=4}),
(\ref{eq:d111-k=5}--\ref{eq:chiMatch-k=5}),
(\ref{eq:d111-k=6}--\ref{eq:chiMatch-k=6}) and
(\ref{eq:d111-k=11}--\ref{eq:chiMatch-k=11}) follows
the procedure in \cite{EW19}, it was necessary in this article 
to determine the vector-valued modular forms $\Phi$ and $\Psi$
for fully general $\{ n_\gamma \}$ in the cases with
$k=4,5 ,6,11$; the prescription in \cite{KKRS, MP} in determining
$\Phi$ and $\Psi$ is not enough, so we employed the
brute force order-by-order method mentioned in
section \ref{sssec:build-f*}. 
} %
So, we will give a summary of the results by stating what has changed
from \cite{EW19}, and what remains the same. 

There is just one major change from \cite{EW19}, which
has already been stated in this section \ref{sec:discussion}. 
It was understood in \cite{EW19} that neither the set of
hypothetical 4D ${\cal N}=2$ branches of Heterotic--Type IIA dual vacua 
nor the set of diffeomorphism classes of real six-dimensional manifolds 
contains the other. Now we know by demanding
that the monodromy matrices should be integer valued that
the set of theoretically consistent hypothetical vacuum branches 
should be smaller
than in \cite{EW19}, to the extent that the former is
contained in the latter, when we close our eyes on the caveat
stated at the beginning of section \ref{ssec:mod-diffeo-summary}. 
In a notation similar to the one in \cite{EW19}, the present situation
may be summarized as follows: 
\begin{align*}
  [M_0^\Z(11-\rho/2,\rho_{\Lambda_S})]^{[EW19,\S2],M_{\tilde{g}}^\Z}
 = [M_0^\Z(11-\rho/2,\rho_{\Lambda_S})]^{\rm r.mfd} \subsetneq 
   [M_0^\Z(11-\rho/2,\rho_{\Lambda_S})]^{[EW19,\S2]}, \\ 
  [M_0^\Phi(13-\rho/2,\rho_{\Lambda_S})]^{[EW19,\S3],M_{\tilde{g}}^\Z}
 = [M_0^\Phi(13-\rho/2,\rho_{\Lambda_S})]^{\rm r.mfd} \subsetneq 
   [M_0^\Phi(13-\rho/2,\rho_{\Lambda_S})]^{[EW19,\S3]}. 
\end{align*}

There are two observations in \cite{EW19} that continue to
be true after looking at the cases with degree-4, 6, 8, 10, 12
and degree-22 lattice-polarized K3 fibrations. To state those
observations, let us recall from \cite{EW19} the following facts. 
Although the set of invariants $\chi(M)$, $b_2(M)$, $\mu$ and $c_2(TM)$
specifies the diffeomorphism class of a spin, orientable and simply 
connected real six-dimensional manifold $M$, one may focus on a subset 
of the information that is sensitive to $\Phi$; we then 
deal with real six-dimensional manifolds with a larger equivalence 
relation (coarse classification) than in the modulo-diffeomorphism 
classification. 
The subset of information is equivalent to the lattice $\Lambda_S$ and
the combination $d'_{abc}$ for a class realized by a Calabi--Yau threefold. 
The set of diffeomorphism
classes with a given $(\Lambda_S, d'_{abc})$ is denoted by
${\rm Diff}^{d'}_{\Lambda_S}$; this set provides a finer classification.
To summarize, the modulo-diffeomorphism classification can be split 
into two stages: the coarse classification first, and then the fine 
classification. The coarse vs fine classification of
diffeomorphism classes can also be characterized in terms of the group
cohomology of the isometry group of the lattice $\widetilde{\Lambda}_S$
(see \cite[appendix B]{EW21}). 

It was observed in \cite{EW19} in the case of $\Lambda_S=\vev{+2}$ and
$\Lambda_S=U$ that Calabi--Yau threefolds realize very small fraction
of all the possible entries in the coarse classification. That remains
true in the cases of $\Lambda_S = \vev{+4}$ (quartic K3 fibration),
and all other $\Lambda_S=\vev{+2k}$-polarized cases with $k=3,4,5,6,11$.
This is because $\chi(M) = 24\times 2 - NL_{1,[0]} = - c_0(0) =
 \chi_{\rm match}$ is determined completely by 
the finite number of integers $\{n_\gamma \}$,  
and the combination $d'_{111}$ is also determined by $\{ n_\gamma \}$. 

It was also observed in \cite{EW19} in the case of $\Lambda_S = U$
and $\Lambda_S = \vev{+2}$ that all the diffeomorphism classes
in ${\rm Diff}_{\Lambda_S}^{d'}$ are realized by Calabi--Yau 
threefolds\footnote{
See footnote \ref{fn:caution2}. This caution applies also to 
similar expressions and logic that appear many times in the rest of this 
section. 
} %
when one diffeomorphism class in ${\rm Diff}_{\Lambda_S}^{d'}$ is. 
This remains to be true in all the coarse classified diffeomorphism 
classes with $\Lambda_S = \vev{+2k}$ for $k=2,3,4,5,6,11$,
as we see below. 

Let us first determine the set of diffeomorphism classes
${\rm Diff}_{\Lambda_S}^{d'}$ of one entry in the coarse classification
with $(\Lambda_S, d'_{111})$ in the $\rho =1$ series. With a fixed value
of $d'_{111}$, both $d_{111}$ and $(c_2)_1$ can remain
integers---the first one of Wall's condition (\ref{eq:cond-Wall})---if
they change by 
\begin{align}
 \Delta d_{111} = (k/G) m + 6k (\delta n_1), \qquad 
 \Delta (c_2)_1 = (4/G) m + 24 (\delta n_1), \qquad m \in \Z; 
\end{align}
here, $G := {\rm gcd}(k,4) \in \{1,2,4\}$, and $m \in \Z$ parametrizes 
the variation; variations of $d_{111}$ and $(c_1)_1$ by $\delta n_1 \in \Z$,
on the other hand, can be absorbed by change of basis of $H^2(M;\Z)$, which
means that the variations by $\delta n_1 \in \Z$ correspond to identical
diffeomorphism classes.
One more condition $2d_{111} + (c_2)_1 \equiv 0$
mod 12---the second one of Wall's condition (\ref{eq:cond-Wall})---reduces
the variations $m \in \Z$ to 
\begin{align}
m = (12/m') \tilde{\ell}, \qquad \tilde{\ell} \in \Z, \qquad  
  m' := {\rm gcd}\left( 12, \; 2(k/G) + (4/G)  \right).   
\end{align}
So, the set ${\rm Diff}_{\Lambda_S}^{d'}$ is parametrized by 
$\Delta (c_2)_1 \in (4/G)(12/m')\Z / 24\Z$, or alternatively, 
by $\tilde{\ell} \in \Z/(Gm'/2)\Z = \Z/(6G/(12/m'))\Z$.
How finely the value of $(c_2)_1$ can change over different
diffeomorphism classes, $(4/G)(12/m')$, has been computed
and is shown in the 4th row of Table \ref{tab:Diff-fine}. 
\begin{table}[tbp]
\begin{center}
\begin{tabular}{c|cccccc|cccc|c|c}
  $k$ & 1 & 2 & 3 & 4 & 5 & 6 & 7 & 8 & 9 & 10 & 11 & 12 \\
\hline 
  $G$ & 1 & 2 & 1 & 4 & 1 & 2 & 1 & 4 & 1 & 2 &  1 & 4 \\
  $(12/m')$ & 2 & 3 & 6 & 4 & 6 & 3 & 2 & 12 & 6 & 1 & 6 & 12 \\
  $(4/G)(12/m')$ & 8 & 6 & 24 & 4 & 24 & 6 & 8 & 12 & 24 & 2 & 24 & 12 \\
  \# ${\rm Diff}_{\Lambda_S}^{d'}$ & 3 & 4 & 1 & 6 & 1 & 4 & 3 & 2 & 1 & 12 & 1 & 2  \\
 \hline 
  $[\Delta (c_2)_1]_{\rm phys.}$ &  8 & 6 & 24 & 4 & 24 & 6 & ? & ? & ? & ? & 24 & ?  
\end{tabular}
\caption{\label{tab:Diff-fine}
On the number of diffeomorphism classes sharing one $\Phi$, from two 
perspectives. On one hand, it follows from Wall's theorem that 
the number of diffeomorphism classes with a give set of 
$(\chi(M), d'=d'_{\rm match})$ (the 5th row in this table) 
is the number of elements of $(4/G)(12/m')\Z / 24\Z$ (24 divided by 
the 4th row), as we explain in the main text. 
The 2nd--5th rows depend only on $k$ mod 12. 
On the other hand, the 6th row is based on our analysis in 
section \ref{sec:classification}, and indicate how finely 
the observable $(c_2)_1$ can be scanned by theoretically consistent 
4D ${\cal N}=2$ vacua. In all the cases we have studied (including the 
case $k=1$ in \cite{EW19}), the 4th row and the 6th row agree. 
}
\end{center}
\end{table}

On the other hand, we may also read out how finely the value of
$(c_2)_1$ can vary over distinct hypothetical branches of the 
vacuum moduli space; that is to read out in the formulae such
as (\ref{eq:c2-k=2}, \ref{eq:c2-k=3}, \ref{eq:c2-k=4}) how $(c_2)_1$
depends on $\ell_{\rm aux}$, while paying attention to the quantization
condition on $\ell_{\rm aux}$. The result is shown in the 6th row
of Table \ref{tab:Diff-fine}. By comparing the 4th and 6th rows of
the table, we find that all the diffeomorphism classes in
${\rm Diff}_{\Lambda_S}^{d'}$ can be realized by Calabi--Yau threefolds,
whenever there is $\Phi$ that realizes the invariant $(\chi(M), d')$ 
of a diffeomorphism class of a real six-dimensional manifold $M$. 
It is tempting to speculate, for a general lattice $\Lambda_S$ with 
$\rho \geq 1$, that all the diffeomorphism classes with a given pair 
of invariant $(\chi(M), d')$ are realized by Calabi--Yau threefolds 
whenever there is $\Phi$ for $(\chi(M), d')$. 

\subsection*{Acknowledgments}

We thank J. Gray, S. Kondo and Y. Tachikawa for useful comments. 
This work is supported by JSPS Fellowship for Young Scientists (YE, YS), 
the FMSP program (YE), the IGPEES program (YS), the WPI Initiative (all), 
and a Grant-in-Aid for Scientific Research on Innovative Areas 6003, MEXT, 
Japan. 

\appendix 

\section{The Graft Method Illustrated}
\label{sec:grafting}

In the case of $k=1$, it is known how to express the weight-6 
meromorphic cuspform $f_*(t)$ for $\Gamma_0(k)$ as a ratio 
\begin{align}
 f_*(t) = \frac{1}{(2\pi i)^3} \frac{{\rm Numr}f_*}{{\rm Denm}f_*}
   \label{eq:f*-as-a-ratio-of-MF}
\end{align}
of a weight-$w_D$ holomorphic modular form ${\rm Denm}f_*$
and a weight-$(6+w_D)$ holomorphic modular form ${\rm Numr}f_*$; 
the expression for ${\rm Denm}f_*$ and ${\rm Numr}f_*$, originally 
due to \cite{KapLusThs}, is given as eq. (3.31) and (3.46) in \cite{EW21}. 
All the poles of order three associated with the massless matter 
fields are given by the zeros of the modular form ${\rm Denm}f_*$. 

We introduced the graft method in section \ref{sssec:build-f*}
for its practical benefit. It allows us to compute the power series 
expansion of $f_*$ efficiently by exploiting the available codes  
computing the power series expansion of ${\rm Numr}f_*$. 
The method, however, involves the determination of the modular 
form ${\rm Numr}f_*$, and hence an expression (\ref{eq:f*-as-a-ratio-of-MF})
is obtained along the way for $k\geq 1$. Although we are no longer able 
to specify the modular forms ${\rm Denm}f_*$ and ${\rm Numr}f_*$ 
within the algebra $\C[E_4, E_6]$ in the cases with $k >1$, one can still 
specify them within the algebra of modular forms for $\Gamma_0(k)$,  
relatively to a basis provided by the Hecke newforms, their oldforms and 
the Eisenstein series. 

We will explain in the appendix \ref{ssec:def-eq-MMS} how to find 
the modular form ${\rm Denm}f_*$ so that all the poles of $f_*$ are 
captured by the zeros of ${\rm Denm}f_*$.  The case $k=2$ is used 
to illustrate how ${\rm Numr}f_*$ is determined in the 
appendix \ref{ssec:fix-Numrf*-k=2}.

In the case of $k=2$ with $n_0 = -2$ and $n_1=n_2=0$, an analytic
expression for $f_*$ is known \cite{AP}:
\begin{align}
 f_*(t) =  - \frac{640}{5(2\pi i)^{6}}
     \frac{(\partial_t h)^3}{(h(t)-h(i/\sqrt{2}))^3} 
     \frac{5h(t) + 3h(i/\sqrt{2})}{h(t)^2},  
  \label{eq:f*-AP95-k=2-n12=0}
\end{align}
where 
\begin{align}
  h(t) := \frac{[\theta_3(t)^4+\theta_4(t)^4]^4}{16[\eta(t)\eta(2t)]^8}
  = q^{-1} + 104 + 4372 q+ \cdots,  
     \label{eq:AP95-def-h}
\end{align}
with the convention $\theta_{3/4} = 1 \pm 2 q^{1/2} + 2 q^2 \pm q^{9/2} + \cdots$.  
Along the presentation in this appendix, we will also explain how the 
expression above can be interpreted in the form 
of (\ref{eq:f*-as-a-ratio-of-MF}).

\subsection{The Defining Equation of Massless Matter Singularities}
\label{ssec:def-eq-MMS}

The infinitely many massless matter singularities 
${\cal P}_{\gamma, D_{|\gamma|}} \subset {\cal H}$ form a finite number of 
orbits of $\Gamma_0(k)$. To find out a function ${\rm Denm}f_*$ on ${\cal H}$
that vanishes at all those points, we should fully exploit the 
$\Gamma_0(k)$ action. The compactification of an open Riemann surface 
$\Gamma_0(k)\backslash {\cal H}$ is denoted by $X_0(k)$, and is called 
a modular curve. The finite number of $\Gamma_0(k)$ orbits of the 
massless matter singularities are regarded as a finite number 
of points\footnote{
Such points $P_{\gamma,D} := \Gamma_0(k) \backslash {\cal P}_{\gamma,D}$
in $X_0(k)$ for an integer $D<0$ are called {\it Heegner points} in math 
literatures \cite{HeegnerPt, GKZ-I}; in this article, we deal with
a special class of them, those with $D=D_{|\gamma|}$ in the range $[-4k,0)$.
} %
 $P_{\gamma,D_{|\gamma|}} = \Gamma_0(k) \backslash {\cal P}_{\gamma,D_{|\gamma|}}$
in $X_0(k)$. So, the question of finding ${\rm Denm}f_*$ is translated 
to a question of finding a section on the curve $X_0(k)$ that vanishes at 
a finite number of points within $P_{\gamma, D_{|\gamma|}}$. 

One can exploit standard reasonings in algebraic geometry when the curve 
$X_0(k)$ is realized as an algebraic variety, not as a compactification 
of a complex analytic manifold. One way to implement $X_0(k)$ as an algebraic 
variety is to use 
\begin{align}
 \phi_{j,j_k}:  {\cal H} \ni t \longmapsto [1: j(t): j(kt)] \in \P^2, 
\end{align}
which is known to factor through $\Gamma_0(k)\backslash {\cal H}$. 
We do not use this $\phi_{j,j_k}(X_0(k))$ implementation in this article, 
because the image $\phi_{j,j_k}(X_0(k))$ is a higher degree curve in $\P^2$ 
with many singularities; there are multiple points in $X_0(k)$ mapped by 
$\phi_{j,j_k}$ to one point in $\P^2$. Instead, we find it useful to use 
the map associated with\footnote{
It is also a very standard technique in algebraic geometry 
to construct a map/embedding/algebraic realization from an algebraic 
variety/Riemann manifold by using the vector space of sections of 
certain bundles. Application of this technique to the modular curves 
includes \cite{MC-defeq}. One of the authors (TW) thanks S. Kondo
for discussions on a related topics. 
} %
 the vector space $M_w(\Gamma_0(k))$ of modular 
forms of a certain weight $w$, 
\begin{align}
  \Phi_{k,w}: {\cal H} \ni t \longmapsto
     [f_{k.w.1}(t): f_{k.w.2}(t): \cdots : f_{k.w.\ell'}] \in \P^{\ell'-1}, 
  \label{eq:idea-MFspace-4-algC-realiztn}
\end{align}
where $\ell' := \dim_\C M_w(\Gamma_0(k))$, and 
$\{ f_{k.w.1},\cdots, f_{k.w.\ell'}\}$ is a basis of the vector space. 

Let us illustrate how one can find ${\rm Denm}f_*$ by
using (\ref{eq:idea-MFspace-4-algC-realiztn}), by working on the
two cases $k=2$ and $11$. It is known that $X_0(2)$ and $X_0(11)$ are
curves of genus 0 and 1, respectively.  

\subsubsection{Level-2} 

We find it useful in the context of this article\footnote{
Because $\dim_\C M_w(\Gamma_0(k))=0$ for an odd weight $w$, 
only an even weight $w$ is an option. 
When $w=2$, the target space of the map $\Phi_{k,w}$ is a point because 
$\dim_\C M_w(\Gamma_0(k))=1$. The map with $w=4$ is just as useful 
as $w=8$ for an algebraic implementation of the curve $X_0(2)$.  
But the version with $w=8$ is more suitable in understanding the 
analytic expression (\ref{eq:f*-AP95-k=2-n12=0}, \ref{eq:AP95-def-h}) 
of \cite{AP}. 
} %
 to use the map $\Phi_{k,w}$ with $w=8$ in the case of the level $k=2$. 
We may use the following three weight-$(w=8)$ level-$(k=2)$ modular forms 
\begin{align}
f_{2,8,1}(t) := E_8(t) & \; = 1+480 (q + 129 q^2 + 2188 q^2 + \cdots), \\
f_{2,8,2}(t) := (E_8|[\diag(2,1)]_8)(t) & \;
     = (1 + 480 (q^2+ 129 q^4+ \cdots )\times 16, \\
f_{2,8,3}(t) := [\eta(t)\eta(2t)]^8 &\; = q-8q^2+12q^3 + 64 q^4 + \cdots
\end{align}
as a basis of the $(\ell'=3)$-dimensional vector space $M_8(\Gamma_0(2))$. 
Here, 
\begin{align}
 (f|[g]_w)(t) := \frac{({\rm det}(g))^{w/2}}{(ct+d)^w} f(t^g), \qquad \quad 
    g = \left( \begin{array}{cc} a & b \\ c & d \end{array} \right)
      \in {\rm GL}_2\R
  \label{eq:def-|opr}
\end{align}
for $t \in {\cal H}$. 
The image $\Phi_{2,8}(X_0(2)) \subset \P^2$ satisfies\footnote{
One may use the dimension formula (or \cite{SAGE}) to find out that 
$\dim_\C M_{16}(\Gamma_0(2)) = 5$. So, there must be one linear relation 
among the six quadratic monomials of three variables $f_{2.8.*}$ here. 
It is enough to check the Fourier coefficients of more than 
$\dim_\C(M_{16}(\Gamma_0(2)))$ terms to be confident with the
relation (\ref{eq:def-eq-mod-C-2-8}). 
}
 the quadratic equation 
\begin{align}
f_{2,8,1}^2  + f_{2,8,2}^2 -(16^{-1} + 16) f_{2,8,1}f_{2,8,2}
+ 2(f_{2,8,1}+f_{2.8.2})(3600f_{2,8,3}) + (3600 f_{2,8,3})^2=0.
  \label{eq:def-eq-mod-C-2-8}
\end{align}
So, the image $C_{2.8} := \Phi_{2,8}(X_0(2)) \subset \P^2$ 
is a conic. One can also verify that this map $\Phi_{2.8}: X_0(2) \rightarrow 
C_{2.8}$ is degree-1 (i.e., not a multiple covering), and the image is 
non-singular, so this is an embedding. $X_0(2) \cong C_{2.8}$. 
The image $C_{2.8}$ being a conic (and hence $g(C_{2.8})=0$) is consistent 
with the fact that $X_0(2)$ is of genus 0. 
The Fricke involution acts as exchange of the two 
homogeneous coordinates $f_{2,8,1}$ and $f_{2,8,2}$ of the ambient space $\P^2$, 
under which $C_{2.8}$ is mapped to itself non-trivially.  

The modular function $h(t)$ for $\Gamma_0(2)$ given by (\ref{eq:AP95-def-h})
is regarded as the ratio of a pair of homogeneous function of degree-1 
on the ambient space $\P^2$, 
\begin{align}
  h(t) = \frac{f_{2,8,1} + f_{2,8,2} + 1152 f_{2,8,3}}{17 f_{2,8,3}},
     \label{eq:def2-haupt-fcn-k=2}
\end{align}
because 
\begin{align}
 \frac{f_{2,8,1} + f_{2,8,2} + 1152 f_{2,8,3}}{17}
 = 1 + 96 q + 3552 q^2 + 62592 q^3 + \cdots 
   = \frac{[\theta_3^4+\theta_4^4]^4}{16}.  
\end{align}
The zero locus of the numerator and the denominator
of (\ref{eq:def2-haupt-fcn-k=2}) are 
lines in the ambient space $\P^2$, which intersect with $C_{2.8} \cong X_0(2)$
at two points. The rational function $h$ on $C_{2.8} \cong X_0(2)$ 
therefore has two zeros and two poles. One can also see 
that both the numerators and the denominator are invariant under 
the Fricke involution $f_{2.8.1} \leftrightarrow f_{2.8.2}$, so 
the two zeros form a pair under the Fricke involution, and so 
do the two poles in $X_0(2)$. Such properties of the function $h(t)$ 
can also be worked out by using the original expression (\ref{eq:AP95-def-h})
and a fundamental region of $\Gamma_0(2)$ in ${\cal H}$, without 
using the language of algebraic geometry here. In a similar study for 
higher level cases, however, the language of algebraic geometry 
makes the analysis easier and understanding more clear cut. 

There massless matter singularities with $\gamma = [0]$, $\pm [1]$ and 
$[2]$ form one, two and one $\Gamma_0(2)$ orbits, respectively; those orbits 
are represented by 
\begin{align}
t^{[0]}=i/\sqrt{2} \in {\cal P}_{0,-8}, & \qquad
     v_{[0]}=(1,1,0) \in {\rm e.ch}_{0,-8}, \\
t^{|[1]|}_{\pm} = \frac{\mp 1+\sqrt{7}i}{4} \in {\cal P}_{\pm 1,-7}, & \qquad
    v_{[ \pm 1]} = (1,1, \pm 1) \in {\rm e.ch}_{\pm 1, -7}, \\
t^{[2]} = \frac{-1+i}{2} \in {\cal P}_{2,-4}, & \qquad 
    v_{[2]} = (1,1,2) \in {\rm e.ch}_{2,-4}.
\end{align}
This $\Gamma_0(2)$-orbit decomposition is consistent 
with \cite[\S I.1 p.505, Prop.]{GKZ-II}.

Any hyperplane of $\P^2$ invariant under the Fricke involution is 
given by the zero locus of 
\begin{align}
  H_{2.8}(t;r) :=
    \frac{f_{2.8.1}(t)+f_{2.8.2}(t) + 1152 f_{2.8.3}}{17}- r f_{2.8.3}(t).
  \label{eq:def-denm-factor-k=2}   
\end{align}
When $H_{2.8}(t;r)$ is restricted on $C_{2.8}$, it vanishes at two points
that form a pair under the Fricke involution. We find by 
numerically\footnote{
\label{fn:Gal}
It is known \cite{HeegnerPt, GKZ-I} that defining equations of the
massless matter singularities 
$\phi_{j,j_k}(P_{\gamma;D_{|\gamma|}})$ (in fact, those of the Heegner points 
$\phi_{j,j_k}(P_{\gamma,D})$ more generally) in the curve $\phi_{j,j_k}(X_0(k))$
can be chosen so that all the coefficients are in $\Q$. 
It is conceivable that the same is true for their images in such 
algebraic realizations of $X_0(k)$ as $C_{2.8}$, and it is even likely 
that an algorithm for computation of the values of the coefficient
(such as $r \simeq 256$, $r \simeq 81$ and $r \simeq 0$ here) by 
exploiting the definition of the Heegner points (analogously 
to \cite[\S3]{StevenH} for CM points). We do not try to establish such
a proof or 
algorithm in this article, and are content with ourselves by determining 
the coefficients numerically and confirming that they are very close to  
rational numbers (such as 256, 81 and 0). 
} %
evaluating the value of the modular forms $f_{2.8.1\sim 3}$ at the 
massless matter singularities above that 
\begin{align}
  H^{(k=2)}_{|[1]|} := H_{2.8}(t;81) = 0
\end{align}
is the defining equation of the two points $P_{\pm 1,-7} \subset C_{2.8}$. 
Similarly, we find that 
\begin{align}
 H^{(2)}_{[0]} := H_{2.8}(t;256)=0, \qquad 
 H^{(2)}_{[2]} := H_{2.8}(t;0) = 0
   \label{eq:set-denm-factor-k=2}
\end{align}
on the curve $C_{2.8}$ have zero at the one point 
$P_{[0],-8}$ and $P_{[2],-4}$, respectively, with multiplicity two. 
They are the defining equations of the massless matter singularities. 
The notation $H^{(k)}_{|\gamma|}$ may be replaced by $H_{|\gamma|}$ without
a reference to the level $k$, when there is small risk of confusion. 

Because the point $t^{[2]} \in {\cal H}$ and its $\Gamma_0(2)$-images 
are fixed under one non-trivial element of $\Gamma_0(2)/\{ \pm 1\}$, 
the modular form $H^{(2)}_{[2]}$ has a zero of order two at 
$\Gamma_0(2)t^{[2]} \in X_0(2)$, but has a zero of order four at 
$t=t^{[2]} \in {\cal H}$. When we set 
\begin{align}
 {\rm Denm}f_*(t) = (H_{[0]}(t))^2 (H_{|[1]|}(t))^3 H_{[2]}(t), 
\end{align}
it is a holomorphic modular form of weight-$(w_D=48)$ for $\Gamma_0(2)$ 
that vanishes at $\Gamma_0(2)t^{[0]}$, $\Gamma_0(2) t^{|[1]|}_\pm$ and 
$\Gamma_0(2)t^{[2]}$ at order 4, 3, and 4, respectively. The modular form 
${\rm Numr}f_*$ should be of weight-54 that has $\Gamma_0(2)t^{[0]}$ 
and $\Gamma_0(2)t^{[2]}$ as zeros of order one in order to reproduce $f_*$; 
we will determine ${\rm Numr}f_*$ in the appendix \ref{ssec:fix-Numrf*-k=2}. 

\subsubsection{Level-11}

The modular curve $X_0(11)$ can be realized as an algebraic curve 
in a weighted projective space in a way almost the same as the 
embedding (\ref{eq:idea-MFspace-4-algC-realiztn}). To get started, 
one can use the dimension formula or SAGE \cite{SAGE} to find out 
\begin{align}
 \dim_\C (M_2(\Gamma_0(11)))=2, \quad 
 \dim_\C (M_4(\Gamma_0(11)))=4, \quad 
 \dim_\C (M_8(\Gamma_0(11)))=8.
\end{align}
We choose the following two modular forms 
\begin{align}
 f_{11.2.1} := [\eta(t)\eta(11t)]^2, \qquad 
 f_{11.2.2} := \frac{E_2|[\diag(11,1)]_2 - E_2}{11-1} - \frac{24}{10}f_{11.2.1}
\end{align}
as a basis of the vector space $M_2(\Gamma_0(11))$. Their quadratic monomials 
can generate 3-dimensional subspace of $M_4(\Gamma_0(11))$; as one more 
independent weight-4 modular form, we choose 
\begin{align}
 f_{11.4}(t):= \frac{E_4|[\diag(11,1)]_4-E_4}{11^2-1}. 
\end{align}
One may think of the following map:
\begin{align}
  \Phi_{11.2+4}: 
    {\cal H} \ni t \longmapsto [f_{11.2.1}(t): f_{11.2.2}(t): f_{11.4}(t)]
     \in WP^2_{[1:1:2]},  
\end{align}
where the target space is the weighted projective space. 

The image $\Phi_{11.2+4}(X_0(11)) \subset WP^2_{[1:1:2]}$ must be a curve, 
which we denote by $C_{11.2+4}$. Its defining equation is 
\begin{align}
 y^2 - (x+4)(x^3 -8x^2 -56 x - 76) = 0, 
\end{align}
where $x = f_{11.2.2}/f_{11.2.1}$ and $y = f_{11.4}/f_{11.2.1}^2$ are 
the inhomogeneous coordinates of $WP^2_{[1:1:2]}$. 
One may be confident with the defining equation of $C_{11.2+4}$ 
by checking the Fourier coefficients of the left hand side times 
$f_{11.2.1}^4$ for more than $\dim_\C(M_8(\Gamma_0(11)))$ terms. 
The curve $C_{11.2+4}$ is non-singular, and the map $\Phi_{11.2+4}: X_0(11)
 \rightarrow C_{11.2+4}$ is an isomorphism. 

All of the modular forms $f_{11.2.1}$, $f_{11.2.2}$ and $f_{11.4}$ 
for $\Gamma_0(11)$ are odd under the Fricke involution. 
So, the inhomogeneous coordinate $x$ is even, and the Fricke 
involution acts on $C_{11.2+4}$ as $y \mapsto -y$. 

The massless matter singularities in ${\cal P}_{\gamma,D_{|\gamma|}}$ and 
${\cal P}_{-\gamma,D_{|\gamma|}}$ combined [resp. $P_{\gamma, D_{|\gamma|}}$ and 
$P_{-\gamma,D_{|\gamma|}}$ combined] are invariant under the Fricke 
involution acting on ${\cal H}$ [resp. on $X_0(k)$]. 
We found\footnote{
We determined the coefficients numerically. For example, 
in (\ref{eq:def-eq-MMS-k=11-gamma-1}), the ratio $f_{11.2.2}/f_{11.2.1}$ 
was evaluated at $t^{|[1]|}_\pm  = (\mp 1 + \sqrt{43}i)/22 
\in {\cal P}_{\pm 1, -43}$ and found a common number that is very close to 12. 
See footnote \ref{fn:Gal}.
} %
 that the defining equations of 
$P_{\gamma, D_{|\gamma|}} \cup P_{-\gamma,D_{|\gamma|}}$ in $C_{11.2+4}$ are 
\begin{align}
 H_{|[1]|}^{(11)}(t) & \; := f_{11.2.2}-12 f_{11.2.1} = 0,
        \label{eq:def-eq-MMS-k=11-gamma-1} \\
 H_{|[4]|}(t) & \; := (f_{11.2.2}-3f_{11.2.1}) H_{|[9]|} = 0, \\
 H_{|[5]|}(t) & \; := f_{11.2.2} = 0, \\
 H_{|[6]|}(t) & \; := f_{11.2.2}+2f_{11.2.1} = 0, \\
 H_{|[9]|}(t) & \; := f_{11.2.2}+3 f_{11.2.1} = 0, \\
 H_{[11]}(t) & \; := f_{11.2.2}+4f_{11.2.1} = 0,
\end{align}
for $\gamma$'s where only Heegner points with class number 1 are involved, 
\begin{align}
 H_{|[2]|} & \; := f_{11.2.2}^2-6f_{11.2.2}f_{11.2.1}-36 f_{11.2.1}^2 = 0, \\
 H_{|[3]|} & \; := f_{11.2.2}^2-4f_{11.2.2}f_{11.2.1}-16 f_{11.2.1}^2 = 0, \\
 H_{|[8]|} & \; := f_{11.2.2}^2+6f_{11.2.2}f_{11.2.1}+12 f_{11.2.1}^2 = 0, \\
 H_{|[10]|} & \; := ( f_{11.2.2}^2+10f_{11.2.2}f_{11.2.1}+26 f_{11.2.1}^2 ) H_{|[6]|} = 0,
\end{align}
for $\gamma$'s where Heegner points with class number 2 are also involved, 
and 
\begin{align}
 H_{|[7]|} := f_{11.2.2}^4 + 9 f_{11.2.2}^3f_{11.2.1}+45 f_{11.2.2}^2f_{11.2.1}^2+117 f_{11.2.2}f_{11.2.1}^3 + 117 f_{11.2.1}^4 = 0
\end{align}
for the 4+4 points $P_{7,-39} \cup P_{-7,-39} \subset C_{11.2+4}$. 

The massless matter singularities $P_{0,-44} \subset C_{11.2+4}$ consist 
of four points that are all characterized by the condition $y=0$. So, 
we can choose the following as the defining equation of $P_{0,-44}$: 
\begin{align}
 H_{[0]} := f_{11.4} = 0.
\end{align}

In the language of algebraic geometry, where
$C_{11.2+4}$ is an elliptic curve in $WP^2_{[1:1:2]}$, it is immediately
clear that those $H_{|\gamma|}$'s do not have a pole on $C_{11.2+4}$, and
have precisely the same number of zeros as
$P_{\gamma,D_{|\gamma|}}\cup P_{-\gamma,D_{|\gamma|}}$. 

Therefore, we may choose 
\begin{align}
  {\rm Denm}f_* = \ell{\rm cm} \left\{ (H_{|\gamma|}^{(k)}(t))^3 \;
    | \; |\gamma| \in G_S/((-1)\times) \; {\rm s.t.} \; n_\gamma \neq 0 \right\};
\end{align}
then all the poles of order 3 of $f_*$ are captured by the zeros 
of this ${\rm Denm}f_*$. 

\subsection{The Numerator in the Level $k=2$ Case}
\label{ssec:fix-Numrf*-k=2}

The weight-6 meromorphic cuspform $f_*$ in the $k=2$ case has the 
following power series expansion for general $n_0$, $n_1$ and $n_2$:
\begin{align}
(2\pi i)^3 f_* & \; = (-320n_0+17n_1+32n_2)q
  + (-161792n_0  + 155716 n_1 - 5120 n_2) q^2 \nonumber \\
 &\; + (-35041536n_0 - 28035 n_1 + 319104 n_2)q^3  \nonumber \\
 &\; + (-5622398976n_0 - 3083501296 n_1 - 14319616 n_2) q^4  \nonumber \\
 &\; + (- 773858608000n_0 - 5918325 n_1 + 544900800 n_2) q^5  \nonumber \\
 &\; + (- 96988412141568n_0 + 30097977494004 n_1 - 18779148288 n_2) q^6
       \nonumber \\
 &\; + ( - 11411884533944832n_0+ 813224335 n_1 + 605762467072 n_2) q^7 + O(q^8); 
\end{align}
the way to compute this is explained in section \ref{sssec:build-f*}.

Let us identify an appropriate modular form ${\rm Numr}f_*$ so that 
the power series above is reproduced through (\ref{eq:f*-as-a-ratio-of-MF}). 
Because we know that 
$f_*$ is linear with respect to $\{ n_\gamma \}$, and that 
poles of $f_*$ at ${\cal P}_{\gamma,D_{|\gamma|}} \cup {\cal P}_{-\gamma,D_{|\gamma|}}$
are due to non-zero $n_\gamma$, we may deal with the expression 
 (\ref{eq:f*-as-a-ratio-of-MF}) in its partial fraction expansion, 
with individual components labeled by a subset of 
$\{ | \gamma | \} = G_S/((-1)\times)$ whose $n_{\gamma}$'s are 
independent.\footnote{
For some values of $k$, there are linear relations among $\{ n_\gamma \}$ 
such as (\ref{eq:lin-rltn-k=11}). 
} %
In the case of $k=2$, all of $|\gamma|=0,1,2$ are independent, so 
we need to determine the $\Gamma_0(2)$ modular forms ${\rm Numr}f_{*}^{[0]}$, 
${\rm Numr}f_{*}^{|[1]|}$ and ${\rm Numr}f_{*}^{[2]}$ of weight 22, 30 and 14, 
respectively, so that  
\begin{align}
 f_* = \frac{1}{(2\pi i)^3 } \left( 
   n_0 \frac{{\rm Numr}f_{*}^{[0]}}{(H_{[0]}(t))^2}
 + n_1 \frac{{\rm Numr}f_{*}^{|[1]|}}{(H_{|[1]|}(t))^3}
 + n_2 \frac{{\rm Numr}f_{*}^{[2]}}{H_{[2]}(t)} \right).
  \label{eq:f*-as-ratio-PFD-k=2}
\end{align}

Let us present a little bit of details in the determination 
of the modular form ${\rm Numr}f_{*}^{[0]}$. It must be 
\begin{align}
 {\rm Numr}f_{*}^{[0]} & \; = (2\pi i)^3 f_*|_{n_0=1,n_{1,2}=0} \times (H_{[0]}(t))^2, 
   \nonumber  \\ 
   & \; = -320 q -59392 q^2 + 4955904 q^3 + 172294144 q^4 + \cdots .
     \label{eq:k=2-f*-gamma0-numerator-series}
\end{align}
The vector space $S_{22}(\Gamma_0(2))$ of weight-22 cuspforms for 
$\Gamma_0(2)$ is of 4-dimensions\footnote{
By using SAGE, we see that $\dim_\C[M_{22}(\Gamma_0(2))] = 6$, 
$\dim_\C[S_{22}(\Gamma_0(2))]=4$, $\dim_\C[S_{22}(\Gamma_0(2))]^{\rm new}=2$. 
} %
 over $\C$, and we can choose the following cuspforms as a basis:
\begin{align}
 \phi_{o.1.\pm}^{(22)} & \; = 1728
     \left( E_4E_6 \eta^{24} \pm (E_4E_6\eta^{24})|[\diag(2,1)]_{22} \right)
\end{align} 
are oldforms, and  
\begin{align}
 \phi_{n.1.\pm}^{(22)} & \; = q \mp 2^{10} q^2 + (65460 \pm 2^{10} \cdot 6)q^3
    + 2^{20} q^4 + \cdots  
\end{align}
are newforms. Both $\phi_{o.1.+}^{(22)}$ and $\phi_{n.1.+}^{(22)}$ are 
even under the Fricke involution, while $\phi_{o.1.-}^{(22)}$ and 
$\phi_{n.1.-}^{(22)}$ are odd. 
One can find by examining more than four coefficients
that ${\rm Numr}f_{*}^{[0]}$ in (\ref{eq:k=2-f*-gamma0-numerator-series})
is within the vector space of weight-22 
cuspforms,\footnote{
\label{fn:cusp-k=2}
In the 2-dimensional space $M_{22}(\Gamma_0(2))/S_{22}(\Gamma_0(2))$, 
we can choose a basis represented by 
\begin{align*}
 \phi_{E.1.\pm} & \; = E_6^3 E_4 \pm (E_6^3 E_4)|[\diag(2,1)]_{22}.
\end{align*}
The Fricke-invariant ${\rm Numr}f_{*}^{[0]}$ could contain 
the $\phi_{E.1.+}$ component, but then ${\rm Numr}f_{*}^{[0]}$ would not 
vanish exponentially at $t \sim i \infty$. So, we see that 
${\rm Numr}f_{*}^{[0]}$ does not have a component outside $S_{22}(\Gamma_0(2))$. 
This argument works in all the cases with
$\dim_\C M_w(\Gamma_0(k)) = \dim_\C S_w(\Gamma_0(k))+2$.
More generally, however, we do not have a top down justification (yet)
that ${\rm Numr}f_*$ should be in $S_{w_D+6}(\Gamma_0(k))$ rather than
$M_{w_D+6}(\Gamma_0(k))$ (cf footnote \ref{fn:cusp});  
There are cases with $\dim_\C M_{w_D+6}(\Gamma_0(k))/S_{w_D+6}(\Gamma_0(k)) > 2$
for $k=4$ and $6$. So, in those cases, we have to determine the power
series coefficients of $(2\pi i)^3 f_*^{|\gamma|} \times {\rm Denm}f_*^{|\gamma|}$
for terms more than $\dim_\C M_{w_D+6}(\Gamma_0(k))$ (rather than
$\dim_\C S_{w_D+6}(\Gamma_0(k))$) to identify it within the vector space
$M_{w_D+6}(\Gamma_0(k))$ (in practice, we can work on the subspace with
an appropriate eigenvalue under the Fricke involution). For all the
$\gamma$'s in the cases with $k=4$ and $6$, ${\rm Numr}f_*^{|\gamma|}$
turns out to be within $S_{w_D+6}(\Gamma_0(k))$. 
} 
and in particular, 
\begin{align}
 {\rm Numr}f_{*}^{[0]} = - \frac{5248}{29}\phi_{n.1.+}^{(22)}
      - \frac{7}{87} \phi_{o.1.+}^{(22)}. 
  \label{eq:Numrf*-k=2-[0]}
\end{align}
It is expected that ${\rm Numr}f_{*}^{[0]}$ is in the Fricke-involution 
invariant subspace, because $(H_{[0]})^2$ is invariant under the 
Fricke involution, and their ratio $(2\pi i)^3f_*$ should also be 
invariant. Now, by combining (\ref{eq:Numrf*-k=2-[0]}, 
\ref{eq:f*-as-ratio-PFD-k=2}) and (\ref{eq:def-denm-factor-k=2}, 
\ref{eq:set-denm-factor-k=2}), we have a conceptually unambiguous 
expression for $f_*$ with $k=2$ and $n_{1,2}=0$ alternative to 
(\ref{eq:f*-AP95-k=2-n12=0}). In both expressions, the poles 
of $f_*(t)$ at the massless matter singularities ${\cal P}_{0,-8} 
\subset {\cal H}$ originate from $h(t)-h(i/\sqrt{2}) \propto H_{[0]}(t)$. 

The same algorithm works for all other $k$ and $\gamma$. 
It is also possible to determine the Fourier coefficients of 
the modular forms forming a basis of the vector space
$M_{w_D+6}(\Gamma_0(k))$ up to very higher order terms 
by using a code \cite{SAGE} available in public, as explained 
in the main text.


\section{Integral Monodromy Matrices and the SU(2) Witten Anomaly}
\label{sec:Witten}

It has been observed \cite{ESW20} that the SU(2) Witten anomaly in the 
4D field theory is not guaranteed to vanish, when we impose the conditions 
that (a) the invariants $\Phi$ and $\Psi$ transform properly under the 
Heterotic string worldsheet modular transformation, (b) all the 
Fourier coefficients of $\Phi$ and $\Psi$ satisfy the integrality 
conditions discussed in \cite{EW19}, and (c) the monodromy matrix 
$M_{\tilde{g}_\infty}$ of the Peccei--Quinn transformation ($t \rightarrow t+1$)
are integer valued. So, this is a sign that some aspects of theoretical 
consistency conditions cannot be captured by the combination 
of (a), (b) and (c). We will see in this appendix \ref{sec:Witten}
that the condition that the monodromy matrix $M_{\widetilde{R}_{v_*}}$ 
of the Weyl reflection $R_{v_*}$ associated with the SU(2) enhanced 
gauge symmetry should be integer valued guarantees that the Witten SU(2) 
anomaly in the 4D effective field theory vanishes.\footnote{
\label{fn:Yuji}
The argument in this appendix \ref{sec:Witten} is due to a discussion 
in early 2021 by two of the present authors (Y.S. and T.W.) 
with Y. Tachikawa, which led to Ref. \cite{STW22}. 
} %
 
Let 
\begin{align}
    M_{\tilde{R}_{v*}}:=\left(\begin{array}{cc}
        R_v & 0\\
        (R_v^T)^{-1}\Lambda_{\widetilde{R}_{v_*}} & (R_v^T)^{-1}
    \end{array}\right)
\end{align}
be the monodromy matrix associated with $R_{v_*}$. 
On one hand, we may compute the matrix $\Lambda_{\widetilde{R}_{v_*}}$
numerically in terms of the BPS classification invariants  
in the case $\rho =1$, as we did in this article. 
On the other hand, it is also possible to see that the SU(2) Witten 
anomaly in the 4D effective theory vanishes, whenever the 
$(\rho+2) \times (\rho+2)$ matrix $\Lambda_{\widetilde{R}_{v_*}}$ 
is integer valued (for a general $\rho$). 
We will do the latter in two different (but much the same) ways 
in the following. When the two approaches are combined, we obtain 
constraints on the BPS classification invariants that was anticipated 
(and hoped for) in \cite{ESW20}. Although sections \ref{sec:PP}
and \ref{sec:classification} of this article do not cover the
$\Lambda_S=\vev{+28}$ case discussed in \cite{ESW20}, we are sure
(because of the argument in this appendix) that an analysis as
in section \ref{sssec:level-6} for the $k=14$ case will yield
the consistency condition on $\{ n_\gamma \}$ so that the
4D spectrum is free from the Witten SU(2) anomaly. 

Now, let us exploit the fact that $R_{v_*}^2= {\rm id}_{(\rho+2)\times (\rho+2)}$
and the monodromy $(M_{\widetilde{R}_{v_*}})^2$ corresponds to a loop that goes 
around the massless matter divisors of the charges proportional to $v_*$
by the phase $+2\pi$. So, we get the relation 
\begin{align}
    \label{Mv2}
    (M_{\widetilde{R}_{v_*}})^2=\left(\begin{array}{cc}
        1&0\\
        \Delta F&1
    \end{array}\right).  
\end{align}
The non-trivial part of the monodromy $\Delta F$ is determined from 
the logarithmic running of the holomorphic gauge
coupling \cite{AFGNT},\footnote{
See Ref. \cite{EW21} for all the subtleties in the normalization conventions. 
An error by a factor of 2 in keeping track of normalization conventions 
would invalidate the whole argument here.  
}\raisebox{4pt}{,}\footnote{
The relation (\ref{eq:FandB}) used as a crucial input in the argument
of this appendix is based on reasonings in 4D field theory. As an alternative,
it may also be possible to derive the expression for $\Delta F$ by dealing
with the period polynomial of $(R_{v_*})^2$ as the residue integral along
the loop $(\widetilde{\gamma}_{\tilde{R}_{v_*}})^{R_{v*}}
\circ \tilde{\gamma}_{\tilde{R}_{v*}}$ in
$D(\widetilde{\Lambda}_S) \backslash X_{\rm singl.}$. 
} %
\begin{align}
    \Delta F=-\frac{b}{2}v_*\otimes v_*; 
  \label{eq:FandB}
\end{align}
we will say more about the normalization convention of the 
1-loop beta function $b$ of the SU(2) gauge coupling later. 
The lower-left block of the matrix relation (\ref{Mv2}) reads 
\begin{align}
    (R_{v*})^T\Lambda_{\widetilde{R}_{v*}} (R_{v*})
 + \Lambda_{\widetilde{R}_{v*}} = \Delta F . 
  \label{eq:bilinForm-involv-DeltaF}
\end{align}

First, it is not difficult to derive that the coefficient $b$ must be 
an integer for the matrix $\Lambda_{\widetilde{R}_{v_*}}$ to be integer valued. 
Indeed, think of (\ref{eq:bilinForm-involv-DeltaF}) as a relation 
among the bilinear forms on $\widetilde{\Lambda}_S$, and 
multiply $\ell_{KM} v_*$ from both sides to obtain one relation among scalars. 
The left hand side is even because the two terms are equal 
(use $R_{v_*}v_*=- v_*$). The right hand side is $-2b$. 
Therefore, $2b$ must be an even integer, which means that $b$ must be an 
integer. 

The 1-loop beta function $b$ used already in (\ref{eq:FandB}) is in 
the normalization 
\[
 \frac{\partial}{\partial \ln (\mu)} \frac{1}{\alpha(\mu)} = \frac{-b}{2\pi}, 
   \qquad b = T_{R(R+R^{cc})} - 2T_G, 
\]
conventional in 4D non-abelian gauge theories; $T_G$ is the dual Coxeter 
number of the gauge group, $R$ ($R+R^{cc}$) are the representation of the 
gauge group in which matter half (full) hypermultiplets are in, and 
$T_R$ the Dynkin indices of those representations.  The SU(2) 
Witten anomaly is absent if and only if the totality of $T_{R(R+R^{cc})}$ 
is an integer. So, the consistency condition that (d) the monodromy over 
the 4D ${\cal N}=2$ Coulomb branch should be realized as an integer-valued 
symplectic transformation on the electric and magnetic charges 
guarantees that the 4D ${\cal N}=2$ effective field theory is free 
from the SU(2) Witten anomaly. 

Secondly, and alternatively, we will have more physical intuition 
on what is being done after (\ref{eq:bilinForm-involv-DeltaF}) 
when we extract a $2\times 2$ monodromy matrix of the Seiberg--Witten 
theory from the $2(\rho+2) \times 2(\rho+2)$ monodromy matrix 
$M_{\widetilde{R}_{v_*}}$ on all the charged states in the ${\cal N}=2$ 
supergravity. In a theory where the 4D SU(2) gauge theory contains 
an SU(2) doublet matter field,\footnote{
In an SU(2) gauge theory without a matter field in a half-odd isospin
representation, there is no Witten SU(2) anomaly in the first place. 
So, we assume that $v_1 = v_*/2$ is the primitive vector 
in $\widetilde{\Lambda}_S^\vee$, not $v_*$ is, in the rest of this 
appendix \ref{sec:Witten}.
} %
 the set of electric charges 
$\Lambda_{\rm el} = \widetilde{\Lambda}^\vee_S$ contains $\Z v_1$, 
where $v_1 := v_*/2$ is the primitive vector in $\widetilde{\Lambda}_S^\vee$
 corresponding to the Cartan-U(1) charge of matter fields in the SU(2) 
doublet representation. The set of 
magnetic charges $\Lambda_{\rm mg} = \widetilde{\Lambda}_S$, on the other 
hand, contains $\Z (\ell_{KM} v_*)$. The charge $\ell_{KM} v_* \in 
\widetilde{\Lambda}_S$ must be a primitive element because 
$(\ell_{KM} v_*, v_1) = -1$; it saturates the Dirac quantization condition; 
there is also a conjecture \cite{BS} (completeness hypothesis) 
that all the charges in a string vacuum are realized by some objects 
(applied to $\ell_{KM} v_* \in \Lambda_{\rm mg}$ in the present context). 
These are a good indication that we can 
interpret $\ell_{KM}v_*$ as the Cartan-U(1) magnetic charge of the 
monopoles/dyons in the Seiberg--Witten theory. 
So, we wish to see how the matrix $M_{\widetilde{R}_{v_*}}$ acts 
(by multiplication from the right) on the rank-2 subset of electric
and magnetic charges $\Z v_1 \oplus \Z(\ell_{KM}v_*) \subset
 \Lambda_{\rm el} \oplus \Lambda_{\rm mg}$.  

It is 
\begin{align}
    \left(\begin{array}{cc}
        -1 & 0\\
        (***)  & -1
    \end{array}\right) = M_{\infty}, \qquad \qquad (***) = - b;
\end{align}
we explain why the lower-left component is $b$. The monodromy by 
$M_{\widetilde{R}_{v_*}}$ yields an additional electric charge 
$v_{m} \cdot (R_{v_*}^T)^{-1} \Lambda_{\tilde{R}_{v*}}$ for a state
with a magnetic charge $v_m$. So, the monopole charge $v_m = \ell_{KM} v_*$ 
is accompanied by the additional electric charge 
$(\ell_{KM} v_*) \cdot R_{v_*}^T \Lambda_{\widetilde{R}_{v_*}}$ after the 
adiabatic motion along the loop $\gamma_{\widetilde{R}_{v_*}}$ in the Coulomb 
branch moduli space. The component of the additional electric charge 
proportional to $v_1$ can be measured by 
\begin{align}
  (***) = 
  (\ell_{KM} v_*) R_{v_*}^T \Lambda_{\widetilde{R}_{v_*}} v_1
       \; \frac{1}{(v_1,v_1)}. 
\end{align}
Now, we may use (\ref{eq:bilinForm-involv-DeltaF}, \ref{eq:FandB}) 
and (\ref{eq:reflection_conditions}) to see that 
\begin{align*}
 -2 (***) & \; = (\ell_{KM} v_*) \left[
     R_{v_*}^T \Lambda_{\widetilde{R}_{v_*}} R_{v_*} + \Lambda_{\widetilde{R}_{v_*}}
     \right] v_1 \; \frac{1}{(v_1,v_1)} = 2b . 
\end{align*}
So, the monodromy matrix $M_{\widetilde{R}_{v_*}}$
of all the electric and magnetic charges contains the $2\times 2$ 
monodromy matrix $M_\infty$ at the SU(2) weak coupling region in the 
Seiberg--Witten theory. The argument right
after (\ref{eq:bilinForm-involv-DeltaF}) is essentially to demand
that the matrix $M_\infty$ is integer valued, and find that $b$ has
to be an integer. 

For the appropriate duality transformation 
of those charges, with all the charges populated in string theory, 
there is no chance for the 1-loop beta function to be non-integral; 
the SU(2) Witten anomaly vanishes as a consequence. The observation
that the consistency condition (d) provides constraints that are not
captured by the conditions (a), (b) and (c) led to the study 
in Ref. \cite{EW21} and in this article.

\end{document}